\documentclass[twocolumn,tighten]{aastex62}
\usepackage{apjfonts}
\usepackage{mathrsfs}
\usepackage{amsmath}
\usepackage{url}
\usepackage[normalem]{ulem}
\usepackage{graphicx}

\usepackage{natbib}
\usepackage{color}
\bibpunct{(}{)}{;}{a}{}{,}

\newcommand\aproxgt{\mathrel{%
      \rlap{\raise 0.511ex \hbox{$>$}}{\lower 0.511ex \hbox{$\sim$}}}}
\newcommand\aproxlt{\mathrel{%
      \rlap{\raise 0.511ex \hbox{$<$}}{\lower 0.511ex \hbox{$\sim$}}}}

\newcommand{\ignore}[1]{}

\def\ir1334{{IRAS\,13349+2438}}
\def\mcg6{{MCG--6-30-15}}

\def\hst{{\it HST}}

\def\apjs{ApJS}
\def\aap{A\&A}

\def\lya{Ly$\alpha$}

\def\cii{{\sc C~ii}}

\def\civ{{\sc C~iv}}

\def\nv{{\sc N~v}}
\def\oi{{\sc O~i}}

\def\siiv{Si~{\sc iv}}

\def\kmps{\ifmmode \rm km~s^{-1} \else $\rm km~s^{-1}$\fi}
\def\psqcm{\ifmmode \rm cm^{-2} \else $\rm cm^{-2}$\fi}
\def\Msun{\ifmmode M_{\odot} \else $M_{\odot}$\fi}
\def\Lsun{\ifmmode L_{\odot} \else $L_{\odot}$\fi}

\newcommand{\qo}{\ifmmode q_{\rm o} \else $q_{\rm o}$\fi}
\newcommand{\Ho}{\ifmmode H_{\rm o} \else $H_{\rm o}$\fi}
\newcommand{\ho}{\ifmmode h_{\rm o} \else $h_{\rm o}$\fi}

\def\fake2{\hphantom{3}}

\shorttitle{AGN STORM. VIII. Modeling the UV Spectrum of NGC 5548}
\shortauthors{Kriss et al.}

\begin{document}

\received{03/29/2019}
\revised{06/27/2019}
\accepted{07/06/2019}

\title{Space Telescope and Optical Reverberation Mapping Project.\\
VIII.\ Time Variability of Emission and Absorption in NGC\,5548 Based on Modeling the Ultraviolet Spectrum
}

\author{G.~A.~Kriss}
\affiliation{Space Telescope Science Institute, 3700 San Martin Drive, Baltimore, MD 21218, USA}
\author{G.~De~Rosa}
\affiliation{Space Telescope Science Institute, 3700 San Martin Drive, Baltimore, MD 21218, USA}
\author{J.~Ely}
\affiliation{Space Telescope Science Institute, 3700 San Martin Drive, Baltimore, MD 21218, USA}
\author{B.~M.~Peterson}
\affiliation{Space Telescope Science Institute, 3700 San Martin Drive, Baltimore, MD 21218, USA}
\affiliation{Department of Astronomy, The Ohio State University, 140 W 18th Ave, Columbus, OH 43210, USA}
\affiliation{Center for Cosmology and AstroParticle Physics, The Ohio State University, 191 West Woodruff Ave, Columbus, OH 43210, USA}
\author{J.~Kaastra}
\affiliation{\ignore{SRON}SRON Netherlands Institute for Space Research, Sorbonnelaan 2, 3584 CA Utrecht, The Netherlands}
\affiliation{\ignore{Leiden}Leiden Observatory, Leiden University, PO Box 9513, 2300 RA Leiden, The Netherlands}
\author{M.~Mehdipour}
\affiliation{\ignore{SRON}SRON Netherlands Institute for Space Research, Sorbonnelaan 2, 3584 CA Utrecht, The Netherlands}
\author{G.~J.~Ferland}
\affiliation{\ignore{UKy}Department of Physics and Astronomy, The University of Kentucky, Lexington, KY 40506, USA}
\author{M.~Dehghanian}
\affiliation{\ignore{UKy}Department of Physics and Astronomy, The University of Kentucky, Lexington, KY 40506, USA}
\author{S.~Mathur}
\affiliation{Department of Astronomy, The Ohio State University, 140 W 18th Ave, Columbus, OH 43210, USA}
\affiliation{Center for Cosmology and AstroParticle Physics, The Ohio State University, 191 West Woodruff Ave, Columbus, OH 43210, USA}
\author{R. Edelson}
\affiliation{\ignore{Maryland}Department of Astronomy, University of Maryland, College Park, MD 20742, USA}
% ...

%
\begin{abstract} 
We model the ultraviolet spectra of the Seyfert 1 galaxy NGC~5548
obtained with the {\it Hubble Space Telescope}
during the 6-month reverberation-mapping campaign in 2014.
Our model of the emission from NGC 5548
corrects for overlying absorption and deblends the individual emission lines.
Using the modeled spectra, we measure the response to continuum variations for
the deblended and absorption-corrected individual broad emission lines,
the velocity-dependent profiles of Ly$\alpha$ and
\ion{C}{4}, and the narrow and broad intrinsic absorption features.
We find that the time lags for the corrected emission lines are comparable to
those for the original data.
The velocity-binned lag profiles of  Ly$\alpha$ and
\ion{C}{4} have a double-peaked structure indicative of a truncated
Keplerian disk.
The narrow absorption lines show delayed response to continuum variations
corresponding to recombination in gas with a density of $\sim 10^5~\rm cm^{-3}$.
The high-ionization narrow absorption lines decorrelate from continuum
variations during the same period as the broad emission lines.
Analyzing the response of these absorption lines during this period
shows that the ionizing flux is diminished in strength relative to the
far-ultraviolet continuum. The broad absorption lines associated with the
X-ray obscurer decrease in strength during this same time interval.
The appearance of X-ray obscuration in $\sim\,2012$ corresponds with an increase
in the luminosity of NGC 5548 following an extended low state.
We suggest that the obscurer is a disk wind triggered by the brightening of
NGC 5548 following the decrease in size of the broad-line region during the
preceding low-luminosity state.
\end{abstract}



\section{Introduction}
\label{sec:intro}

Quantitatively measuring the geometry, kinematics, and physical conditions
in the structures at the centers of active galactic nuclei (AGN) is essential
for understanding how their activity is fueled by inflowing gas, and how
outflows from the central engine may influence the surrounding host galaxy.
Even in the nearest AGN the broad-line region (BLR) and the accretion
disk are nearly impossible to resolve at optical, UV, or X-ray wavelengths,
although the \cite{Sturm18} have used near-IR interferometry to detect the
spatial extent of the BLR of the quasar 3C 273 at an angular size
of $\sim10$ micro-arcseconds.
Reverberation mapping \citep{BM82, Peterson93} provides a powerful technique for
resolving features on physical scales ten times smaller.
A resolution of light days, corresponding to angular scales of micro-arcseconds,
suffices to probe the detailed structure of the BLR and accretion disks of
nearby AGN.

The Seyfert galaxy NGC 5548 has been a prime target of several successful
reverberation mapping campaigns, both in the ultraviolet from space
\citep{Clavel91, Korista95}
and in the optical from the ground
\citep[see][and references therein]{Peterson02}.
These campaigns ascertained the typical size of the BLR is
several light days, and established that it was likely dominated
by virial motions since the size, ionization stratification, and line widths
are all consistent with motions in a gravitational field
\citep{Peterson99}.
These initial campaigns measured only the mean lag and line width for
selected emission lines.
Spurred by these successes and the promise of higher quality data,
more recent campaigns have explored the possibility of measuring lags in
two dimensions with velocity-resolved reverberation mapping
across strong emission-line profiles \citep{Horne04, Bentz10, Grier13}.

Early efforts at velocity-resolved reverberation mapping from the ground were
followed in 2014 by the first attempt to determine high-quality
velocity-resolved delay maps for NGC 5548 by the
AGN Space Telescope and Optical Reverberation Mapping program
\citep[AGN STORM, PI Peterson,][]{DeRosa15}.
This program monitored NGC 5548 on a nearly daily basis for approximately
six months using the {\it Hubble Space Telescope} (\hst), the
{\it Neil Gehrels Swift Observatory}, and several ground-based facilities,
producing a data set of unparalleled quality.
The campaign so far has determined mean lags for the usual bright emission
lines in both the UV \citep{DeRosa15} and the optical \citep{Pei17}, as well
as for the continuum emission from the accretion disk
\citep{Edelson15, Fausnaugh16, Starkey17}.

While the AGN STORM data have exquisite quality, NGC 5548 exhibited some
rather anomalous behavior over the course of the campaign.
As described by \cite{DeRosa15},  the first and second
halves of the campaign showed different mean lags for the emission lines.
\cite{Goad16} trace this behavior in a more detailed way to a decoupling
in the response of the broad-line fluxes from the
variations in the far-ultraviolet (FUV) continuum, meaning that the line
fluxes stopped exhibiting the usual linear correlation with the continuum flux.
This decoupling began $\sim$75 days into the campaign, and lasted for
another $\sim64$ days.
We refer to this time period when the broad emission lines failed to respond to
variations in the continuum flux as the BLR ``holiday".

Straightforward interpretation of data from a reverberation mapping campaign
rests on four basic assumptions:

\begin{enumerate}
\item The illuminating continuum originates from a centrally located point
that is much smaller than the BLR, and it radiates isotropically.
\item The central source and the illuminated gas occupy a small fraction of the
volume encompassed by the BLR, and the continuum radiation propagates freely
at the speed of light throughout this volume.
\item The observed continuum and its variations are an accurate proxy for
the ionizing radiation illuminating the BLR.
\item The light travel time across the BLR is the most important timescale.
\end{enumerate}
The AGN STORM data have revealed potential issues with all of these
assumptions.
First,
the inter-band continuum lags of up to a few days \citep{Edelson15, Fausnaugh16}
indicate a
continuum region that is not point-like, and even comparable in size to the
shortest emission-line lags exhibited by \ion{He}{2} $\lambda 1640$
\citep{DeRosa15, Pei17}.
Second,
heavy intrinsic absorption
affects the blue wings of the most prominent emission lines--
Ly$\alpha$ $\lambda 1216$, \ion{C}{4} $\lambda\lambda 1548,1550$,
\ion{N}{5} $\lambda 1238,1242$ and \ion{Si}{4} $\lambda 1393,1402$.
This absorption consists of a broad component
(full width at half maximum of $\sim2500~\rm km~s^{-1}$)
associated with the X-ray obscurer discovered by \cite{Kaastra14}, plus the
known narrow UV absorption features in NGC~5548 \citep{Crenshaw09}.
The obscurer and the associated broad UV absorption are variable
\citep{Kaastra14, DiGesu15}, and they appear to shadow the more distant gas
producing the narrow UV absorption \citep{Arav15}, rendering those
absorption lines variable as well.
Third,
the decoupling of the emission-line responses from the continuum variations
\citep{Goad16} indicates a potential conflict with the third assumption.
Fourth,
changes in the covering factor of the absorbing gas occur on timescales
of days \citep{DiGesu15},
comparable to the light-travel time within the BLR.

The absorbing gas outflowing from the central engine of NGC 5548 is also a key
element of its nuclear structure.
The blue-shifted narrow intrinsic UV absorption lines associated with the
X-ray warm absorber have been studied in detail for decades.
A close association between the UV absorption features and the X-ray warm
absorber was first proposed by \cite{Mathur95}.
\cite{Mathur99} resolved the UV absorption into six distinct components
ranging in velocity from $+250~\rm km~s^{-1}$ to $-1165~\rm km~s^{-1}$, and
with full-width at half maximum (FWHM) ranging from
$40~\rm km~s^{-1}$ to $300~\rm km~s^{-1}$.
Following their convention, these six components are enumerated starting at the
highest blueshift as Component \#1 to Component \#6.
Subsequent observations noted the variability of these features in response to
changes in the UV continuum flux \citep{Crenshaw03,Crenshaw09,Arav15}, and
\cite{Arav15} used the variability, the density-sensitive absorption lines
of \ion{C}{3}* and \ion{P}{3}*, and photoionization modeling to locate
the UV absorbing gas at distances ranging from 3 pc to $>$100 pc.
The X-ray absorbing gas is similarly complex, both kinematically and in its
ionization distribution \citep{Kaastra02,Steenbrugge05,Kaastra14,Ebrero16}.
Although it has been difficult to link the X-ray absorbing gas to the UV
absorbing gas definitively \citep{Mathur95,Crenshaw03,Crenshaw09},
the apparently common spatial locations \citep{Krongold10} and
close kinematic correspondence \citep{Arav15} suggest they are part of the
same outflow.

The fast, broad obscuring outflow discovered by \cite{Kaastra14} is a new
component of the nuclear structure of NGC 5548.
The initial five observations of the {\it XMM-Newton} campaign
suggested that the obscurer produced both the X-ray absorption and the
broad UV absorption, and that it was located in or near the BLR
\citep{Kaastra14}. This location permits it to shadow the more distant
absorbing gas producing the narrow UV absorption lines \citep{Arav15}
and the X-ray warm absorbers \citep{Ebrero16}.
These hypotheses can be studied in greater depth with the extensive data from
AGN STORM.

The AGN STORM campaign provides detailed monitoring of the variations of the
UV absorption components, both broad and narrow,
 in response to changes in the UV and X-ray flux.
To measure these variations in absorption and to mitigate their influence
on the measurement and interpretation of the emission lines and continuum of
NGC~5548 during our campaign, we have modeled the UV spectra.
Our model has the additional virtue that it deblends some of the more closely
spaced emission lines, e.g., \ion{N}{5} from Ly$\alpha$, and \ion{He}{2}
from \ion{C}{4}, so that we can produce two-dimensional reverberation maps
over a wider range in velocity in each of the overlapping wavelength regions.
In the process of removing the absorption, we also measure its strength,
giving us an additional probe of continuum behavior
along our line of sight. Since the absorption lines respond directly to
changes in the ionizing flux, they provide an independent measure of its
strength, which we use to reconstruct the true behavior of the
ionizing continuum during the period of the BLR holiday.

The modeled spectrum and the measurements we extract from it enable a wealth of
studies which we only begin to touch in this paper.
In \S2 we briefly describe the observations and initial data reduction.
The spectral model we develop in \S3 is time dependent.
We first describe the static model derived from the high signal-to-noise ratio
(S/N) mean spectrum in \S3.1. Then in \S3.2 we describe how we adapt this to
model the whole time series of observations comprising the campaign.
In \S3.3 we describe how we estimate the uncertainties we measure using our
time-dependent models.
In \S3.4 we describe various tests we made to assess the quality and
reliability of our procedures.
With the modeled spectra in hand, we then describe how we make
measurements using the models.
This includes describing the absorption-corrected spectra in \S3.5,
how we measure fluxes in the deblended emission lines in \S3.6, and
how we measure the absorption lines in \S3.7.
Using these measurements, in \S4 we perform an initial analysis of our results,
including velocity-resolved light curves for the deblended emission lines
in \S4.1, and
the physical characteristics we infer for the gas producing the intrinsic
narrow and broad absorption lines based on the mean spectrum in \S4.2.
In \S4.3 we analyze the variability of the intrinsic narrow absorption lines,
and in \S4.4 the
variability of the intrinsic broad absorption features associated with the
obscurer.
In \S5 we discuss the implications of our results for the structure and
evolution of the BLR in NGC 5548, and
in \S6 we summarize our major results.

\section{Observations and Data Reduction}

Our observational program is described in detail by \cite{DeRosa15}.
Summarizing briefly,
we observed NGC 5548 using the Cosmic Origins Spectrograph
\citep[COS,][]{Green12} on HST using 
daily single-orbit visits from 2014 February 1 through July 27.
Out of 179 observations, 171 executed successfully.
To cover the full spectral range of the COS medium-resolution gratings, we
used multiple central wavelength and FP-POS settings.
Each visit used two different settings for gratings G130M and G160M.
These settings covered the wavelength range  1153--1796\,\AA\ in all visits.
Different settings on each day over an 8-visit cycle enabled us to
sample a broader spectral range regularly, 1130--1810\,\AA, over the course
of the whole program.
This strategy also minimized damage and charge extraction from the COS
detectors over the duration of our program.
This broader spectral coverage, particularly on the blue wavelength end,
enabled us to sample the \ion{P}{5} $\lambda 1128$ absorption line, an
important tracer of high-column density absorbing gas \citep{Arav15}.
Our spectra on each visit exceeded a minimum signal-to-noise ratio (S/N) of
$>20$ in the continuum at 1367 \AA\ when measured over $100~\rm km~s^{-1}$ bins.
\noindent
$\phantom{0000000000}$\\
$\phantom{0000000000}$\\

\subsection{Data Reduction}
\label{sec:data_reduction}
As described by \cite{DeRosa15}, we used
the {\tt CalCOS} pipeline v2.21 to process our data, but we made a special
effort to enhance the calibration files for our particular observations.
We developed special flat-field files, enhanced the wavelength calibration
through comparison to prior STIS observations of NGC~5548, and tracked the
time-dependent sensitivity of COS to achieve higher S/N and
better flux reproducibility.
For each day, all four exposures were calibrated, aligned in wavelength
and combined into a single spectrum for each grating for each visit.
We binned these spectra by 4 pixels (approximately half a resolution element)
to reduce residual pattern-noise features and achieve higher S/N.
Ultimately, our wavelength scale achieves a root-mean-square precision of
$< 6~\rm km~s^{-1}$, and our flux reproducibility for G130M is
better than 1.1\%, and better than 1.4\% for G160M.

\section{The Time-Dependent Spectral Model}
\label{sec:model}

\subsection{Modeling the Mean Spectrum}
\label{sec:meanmodel}
To produce a spectral model that can be adjusted to each individual observation
in the reverberation campaign, we start with the high-quality mean spectrum
produced from the whole campaign data set.
This reveals individual weak features which we cannot precisely track in
individual observations, but which we must include to avoid biasing our
results for stronger features of interest.
The model of the mean spectrum is described in detail by \cite{DeRosa15}, but
we also describe it here once again as a key reference for understanding the
components that we vary in the fits to the individual spectra in our
campaign's time series.

The basic model starts with the one developed for the {\it XMM-Newton}
campaign of \cite{Kaastra14}, where the soft X-ray obscuration and broad UV
absorption was first discovered. This model used a powerlaw continuum and
multiple Gaussian components for both the emission and the broad absorption
features.
The high S/N of the mean spectrum from the reverberation mapping campaign
requires the addition of more weak emission features, additional weak
broad absorption features associated with all permitted transitions in the
spectrum, and more components in the bright emission lines.
We emphasize that the model is not intended as a physical characterization of
the spectrum, but rather as an empirical tool that enables us to deblend
emission-line components and correct for absorption by using some simple
assumptions about the shape of the spectrum.

Starting with the continuum, we use a powerlaw with
$F_\lambda(\lambda)  = F_\lambda(1000\, \mbox{\AA}) (\lambda / 1000\,\mbox{\AA})^{- \alpha}$.
Although \cite{Kraemer98} find a modest amount of internal extinction in the
narrow-line region of NGC\,5548, $E(\bv)=0.07^{+0.09}_{-0.06}$, the continua of
Type 1 AGN in general show little extinction \citep{Hopkins04}, and none was
required in fitting the broad-band spectral energy distribution of
NGC 5548 \citep{Mehdipour15}.
Given that our continuum model is simply an empirical characterization of its
shape, we assume there is no internal extinction in NGC\,5548, and we
redden the powerlaw with Galactic foreground extinction of $E(\bv)$ fixed at
0.017\,mag \citep{Schlegel98,Schlafly11}
using the mean Galactic extinction curve of
\cite{CCM89} with $R_V = 3.1$.
Weak, blended \ion{Fe}{2} emission is expected at $\lambda > 1550$ \AA,
which we include as modeled by
\cite{Wills85} and broadened with a Gaussian with full-width at half-maximum 
${\rm FWHM} = 4000\,{\rm km~s}^{-1}$.
This model component is essentially a smooth, low-level addition to the
continuum at long wavelengths, and it has no predicted or observed
spectral features associated with it. Also, as revealed in prior
monitoring campaigns on NGC 5548, the \ion{Fe}{2} varies only weakly
\citep{Krolik91,Vestergaard05} on timescales of weeks.
During AGN STORM, the optical \ion{Fe}{2} emission varied by at most 10\% from
its mean value \citep{Pei17}.
In the UV model we therefore keep its intensity fixed,
and we normalize its flux using the modeled \ion{Fe}{2} emission from the
mean {\it XMM-Newton} Optical Monitor grism spectrum of \cite{Mehdipour15}.

For the emission lines we use multiple Gaussian components.
We do not assign any particular physical significance to most of these
individual kinematic components, especially since there is not a unique
way to decompose these line profiles using such non-orthogonal elements.
However, the narrow components and the intermediate-width components of
Ly$\alpha$, \ion{N}{5}, \ion{C}{4} and \ion{He}{2} are discernible as discrete
entities in prior observations of NGC~5548 in faint states \citep{Crenshaw09}.
Although narrow \ion{Si}{4} was not present in the 2004 STIS spectrum of
\cite{Crenshaw09}, our high S/N mean spectrum requires it, and we include it
in our fit.
Similarly, there is a non-varying intermediate-width component with
FWHM$\sim 800~\rm km~s^{-1}$ in 
Ly$\alpha$, \ion{N}{5}, \ion{Si}{4}, \ion{C}{4} and \ion{He}{2}.
We call this the Intermediate-Line Region (ILR).
In weaker emission lines such as
\ion{C}{3}*\,$\lambda$1176, \ion{Si}{3}\,$\lambda$1260,
\ion{Si}{2}$+$\oi\,$\lambda$1304, \cii\, $\lambda$1335,
\ion{N}{4}]\,$\lambda$1486, \ion{O}{3}]\,$\lambda$1663,
and \ion{N}{3}]\,$\lambda$1750,
this intermediate-width component is the only one observed in the spectrum.

\cite{Crenshaw09} saw little or no evidence for variability of the NLR and ILR
components. We allow these to vary freely in determining our best fit to the
mean spectrum, but for the bright emission lines
(Ly$\alpha$, \ion{N}{5}, \ion{Si}{4}, \ion{C}{4} and \ion{He}{2})
we keep these components fixed when fitting the individual spectra
from the campaign.
For the weaker, lower-ionization emission lines listed above, however,
we allow their flux, central wavelength, and FWHM to vary
since these lines are not heavily blended with other components,

The strongest emission lines (Ly$\alpha$, \ion{N}{5}, \ion{Si}{4}, \ion{C}{4}
and \ion{He}{2}) all require up to three additional broad components.
These have approximate widths of 3000 $\rm km~s^{-1}$ (Broad, or B),
8000 $\rm km~s^{-1}$ (Medium Broad, or MB),
and 15,000 $\rm km~s^{-1}$ (Very Broad, or VB).
The \nv, \siiv, and \civ\ emission lines are all doublets.
We allow for independent narrow, intermediate, broad, and medium-broad
components for each doublet transition. In each case we link their wavelengths
at the ratio of their vacuum values, assign them the same FWHM, and  we assume
their relative fluxes have an optically thick 1:1 ratio.
For the Very Broad component, however, which is much broader than the doublet
separations, we use only a single Gaussian for each ion.

A final set of empirical emission components in our model accounts for weak
bumps on the red and blue wings of \ion{C}{4} and on the red wing of Ly$\alpha$.
These bumps have an interesting variability pattern that we discuss later,
and they are especially visible in the RMS spectrum shown by
\cite{DeRosa15} in their Figures 1 and 2.
We use Gaussian components for each of these features.

We do not model the narrow, intrinsic absorption lines in NGC 5548,
nor the foreground interstellar lines, but we do model the variable intrinsic
broad absorption associated with the obscurer discovered by \cite{Kaastra14}.
The strongest, most easily modeled broad absorption features are on the blue
wings of the Ly$\alpha$, \ion{N}{5}, \ion{Si}{4} and \ion{C}{4} emission lines.
For these, like \cite{Kaastra14},
we use an asymmetric Gaussian with negative flux.
To specify the asymmetry we use a
larger dispersion on the blue side of the central wavelength
than on the red side. We allow the ratio of blue to red dispersion to vary
as a free parameter.
This parameterization produces a rounded triangular shape with the deepest
point in the trough near the red extreme, and a blue wing that extends far out
along the blue wing of the emission line.
\citep[][show these profiles in their Figure 2]{Kaastra14}.
We emphasize that this absorption profile is strictly an empirical
characterization of the observed flux in the spectrum that has no
independent physical meaning.
Deriving physical information requires making further assumptions about which
emission components are covered, producing a transmission profile, and
integrating that profile to obtain the actual opacity.
We discuss these measurements later in \S\ref{sec:measuring_absorbers}.

In addition to these main troughs, additional small depressions appear
further out on the blue wings. We model these with additional symmetric
Gaussians in negative flux.
As with the emission lines, the absorption in 
\nv, \siiv, and \civ\ is due to doublets. These are unresolved, and we assume
they are optically thick. We model each line in the doublet using the same
shape and depth.
Finally, all the UV resonance lines in the spectrum have weak, blue-shifted
absorption troughs at velocities comparable to the main portion of the troughs
observed in Ly$\alpha$, \ion{N}{5}, \ion{Si}{4}, and \ion{C}{4}.
These cannot be modeled in the detail that we apply to the
strongest absorption troughs, and they are only readily apparent in the mean
spectrum. For these troughs we use symmetric Gaussians in negative flux.

The final component of our model is absorption of all model components by
damped \lya \ from the Milky Way. We fix the column density at
$\rm N(HI) = 1.45 \times 10^{20}\,{\rm cm}^{-2}$ \citep{Wakker11}.

Figure \ref{fig:bestfit}\footnote{Figure \ref{fig:bestfit} appears at the end
of the paper to facilitate the formatting.}
gives a detailed view of the best-fit model overlaid on the data.
We illustrate all individual components of the model, the best-fit model
overlaid on the data, and the absorption-corrected model
(which is the ultimate goal of our efforts).
Note that the \ion{C}{4} absorption is not as deep in the mean spectrum from the
reverberation campaign as it was during the deepest phase of the obscuration
as observed in the {\it XMM-Newton} campaign of
\citep[See Figure S1 in][which shows the individual components of the region
surrounding the \ion{C}{4} emission line.]{Kaastra14}
For further illustrations of our model of the mean spectrum, see
Figures 1 and 2 of \cite{DeRosa15}, which compare the model to the full
G130M and G160M spectra.

All the model parameters are listed in Table \ref{tab:parameters}\footnote{Table \ref{tab:parameters} appears at the end
of the paper to facilitate the formatting.}.
The model consists of a total of 97 individual components, each with 2--4
parameters. Although the total number of parameters is 383, many of these
are fixed, or linked to other parameters. Free parameters in the fit total 143.
Although this is a large number, the fit is tightly constrained.
The fitted regions in the mean spectrum comprise $\sim12\,598$ points, each
approximately one-half of a resolution element.
Thus each spectrum has $\sim6\,000$ spectral elements included in the fit,
and that is described by only 143 parameters.

To optimize the parameters of the model and obtain the best fit, we use a
combination of minimization algorithms.
After determining initial guesses by visual inspection, we start the
optimization process using a {\tt simplex} algorithm \citep{Murty83}.
This works well for problems with many parameters that are not initially well
tuned.
Once the fit is nearly optimized, this algorithm generally loses
efficiency in approaching full convergence.
At that point (usually after several tens of iterations) we switch to a 
Levenberg-Marquardt algorithm \citep[as originally coded by][]{Bevington69}.
When close to an optimum fit, this algorithm converges rapidly, but with the
large number of parameters in our fit, it can get stuck in false minima.
To escape these pitfalls, we then alternate sets of 5--10 iterations
using the {\tt simplex} algorithm and the Levenberg-Marquardt algorithm
until the fit has fully converged. We define full convergence as
$\Delta \chi^2 < 0.01$ and a change in each parameter value of $<$1\% after
a set of iterations.

\subsection{Modeling the Whole Time Series}
\label{sec:modelintime}

As we noted above, our Gaussian decomposition of the emission lines is not
unique. Therefore, unless one takes care to preserve the overall character of
the spectral shape of the model from visit to visit among the individual
observations, best fits and parameters can wander far from the character of our
fit to the mean spectrum. One could try to avoid this by tailoring initial
guesses for fits to each individual spectrum interactively,
but this would introduce a unsatisfying degree of subjectivity
into our final results. We therefore employed an approach that used quantitative
characteristics of the spectra to tune each individual fit and guide it to an
optimum result. We verified the soundness of this approach through multiple
trials and experiments before converging on a process that produced
consistent results from spectrum to spectrum without drastic changes in
parameters that might signify unphysical solutions.

In our first trials, we noticed that most weak emission and absorption features
were often too weak to be effectively constrained in a single observation.
We therefore produced a series of grouped spectra that improved the S/N
for our measurements of these weak spectral features.
As described in \cite{DeRosa15}, the observations were done in a cycle of 8,
where central wavelength settings and FP-POS positions were changed on a daily
basis. Combining spectra according to these natural groups produces better S/N,
reduces pattern noise, and provides the full spectral coverage that extends
down to the \ion{P}{5} region at the short wavelength end.
In the fits to the time series we discuss below, we use the values for the
weak features determined from these grouped spectra as our initial guess
for starting parameters when fitting the individual spectra in a group.

Another outcome of our trials, perhaps an obvious one, is that good initial
choices for parameters led to quicker convergence and less chance of solutions
where parameters strayed into unphysical regions of parameter space.
Thus, given our exquisite fit to the mean spectrum, its parameters provide the
best guess for spectra that are similar.
To strengthen this similarity from fit to fit, we tried two different methods.
First, we ordered the spectra by the flux level in the 1367 \AA\ continuum
window. We then chose the spectrum near the middle of this distribution with
a flux most comparable to the flux in the mean spectrum as the first one to
fit. 
The best fit from this spectrum was then used as the initial guesses for
parameters in fitting the next spectra in the series.
The series of fits followed two parallel paths moving both higher and lower
in flux from this initial middle spectrum.
Our second method was to keep the spectra ordered by time, and then pick a
spectrum close to the mid-point of the campaign with a  mean continuum
close to that of the mean spectrum.
Fits then progressed both forward and backward in time from this middle point.

These experiments validated our intentions to develop an objective process for
determining the best fit to each spectrum. Both methods achieved good fits for
all spectra, and the parameters from each method were close in value to
each other (typically within the 1-$\sigma$ uncertainties).
Figure \ref{fig:timevsflux} compares the values obtained as a function of time
for the two different methods for three selected parameters from the model.
In Figure \ref{fig:timevsflux} and in the remainder of the paper, we define
times by the Truncated Heliocentric Julian Date (THJD),
THJD = HJD $-$ 2400000.

\setcounter{figure}{1}
\begin{figure}
\centering
\resizebox{\hsize}{!}{\includegraphics[angle=0, width=0.9\textwidth]{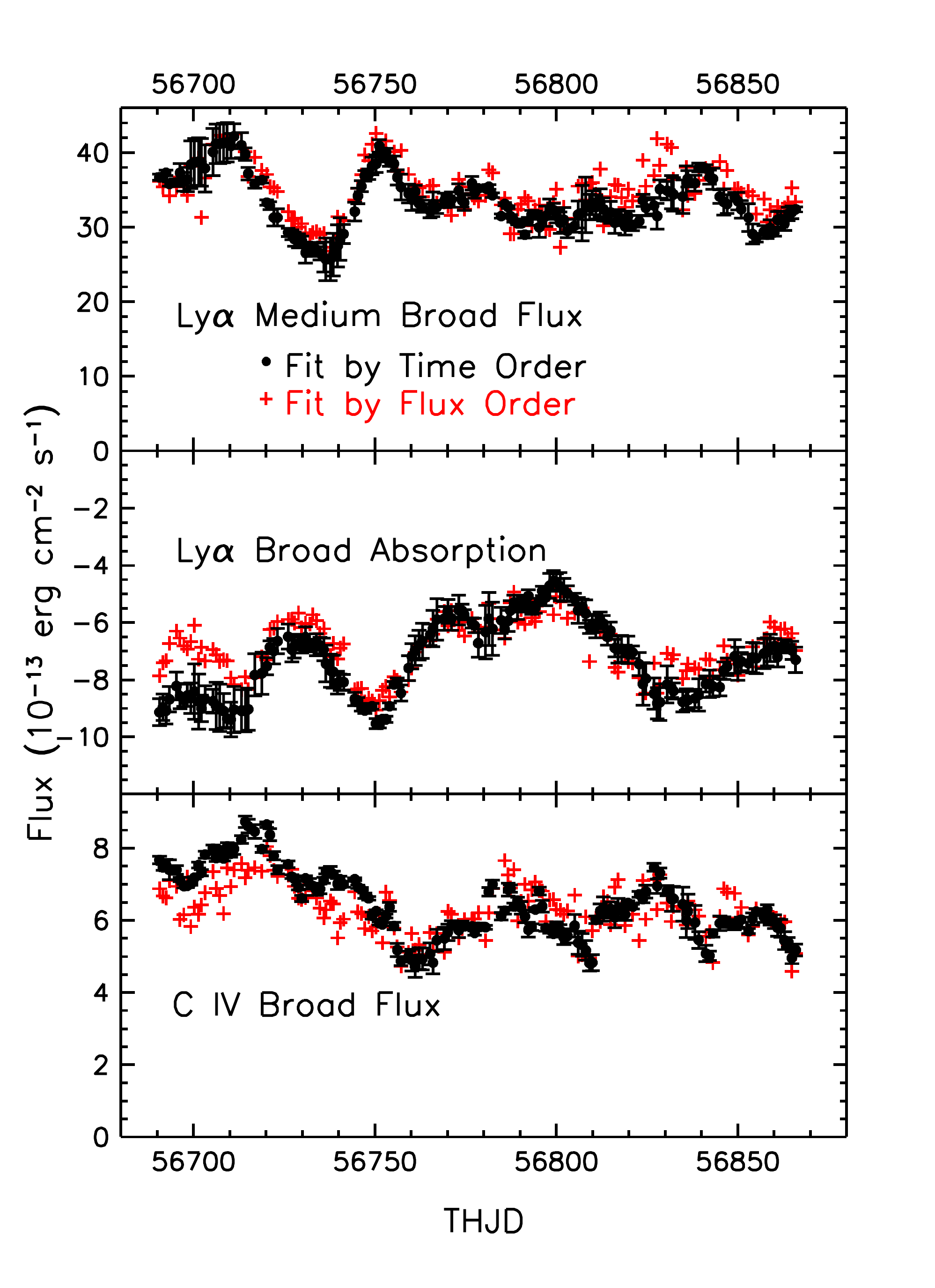}}
\caption{Examples of the evolution of selected parameters comparing values
obtained from fits using time-ordered spectra (black points with error bars)
to values from fits using flux-ordered spectra (red crosses).
Top panel shows the Ly$\alpha$ Medium Broad Flux;
the middle panel shows the Ly$\alpha$ Broad Absorption;
the bottom panel shows the \ion{C}{4} Broad Flux.
}\label{fig:timevsflux}
\end{figure}

\begin{figure}
\centering
\resizebox{\hsize}{!}{\includegraphics[angle=-90, width=0.9\textwidth]{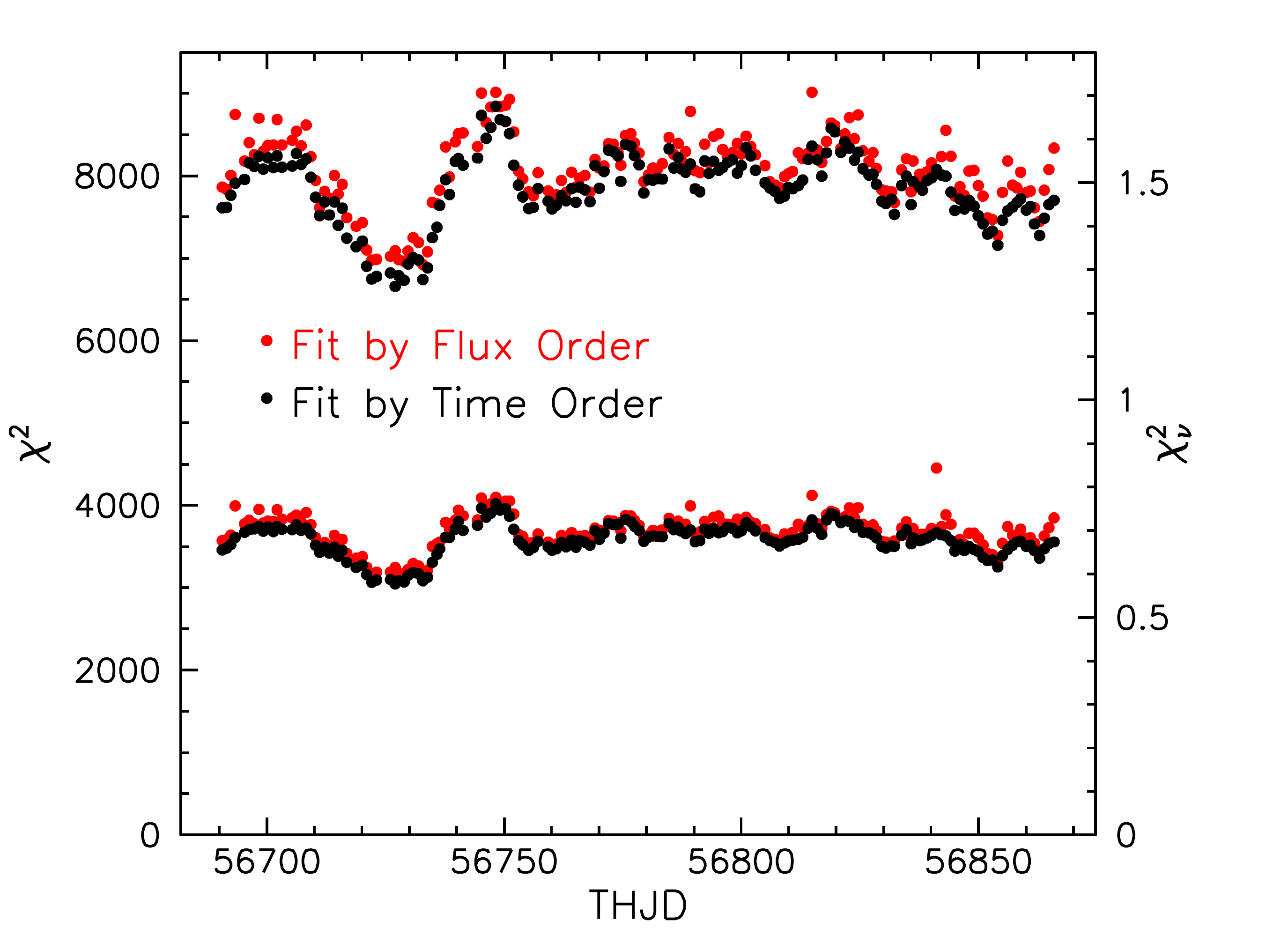}}
\caption{Minimum $\chi^2$ achieved in the fit to each spectrum in the campaign
(upper curves and left axis).
Black points show the results for our adopted method of fitting the sequence of
spectra in time order.
Red points show the results for fitting the spectra when ordered by flux.
Lower curves and the right axis show the reduced $\chi^2$.
The reduced $\chi^2$ is considerably less than one, indicating that our
errors are overestimated, likely due to correlated errors,
as discussed in \S3.4.}\label{fig:chi2}
\end{figure}

\begin{figure*}
\centering
\resizebox{0.9\hsize}{!}{\includegraphics[angle=0, width=0.8\textwidth]{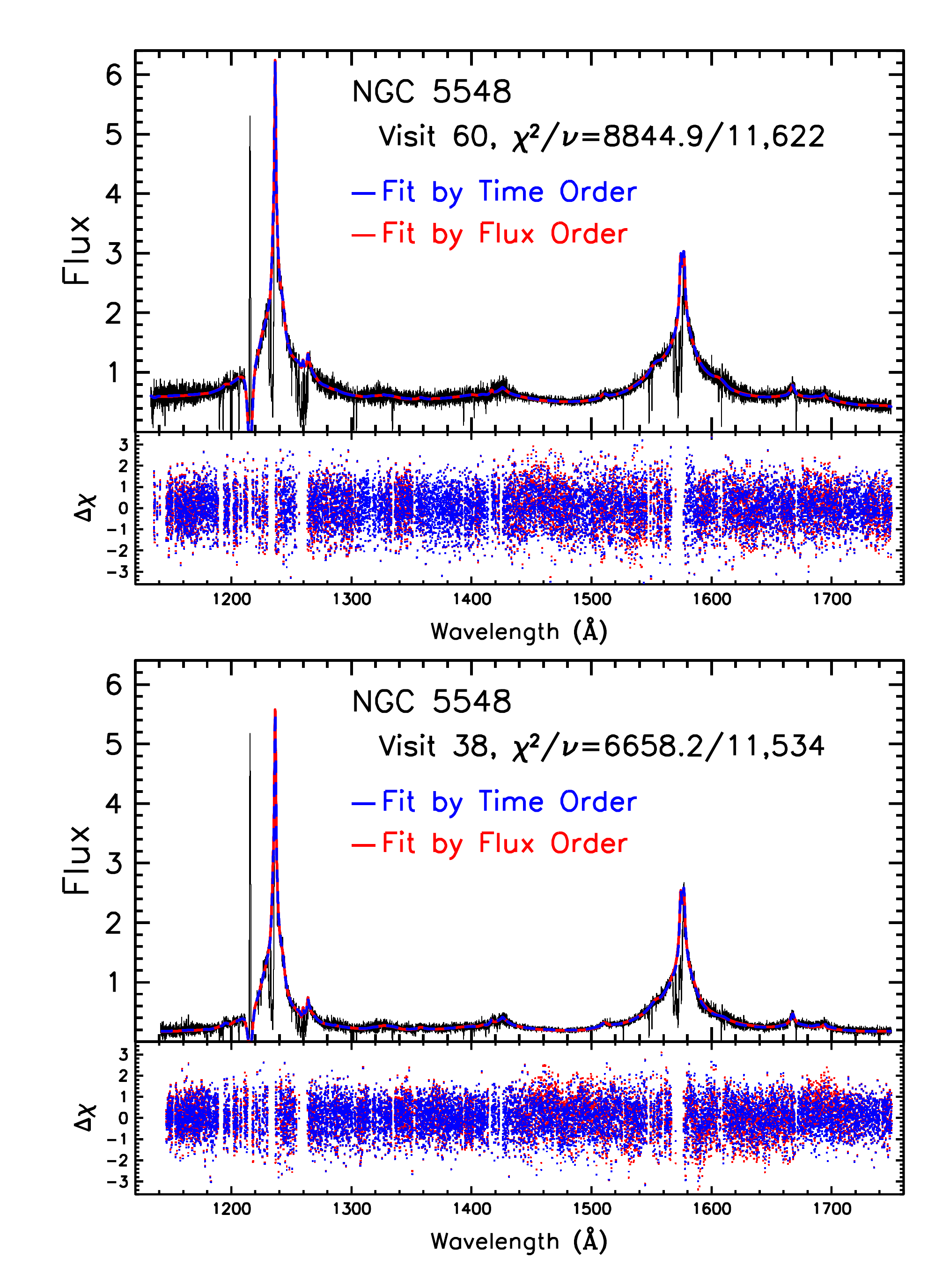}}
\caption{(Top Panel) 
Best-fit models overplotted on the Visit 60 spectrum. This model has the worst
$\chi^2$ for the ensemble of fits. The adopted best fit model curve
(from fits done in time order) is in blue; the best-fit model for the
fits done when ordered by flux is in red.
Flux is in the observed frame in units of
$10^{-14}~\rm erg~cm^{-2}~s^{-1}~\mbox{\AA}^{-1}$.
Residuals to the fits are shown as points ("+") scaled by the 1-$\sigma$
uncertainties, e.g.,
$\rm \Delta\chi = (Data - Model) / Error$.
(Bottom Panel) Same as top panel for the Visit 38 spectrum, which has the
best $\chi^2$.}\label{fig:fit_residuals}
\end{figure*}

Despite the good agreement in the quality of the fits for the two different
methods, our experiments also showed that the second method, ``ordered by time",
produced better results than ``ordered by flux" in the sense that
variations in parameter values were smoother, and, as shown in
Figure \ref{fig:chi2}, the best-fit $\chi^2$ was typically slightly less.
Despite the significant reduction in $\chi^2$ we achieved for the fits
ordered by time, the differences in the fits are not obvious.
We note that for each method $\chi^2$ varies systematically with time
during the campaign. The variations loosely correspond to the overall
variations in brightness of NGC 5548 (as one can see in later figures
showing light curves for the continuum and emission line).
Our inference for these systematic variations is that the brighter spectra
have higher signal-to-noise ratios per pixel, and that subtle residual
pattern noise in the flat-field properties of the COS detectors degrade the
quality of our fits.
Figure \ref{fig:fit_residuals} compares the best-fit models to the data for two
extreme cases, our overall lowest $\chi^2$ solution for Visit 38, and our
overall highest $\chi^2$ obtained for Visit 60.
One is hard pressed to see the differences between the models fit using
either method, or even why $\chi^2$ is significantly higher for Visit 60
compared to Visit 38.

Our best inference for why the ``ordered by time" sequence produces better
results is that the spectrum evolves on timescales of a few days.
This is measured in the lags of the emission lines, and, as we will show later
in \S5.2, it is also true for the absorption features.
Therefore, although one spectrum might have the same continuum flux as another,
if it is separated in time by more than several days, parameters of major
features may differ significantly, making it harder for the minimization
algorithm to converge on the best solution.

To fit an individual spectrum, we first determined the best first approximation
to the normalization and spectral index of the powerlaw continuum by fitting
only those points identified by \cite{DeRosa15} as continuum windows.
To avoid ``contaminating" these windows with broad-line flux, we also set the
flux of all the Very Broad components to zero.
In our experiments, we found that if we did not do this, the Very Broad
components developed a tendency to grow in width until they formed their own
pseudo-continuum across the whole spectrum.
In this first pass, only the
powerlaw normalization and the index were allowed to vary freely.

After this step, we then optimized the strength of the brightest emission
lines--Ly$\alpha$, \ion{N}{5}, \ion{Si}{4}, \ion{C}{4} and \ion{He}{2}, 
and let the fluxes of their Broad and
Medium Broad components and the powerlaw normalization vary freely,
while keeping the powerlaw index fixed and the Very Broad component turned off.
Next, we restored the Very Broad fluxes to their
original values, and let the fluxes of all broad components of the above lines
vary freely, as well as the powerlaw normalization.

At this point the fit formed a remarkably good representation of an individual
observation, but it was still far from the best fit. In the next steps, we
turned our attention to optimizing the fits for each individual bright
emission line. In these separate steps, we kept all parameters not related to
the specific spectral region fixed, including the continuum parameters, the
\ion{Fe}{2} flux, the Narrow and Intermediate line components for all lines,
and any weak blended lines.

For the Ly$\alpha$ region, we did separate optimization steps in this order:
\begin{enumerate}
\item Fit Ly$\alpha$ only (Broad, Medium Broad, Very Broad)
      using only wavelengths 1150--1245 \AA.
      As before, let the fluxes vary freely first, then both the widths and
      fluxes. Keep all components of \ion{N}{5} fixed throughout this.
\item Fit the red wing of \ion{N}{5} using only wavelengths 1263--1300 \AA.
      Keep Ly$\alpha$ and all other components fixed.
      Let the two Broad components vary freely first, then fix them, and let
      the Medium Broad Components vary.
      Finally, fix those, and let the Very Broad component vary.
\item Fit the \ion{N}{5} absorption on its blue wing, keeping the doublet's
      fluxes tied at a 1:1 ratio, using only wavelengths 1245--1264 \AA.
\item Free the flux and width of the Broad, Medium Broad, and Very Broad
components of Ly$\alpha$.
\item Free the flux and width of the Broad, Medium Broad, and Very Broad 
components of \ion{N}{5}.
\item Free the flux of the Ly$\alpha$ Broad Absorber.
\item Free flux, width, and asymmetry of the Ly$\alpha$ Broad Absorber.
\end{enumerate}

We next did the \ion{Si}{4} region. All parameters that we varied freely above
for the optimization of the Ly$\alpha$ region were fixed, and we then:
\begin{enumerate}
\item Free the flux of the Broad, Medium Broad, and Very Broad
components of \ion{Si}{4}.
\item Free the flux and width of the Broad, Medium Broad, and Very Broad 
components of \ion{Si}{4}.
\item Free the flux of the \ion{Si}{4} Broad Absorber.
   (Since the \ion{Si}{4} absorption is so weak, we linked the width
    and asymmetry of the \ion{Si}{4} Broad Absorber
    to that of \ion{C}{4}.)
\end{enumerate}

We fit \ion{C}{4} and \ion{He}{2} together since they are tightly blended.
For optimizing this region, we fix all the previous freely varying parameters,
then:
\begin{enumerate}
\item Free the flux and width of the Broad, Medium Broad, and Very Broad 
components of \ion{C}{4}.
\item Free the flux (but not the width) of the Broad, Medium Broad, and
Very Broad components of \ion{He}{2}.
\item Free the flux of the \ion{C}{4} Broad Absorber.
\item Free the flux, width, and asymmetry of the \ion{C}{4} Broad Absorber.
\end{enumerate}

Once the major emission and absorption components are tuned up, we then allow
the weaker emission-line features to adjust. We keep the continuum fixed,
as well as all the parameters associated with
Ly$\alpha$, \ion{N}{5}, \ion{Si}{4}, \ion{C}{4} and \ion{He}{2}.
The fluxes of all other weak emission features are then allowed to vary.

To complete the optimization, all parameters designated as free to vary in
Table \ref{tab:parameters} are freed,
and we iterate the $\chi^2$ minimization process
until it converges.
The best fit parameters for this spectrum are then used as the initial
guesses for doing the fit to the next spectrum in the series
(with the exception that the initial guesses for the weak features are taken
from the best fit to the grouped spectrum corresponding to that spectrum).
Final values of all components as a function of wavelength for each
spectrum are available as a high-level science product in the
\href{http://archive.stsci.edu}{{\it Mikulski Archive for Space Telescopes} (MAST)} as the data set identified by
\dataset[https://doi.org/10.17909/t9-ky1s-j932]{https://doi.org/10.17909/t9-ky1s-j932}.

\subsection{Propagating the Uncertainties}
\label{sec:errors}

In principle one can obtain  1-$\sigma$ uncertainties on each of the parameters in our model
from the best-fit covariance matrix. However, given the 383 parameters,
143 of which are freely varying, this is computationally impractical.
An alternative is to assume that parameter space can be approximated by a
parabola near the best-fit minimum in $\chi^2$. Using numerically
calculated first and second derivatives, one can then extrapolate from the
minimum changes in each parameter to achieve $\Delta \chi^2 = 1$,
which corresponds to a 1$\sigma$ uncertainty for a single interesting
parameter \citep{Bevington69}.
Unfortunately, owing to the high dimensionality of our parameter space and
its poor sampling in our calculations,
this method proved inadequate.

All the main quantities of interest to be extracted from our models are the
fluxes of individual features, either in emission or in absorption.
We therefore calculate uncertainties for these quantities using the data and
associated uncertainties in each original spectrum,
and scaling them in proportion to the
quantities that we integrate from our models.
To explain this quantitatively, first we define these quantities:\\
\indent
     $\rm f_{data,i} =$ flux in pixel $i$ of the original spectrum\\
\indent
     $\rm \sigma_{data,i} =$ 1-$\sigma$ uncertainty for pixel $i$ in the original spectrum\\
\indent
     $\rm f_{mod,i} =$ flux in pixel $i$ of the modeled spectrum\\
\indent
     $\rm \sigma_{mod,i} =$  1-$\sigma$ uncertainty for pixel $i$ in the model spectrum\\
\indent
     $\rm f_{c,j,i} =$ flux in pixel $i$ of the component $j$ (of 97 total)\\
\indent
     $\rm \sigma_{c,j} =$  1-$\sigma$ uncertainty for component $j$.\\
\noindent
The flux of the model in pixel $i$ can be decomposed into the sum of the
contributions of the individual components:
\begin{equation}
\rm f_{mod,i} = \sum_{j} f_{c,j,i}. 
\end{equation}
The variance of the model in pixel $i$ is then
\begin{equation}
\sigma_{mod,i}^2 = \sum_{j} \sigma_{c,j}^2. 
\end{equation}
In the limit of very good statistics, the variance predicted by the model should simply be the variance in the data themselves, i.e.,
\begin{equation}
\sigma_{mod,i}^2 \sim \sigma_{data,i}^2, 
\end{equation}
so the variance in the total flux of any individual component is then
\begin{equation}
 \sigma_{c,j}^2 = \sum_{i} \sigma_{data,i}^2 \times  f_{mod,i}^2 / (\sum_{k \neq j} f_{c,k,i}^2). 
\end{equation}
If the quantity of interest is the sum of multiple components (i.e., the
several components of an emission line contributing to the flux in a given
velocity bin), then the 1-$\sigma$ uncertainty we derive for that quantity is
\begin{equation}
\rm \sigma_{tot} = \sqrt{\sum_{j} \sigma_{c,j}^2}. 
\end{equation}

\begin{figure}
\centering
\resizebox{\hsize}{!}{\includegraphics[angle=-90, width=0.9\textwidth]{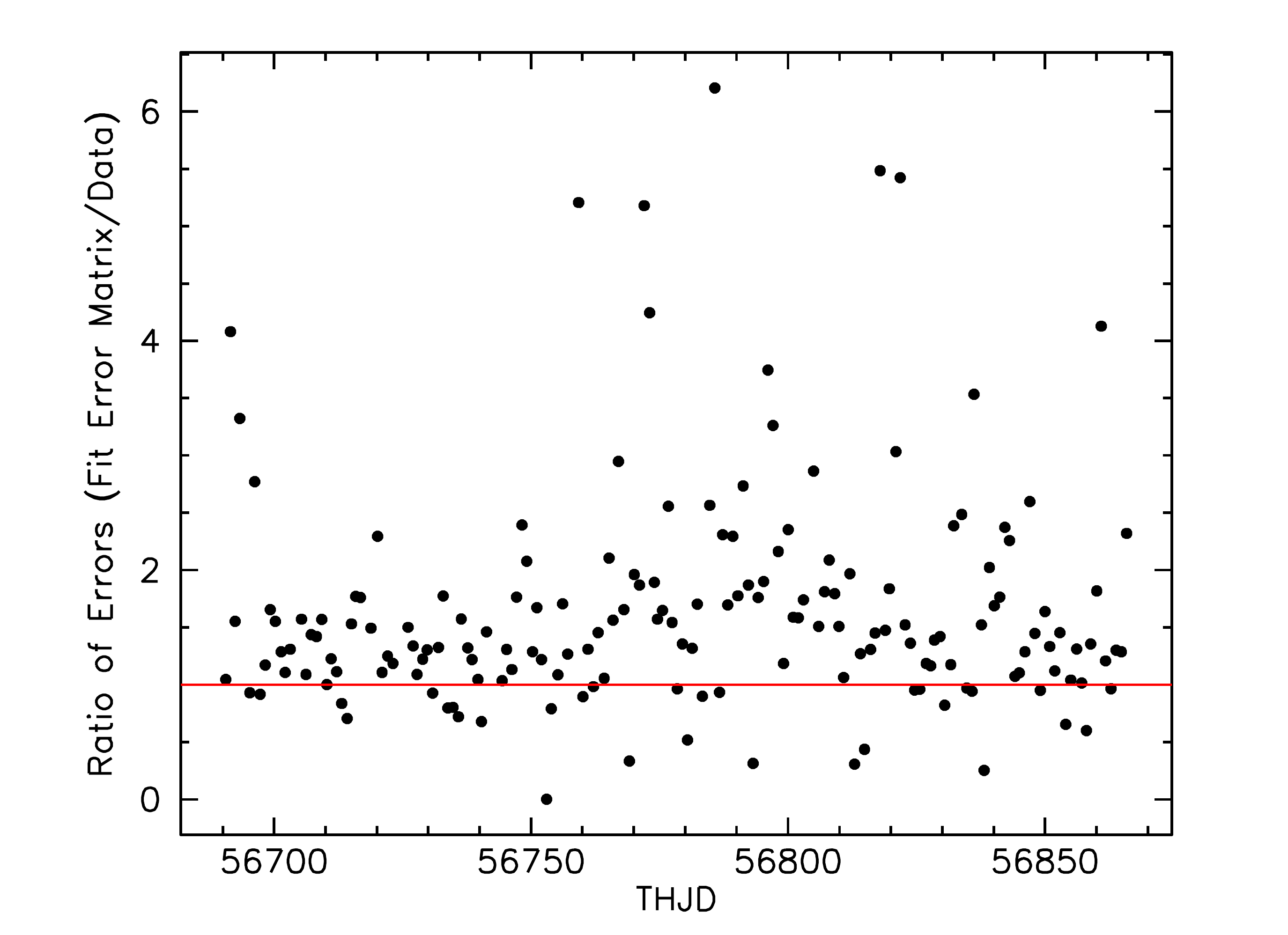}}
\caption{Comparison of uncertainties for the broad Ly$\alpha$ absorption feature
in our model. Black points show the ratio of the uncertainty derived from the
error matrix
of the fit for $\Delta\chi^2=1$ divided by the  1-$\sigma$ uncertainty derived
as described in \S\ref{sec:errors}.
The ratio scatters about unity (the red line), with large excursions.
}\label{fig:error_ratio}
\end{figure}

Figure \ref{fig:error_ratio} shows the consequences of this method for
calculating the  1-$\sigma$ uncertainties.
When calculated directly from the data, as described above, the uncertainties
are more uniform. The numerical instabilities in our method of interpolating in
the error matrix of the fit give uncertainties that largely cluster around the
calculation based on the data, but show large excursions, both higher and
lower.

\subsection{Quality Checking the Fits}
The resulting best-fit $\chi^2$ for each spectrum is shown as a time series
in Figure \ref{fig:chi2}.
The number of points in each spectrum varies slightly since the spectrum is
moved to multiple positions on the detector using different central
wavelength settings.
Because of the differing central wavelength settings, not all spectra cover
the same range in wavelength as the mean spectrum; the settings tend to lose
several hundred points on the blue and red ends of each spectrum.
On average, individual spectra have $\sim$11\,600 points in each fit.
With 143 freely varying parameters, given the $\chi^2$ values ranging from
$\sim$6800 to $\sim$9000 in Figure \ref{fig:chi2}, one can see that our
uncertainties are too large.
This is most likely because in aligning and merging each
spectrum, we have resampled the original pixels, introducing correlated errors
in adjacent bins after rebinning.
Note also that $\chi^2$ varies systematically with time in the campaign,
largely following the light curve of total brightness.
A likely explanation for this trend is that it is easier to get a good fit
to our complex model when
the fluxes are lower and uncertainties are larger.

\begin{figure*}[htb]
\centering
\resizebox{\hsize}{!}{\includegraphics[angle=-90,width=0.9\textwidth]{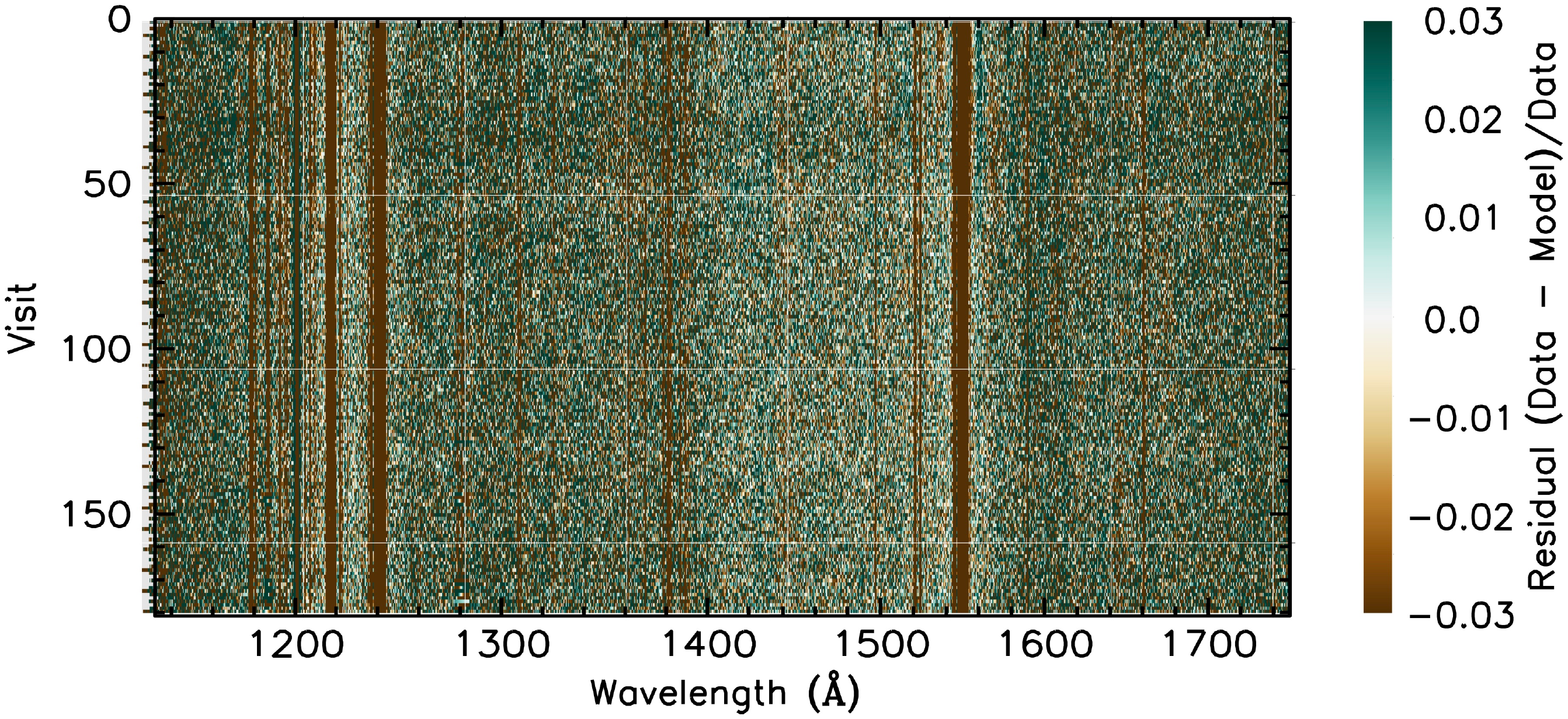}}
\vskip -36pt
\caption{Normalized residuals from the best fit to each spectrum
from the campaign.
The horizontal axis is in pixels for each spectrum, spanning a wavelength range
of 1130--1805 \AA.
The vertical axis is each individual visit in the campaign.
}\label{fig:carpet}
\vskip 14pt
\end{figure*}

While the $\chi^2$ shows that we have good fits overall, we visually examined
each individual fit to see if there were any points or regions where there
were systematic deviations of the model from the data.
It was these inspections in our early experiments that led us to develop the
fitting strategies we documented above.
The final fits show no gross or systematic residuals.
This is illustrated visually by the image in Figure \ref{fig:carpet} that
shows the residuals of Data$-$Model for each fit stacked into a two-dimensional
spectrogram.
The only significant features visible are the vertical lines at the positions
of interstellar and intrinsic narrow absorption lines,
which are not part of our model.

\begin{figure*}
\centering
\resizebox{\hsize}{!}{\includegraphics[angle=0, width=0.9\textwidth]{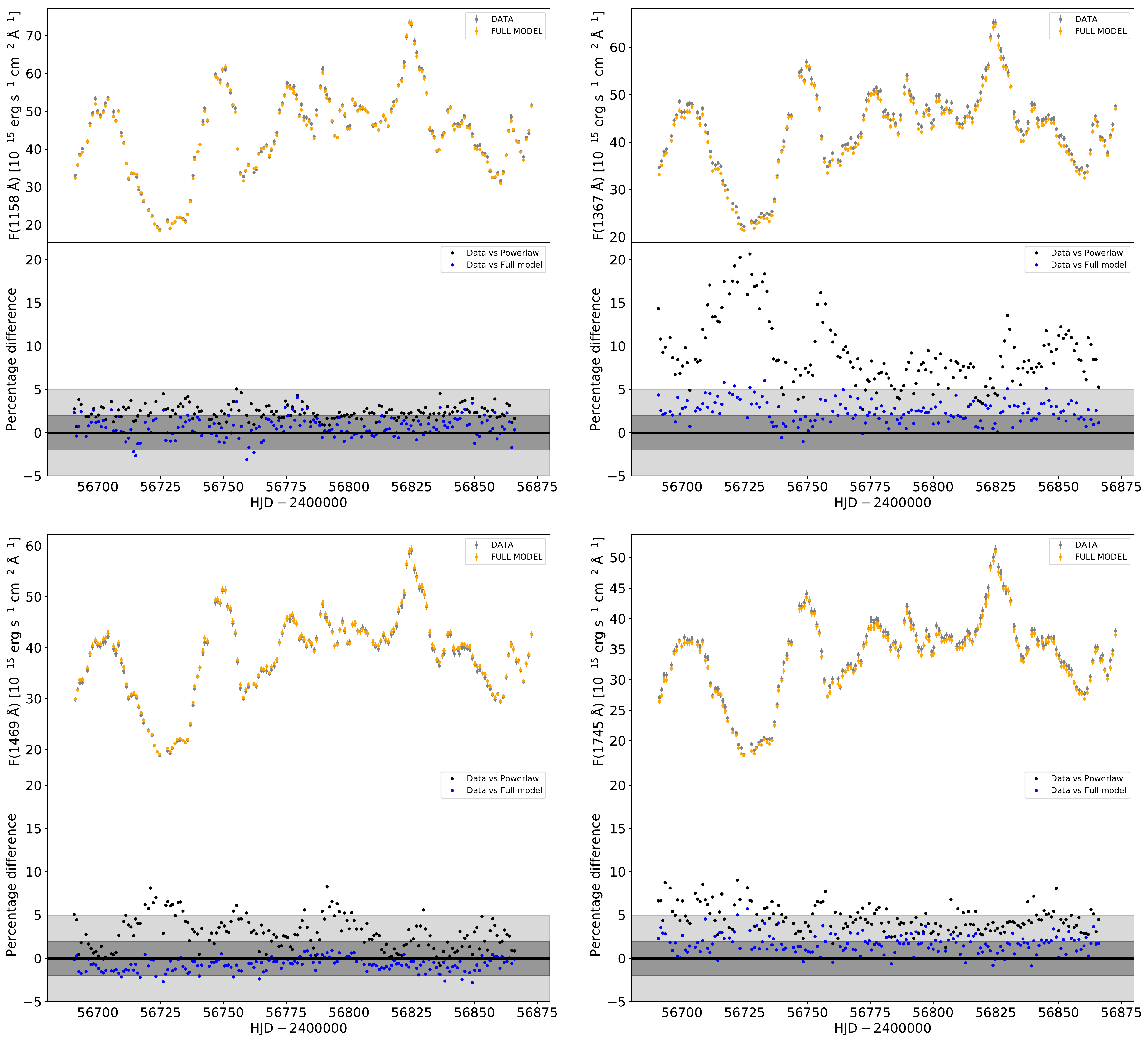}}
\caption{Comparisons of the continuum light curves
integrated from the data (green points with error bars) to the fluxes
integrated from our best-fit models (gold points with error bars).
The top left panel shows the flux at 1158 \AA, F(1158 \AA), the top right panel 
F(1367 \AA), the bottom left panel F(1469 \AA),
and the bottom right panel F(1745 \AA).
The dark gray region in each panel highlights $\pm2$\% errors,
and the light gray region shows $\pm5$\% errors.\label{fig:compare1157}}
\vskip 14pt
\end{figure*}

Our final set of tests compared light curves of fluxes extracted from our
model fits to integrations of the same regions of data used in
\cite{DeRosa15}.
Again here we see no systematic deviations, but these figures do illustrate
how the continuum regions are slightly contaminated by
wings of the broad emission lines.
Figure \ref{fig:compare1157} compares light curves for continuum windows
integrated from the raw data, as described by \cite{DeRosa15},
to integrations of the same wavelength regions in our models,
both for the full model, and for just the power-law continuum.
The lower half of each panel in the figure shows the differences between the
two curves along with uncertainties from the raw data.
These uncertainties include the systematic repeatability errors
described by \cite{DeRosa15} that apply to the 
time-series analysis of fluxes from the campaign.
These are $\delta_P = 1.1$\% for data with $\lambda < 1425$ \AA, and
$\delta_P = 1.4$\% for $\lambda > 1425$ \AA.
One can see here that the cleanest continuum window, i.e., the one with
the least contamination by surrounding emission lines, is the
shortest wavelength window surrounding 1158 \AA.
Although this is the cleanest window in terms of total flux, we
use the modeled continuum flux at 1367 \AA\ in our subsequent analysis.
Since this is deterministically connected to the modeled flux at 1158 \AA\ 
through the continuum model, there is no difference between using one or
the other.

\subsection{The Absorption-corrected Spectra}
One of our main goals for these fits to the emission model of each spectrum is
to correct for the effects of intrinsic broad absorption, and also to bridge
the regions affected by foreground interstellar absorption lines and the
narrow intrinsic absorption features in NGC 5548 itself.
To correct for the broad absorption, we simply apply the inverse of these
model elements to the data. For each pixel corrected in this way, we apply
the same scaling to the associated uncertainty as well as the data themselves.
To correct for the narrow absorption features (both foreground and intrinsic),
since these are not modeled, we replace the data in the wavelength regions
affected by the absorption with the emission model.

More specifically, we first visually examined the fit to the mean spectrum
to identify points that were significantly affected by narrow absorption lines.
These intervals and their identifications are listed in
Table \ref{tab:correction_intervals}\footnote{Table \ref{tab:correction_intervals} appears at the end of the paper to facilitate the formatting.}.
Next, we define the following quantities:\\
\indent
     $\rm f_{orig} =$ flux in the original spectrum\\
\indent
     $\rm f_{mod} =$ flux in the model spectrum (including broad absorption)\\
\indent
     $\rm f_{abs} =$ model flux (negative) in the broad absorption lines\\
\indent
     $\rm t_G =$ transmission profile of Galactic damped Ly$\alpha$\\
\indent
     $\rm f_{cor} =$ flux in the corrected spectrum.\\
\noindent
The corrected spectrum is then computed in two steps. First, we replace all
pixels in the original spectrum with $\rm f_{mod}$ if they fall within the
wavelength intervals defined in Table \ref{tab:correction_intervals}.
We then compute
\indent
     $\rm f_{cor} = (f_{orig} - f_{abs}) / t_G$.
\noindent
To calculate the 1-$\sigma$ uncertainties, we scale the original uncertainties in each
pixel by the ratio of the corrected flux to the original flux:\\
\begin{equation}
\rm \sigma_{cor} = \sigma_{orig} \times (f_{cor} / f_{orig}). 
\end{equation}

\vskip 36pt
\subsection{Fluxes in Deblended Emission Lines}
\label{sec:deblended_fluxes}

With our model fits to the entire series of spectra from the STORM 
campaign, we can now extract absorption-corrected spectra for all 
emission lines across their full, deblended velocity profiles.
Our models also allow us to separate the variable and non-variable
components of the strong emission lines as well as to deblend adjacent lines.
For example, \cite{Crenshaw09} were able to use the faint state of NGC 5548
in 2004 to separate and measure the narrow-line and intermediate-line width
components of the Ly$\alpha$ and \ion{C}{4} emission lines. They demonstrated
that these components vary only slightly over timescales of years.
Using our model, we are able to exclude these non-varying components of the
emission lines from the overall broad-line profile.
Likewise, we can separate the contributions of blended lines from the
wings of Ly$\alpha$ and \ion{C}{4}, and measure lines such as \ion{N}{5}
as individual species.

For each individual broad emission line, we construct a model profile at each
individual pixel $i$ that includes only the contributions of the relevant
broad-line components from our model.
As described in \S\ref{sec:errors}, the net flux associated with a given
emission feature is

\begin{equation}
\rm f_{totem,i} = \sum_{j} f_{c,j,i} ,
\end{equation}

\noindent
where the index $j$ runs over all components associated with the desired
emission feature.
As described in \S\ref{sec:errors},
we calculate the associated 1-$\sigma$ statistical uncertainty as the fraction
in quadrature (relative to all model components in that pixel)
of the 1-$\sigma$ uncertainty on the data in that pixel:
\begin{equation}
\rm \sigma_{totem,i} = \sigma_i \sqrt{\frac{\sum_{j} f_{c,j,i}^2}
{\sum_{k} f_{c,k,i}^2}} ,
\end{equation}
where the index $j$ runs over all components contributing to the desired
emission feature, and the index k runs over all components of the model
contributing to the flux in pixel $i$.
As described in detail by \cite{DeRosa15}, systematic repeatability errors
affect the data when considering time-series analysis of these quantities.
We therefore add in quadrature the same errors in precision, namely
$\delta_P = 1.1$\% for data from grating G130M ($\lambda < 1425$ \AA), and
$\delta_P = 1.4$\% for data from grating G160M ($\lambda > 1425$ \AA).
These errors in reproducibility actually dominate the uncertainties for all
quantities at $\lambda < 1180$ \AA\ and $\lambda > 1425$ \AA.

\begin{figure*}
\centering
\resizebox{\hsize}{!}{\includegraphics[angle=0, width=0.7\textwidth]{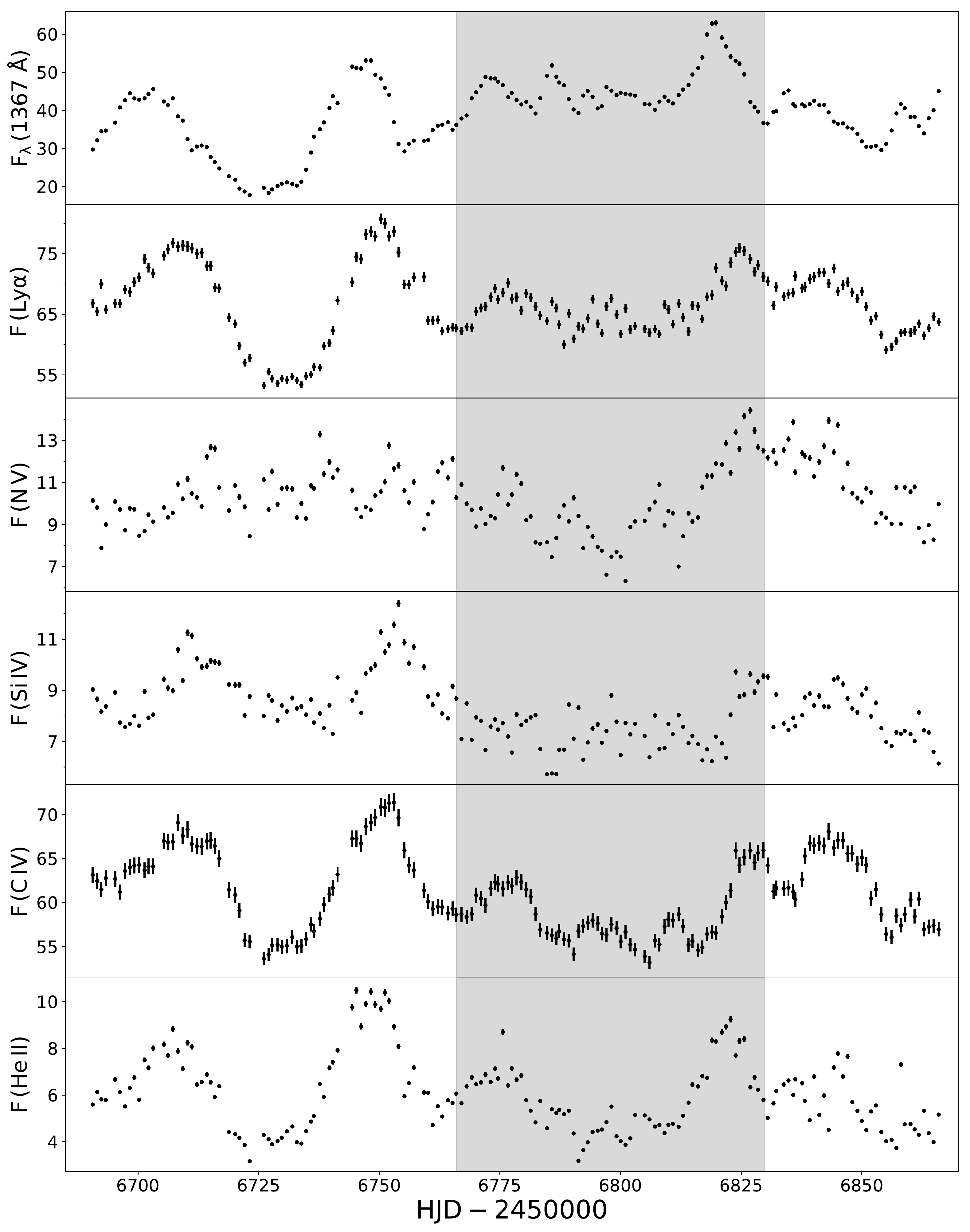}}
\caption{
Light curve for the modeled continuum flux at 1367 \AA\ (top panel) in units of
$10^{-15}~\rm erg~cm^{-2}~s^{-1}~\mbox{\AA}^{-1}$ vs.\,the Heliocentric Julian Date (HJD) $-$ 2450000.
Subsequent panels show light curves for the integrated fluxes of the
absorption-corrected deblended emission lines (as labeled) in units of
$10^{-13}~\rm erg~cm^{-2}~s^{-1}$.
The shaded region shows the time interval of the BLR holiday.
}\label{fig:light_curves_int}
\end{figure*}

Figure \ref{fig:light_curves_int} compares light curves for the
modeled continuum flux at 1367 \AA\ to the deblended broad emission lines of
Ly$\alpha$, \ion{N}{5}, \ion{Si}{4}, \ion{C}{4}, and \ion{He}{2}. 
Portions of these light curves are tabulated in
Table \ref{tab:light_curves_int}\footnote{Table \ref{tab:light_curves_int} appears at the end of the paper to facilitate the formatting.}, with full tabulations of these quantities and the other continuum
windows (1158 \AA, 1430 \AA, and 1740 \AA) available online.
The light curves in Figure \ref{fig:light_curves_int} closely resemble those
derived from the original data shown in Figure 3 of \cite{DeRosa15}.
In general, Ly$\alpha$, \ion{Si}{4} and \ion{C}{4} are brighter
due to the corrections for absorption, the additional flux from the wings of
the very broad emission components, and the elimination of contaminating
emission-line flux from the original continuum windows.
For \ion{He}{2} the flux levels are roughly the same; additional flux from the
very broad component of the emission line and a less contaminated continuum are
offset by subtraction of blended emission from \ion{C}{4}.
Overall, the error bars are smaller since the modeled flux for any given
component is determined by many more pixels than the limited wavelength range
used in the original integrations.
The light curve for \ion{N}{5} is a new addition enabled by the
deblending from Ly$\alpha$ in our model.
In some respects \ion{N}{5} differs in character from the other emission lines,
especially during the first 75 days of the campaign prior to the BLR holiday.
However, starting with the BLR holiday, its behavior is very similar to that
of \ion{C}{4} and \ion{He}{2}.
We will quantify these similarities and differences in \S\ref{sec:lags}
when we discuss the emission-line lags.

\subsection{Measuring the Absorption Lines}
\label{sec:measuring_absorbers}

\begin{figure}
\centering
\resizebox{\hsize}{!}{\includegraphics[angle=0, width=0.9\textwidth]{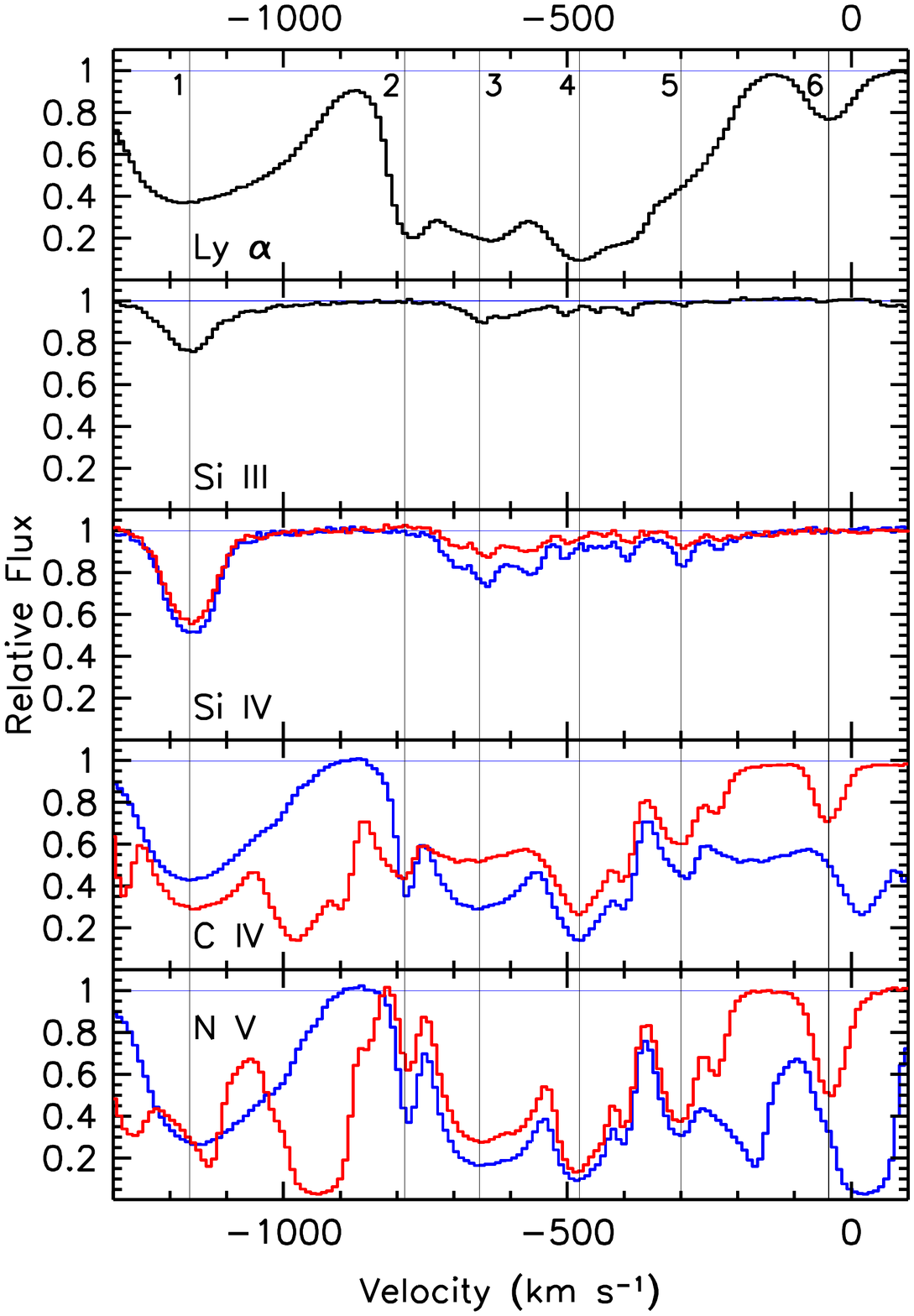}}
\caption{Intrinsic narrow absorption features in NGC 5548.
Normalized relative fluxes are plotted as a function of velocity relative to
the systemic redshift of $z=0.017175$.
The top panel shows Ly$\alpha$, the second panel \ion{Si}{3} $\lambda1206$,
the third panel \ion{Si}{4} $\lambda1393$ (blue) and
\ion{Si}{4} $\lambda1402$ (red),
the fourth panel \ion{C}{4} $\lambda1548$ (blue) and
\ion{C}{4} $\lambda1550$ (red), and
the bottom panel \ion{N}{5} $\lambda1238$ (blue) and
\ion{N}{5} $\lambda1242$ (red).
Thin vertical lines indicate the velocities of the six intrinsic absorbers.
}\label{fig:narrow_abs_norm}
\end{figure}

The intrinsic narrow absorption lines comprise six discrete velocity components.
We adopt the nomenclature of \cite{Mathur99}, numbering each component in order
starting at the highest blue-shifted velocity. To illustrate this kinematic
structure, Figure \ref{fig:narrow_abs_norm} shows normalized absorption
profiles for the most prominent intrinsic absorption lines
as a function of velocity relative to the systemic velocity of the host
galaxy NGC 5548. For this we adopt the \ion{H}{1} 21-cm redshift of
$z = 0.017175$ \citep{deVaucouleurs91}.

Measuring the strengths of the intrinsic narrow absorption lines is
straightforward. Using the complete model of each spectrum
(including the broad absorption components, since they help define the local
continuum surrounding the narrow intrinsic absorption lines), we measure the
equivalent widths by integrating across each absorption-line profile
in the normalized spectrum.
These integrations are performed as discrete sums over pixels lying within the
wavelength regions defined for each feature in
Table \ref{tab:absorption_intervals}\footnote{Table \ref{tab:absorption_intervals} appears at the end of the paper to facilitate the formatting.}:

\begin{equation}
 EW = \sum_{i} (f_{orig,i} - f_{mod,i}) / f_{mod,i} \times \Delta\lambda .
\end{equation}

\noindent
The 1-$\sigma$ uncertainty for EW is obtained by simply propagating the
uncertainty associated with each data point in the original spectrum used
in the sum:

\begin{equation}
 \sigma_{EW} = \sqrt{ \sum_{i} (\sigma_{orig,i}^2 / f_{orig,i}^2) \times \Delta\lambda^2 } .
\end{equation}

\begin{figure}
\centering
\resizebox{\hsize}{!}{\includegraphics[angle=0, width=0.9\textwidth]{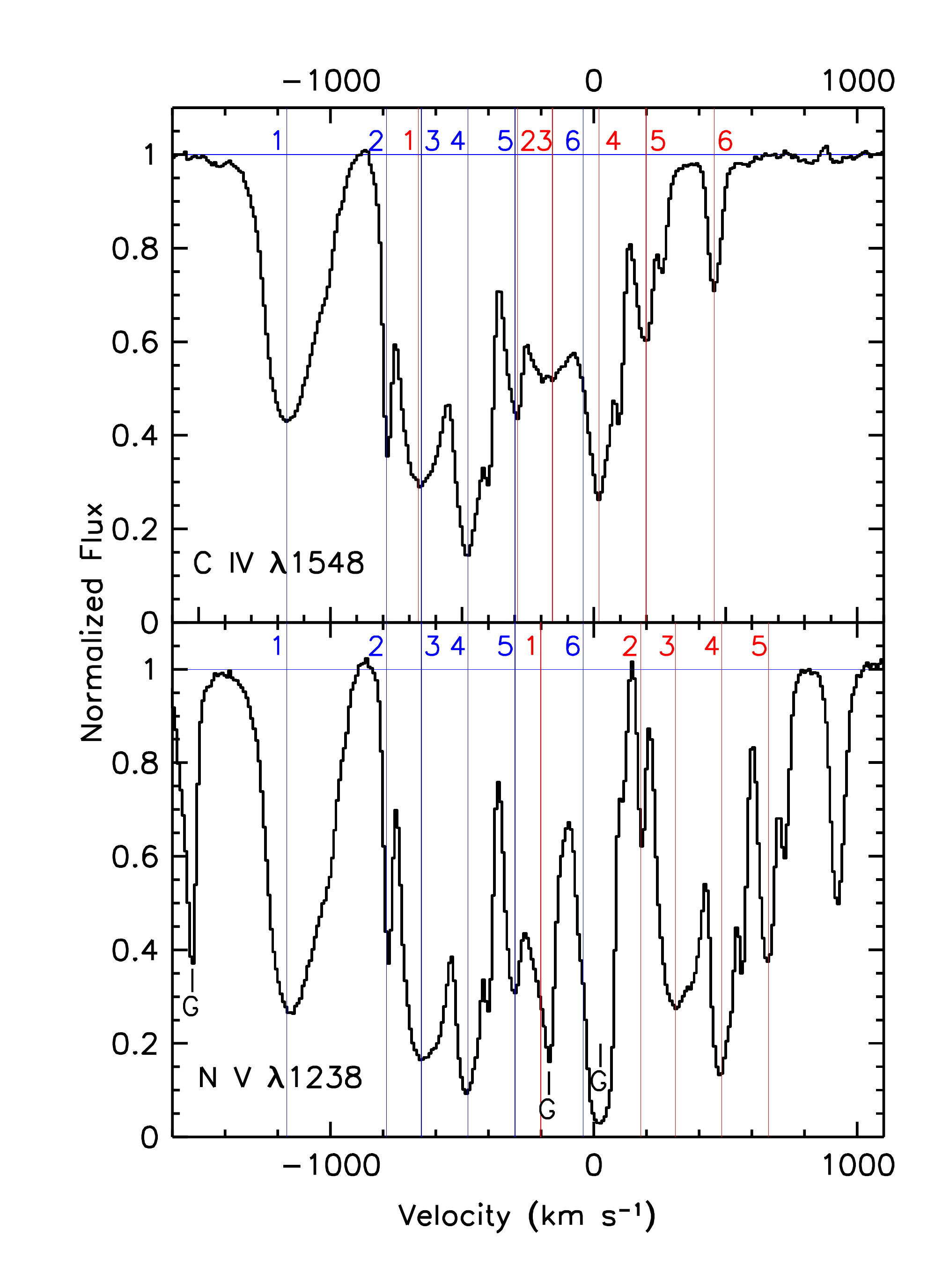}}
\caption{Illustration of blending in the absorption lines of the
 \ion{C}{4} and \ion{N}{5} doublets in NGC 5548.
Normalized relative fluxes are plotted as a function of velocity relative to
the systemic redshift of $z=0.017175$.
The top panel shows \ion{C}{4} $\lambda1548$,
and the bottom panel \ion{N}{5} $\lambda1238$.
Thin vertical blue lines indicate the velocities of the blue components of
the doublets for the six intrinsic absorbers.
Vertical red lines show the locations of the corresponding red components.
Foreground Galactic interstellar absorption lines are marked with a ``G".
}\label{fig:c4n5_abs_norm}
\end{figure}

We take caution in performing these integrations to avoid
features blended with Galactic absorption lines, or with other transitions.
For example, the close velocity spacings of the \ion{C}{4} and \ion{N}{5}
doublets ($498~\rm km~s^{-1}$ and $964~\rm km~s^{-1}$, respectively)
cause components \#1, \#2, \#3, and \#5 in \ion{C}{4} to overlap, and
Components \#1 and \#5 in \ion{N}{5} to overlap, as shown in
Figure \ref{fig:c4n5_abs_norm}.
The red transition of Component \#1 in the \ion{N}{5} doublet 
is blended with Galactic \ion{S}{2}$\lambda 1259$, and the blue
transition of Component \#6 is blended with \ion{Si}{2}$\lambda 1260$.
Therefore Table \ref{tab:absorption_intervals} gives measurements only for
clean, unblended features.

The equivalent width of each broad absorption feature (EW) is calculated
from the normalized modeled spectrum,
\begin{equation}
\rm f_{norm,i} = f_{mod,i} / (f_{mod,i} - f_{abs,i}),
\end{equation}
as
\begin{equation}
 EW = \sum_{i} (1 - f_{norm,i}) \times (\lambda_{i+1} - \lambda_{i}), 
\end{equation}
where $\lambda_{i}$ is the wavelength of pixel $i$.
Since our spectra are linearized,
$\Delta\lambda = \lambda_{i+1} - \lambda_{i}$, is actually a constant.
The corresponding 1-$\sigma$ uncertainty is
\begin{equation}
 \rm \sigma_{EW} = (f_{totabs} / EW) \times \sigma_{totabs},
\end{equation}
where $f_{totabs}$ and $\sigma_{totabs}$ are defined below.

The broad UV absorption troughs associated with the obscurer in NGC 5548 are
shown in Figure 2 of \cite{Kaastra14}.
These broad troughs are asymmetric, and they extend from near zero velocity in
the systemic frame of the host galaxy to $\sim -5500~\rm km~s^{-1}$.
The time-varying strengths of the intrinsic broad absorption lines in NGC 5548
that are associated with the obscurer are part of the models we have fit to
all the spectra. These all have one main component, but there are also weaker
components on the high-velocity blue wing of the absorption profile.
These individual weak components are often not well constrained
by the model fits.
We therefore calculate the total absorption, $\rm f_{totabs}$ for the sum of
all components associated with a given spectral transition.
The total flux is then
\begin{equation}
\rm f_{totabs} = \sum_{j} f_{c,j} ,
\end{equation}
where the index $j$ runs over all components associated with the desired line,
and the associated uncertainty is calculated as the quadrature sum of the
1-$\sigma$ uncertainties of each component $j$:
\begin{equation}
\rm \sigma_{totabs} = \sqrt{\sum_{j} \sigma_{c,j}^2} .
\end{equation}

\begin{figure}
\centering
\resizebox{\hsize}{!}{\includegraphics[angle=0, width=0.9\textwidth]{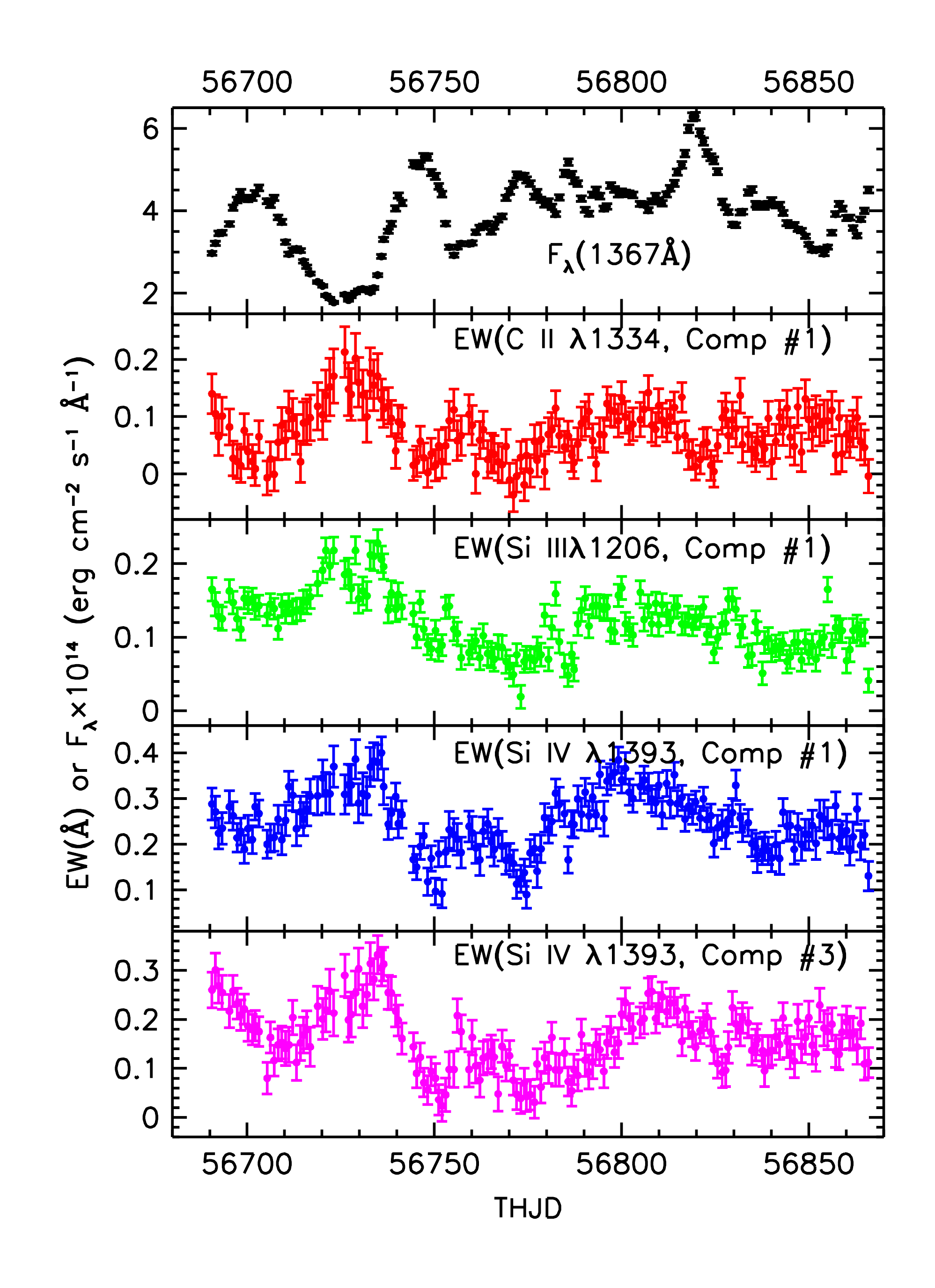}}
\caption{(Top Panel) Light curve for the UV continuum at 1367 \AA\ in units
of $10^{-14}~\rm erg~cm^{-2}~s^{-1}~{\mbox{\AA}}^{-1}$.
Lower panels show the absolute value of the equivalent width in \AA\ vs.
time for selected absorption lines.
(Second Panel) \ion{C}{2} $\lambda 1334$, narrow absorption Component \#1.
(Third Panel) \ion{Si}{3} $\lambda 1206$, narrow absorption Component \#1.
(Fourth Panel) \ion{Si}{4} $\lambda 1393$, narrow absorption Component \#1.
(Bottom Panel) \ion{Si}{4} $\lambda 1393$, narrow absorption Component \#3.
}\label{fig:sample_ew_lcs}
\end{figure}

\begin{figure}
\centering
\resizebox{\hsize}{!}{\includegraphics[angle=0, width=0.9\textwidth]{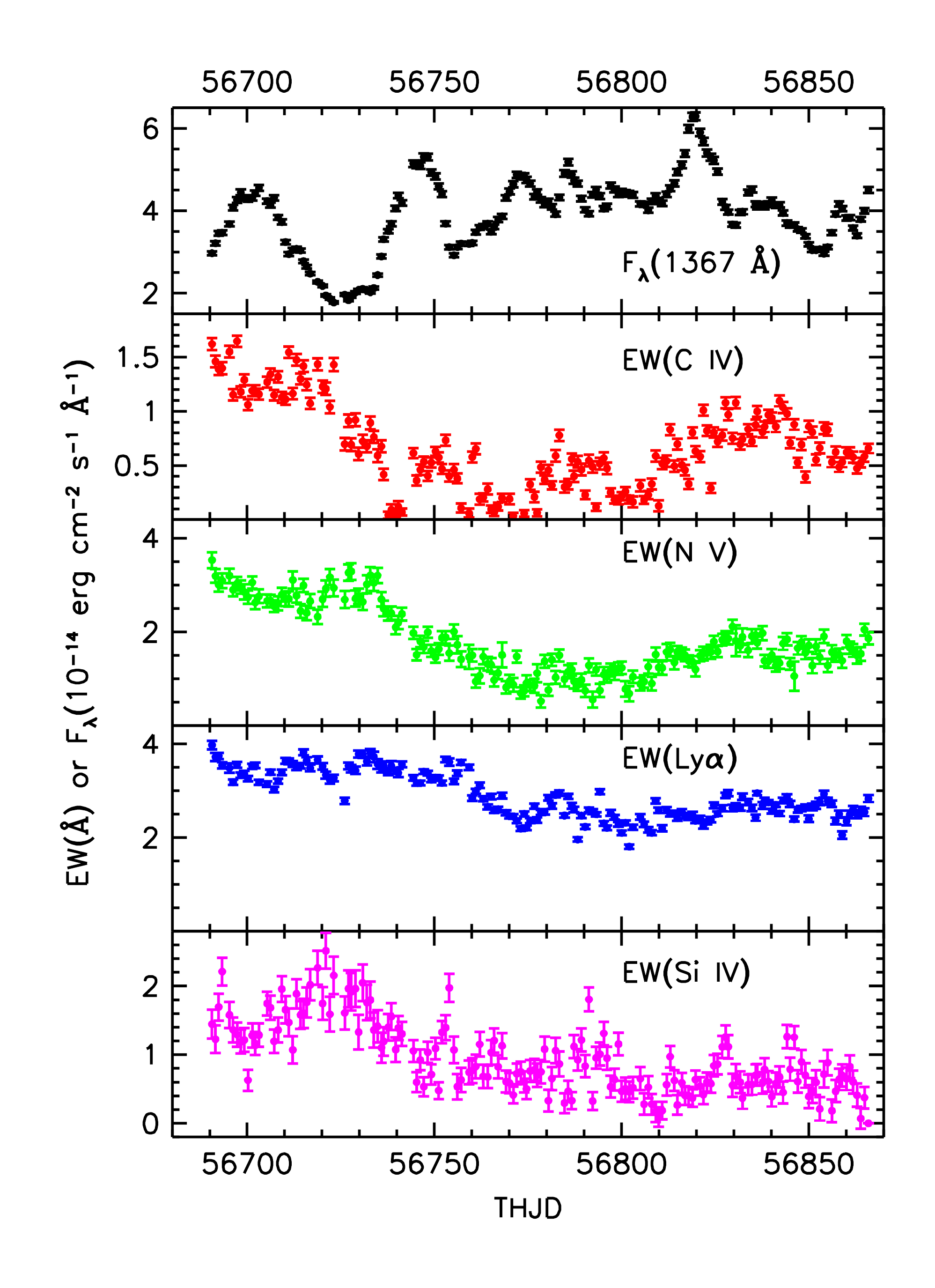}}
\caption{(Top Panel) Light curve for the UV continuum at 1367 \AA\ in units
of $10^{-14}~\rm erg~cm^{-2}~s^{-1}~\mbox{\AA}^{-1}$.
(Second Panel) Absolute value of the equivalent width in \AA\ vs.\,time for the
broad \ion{C}{4} absorption feature.
(Third Panel) Same as second panel for \ion{N}{5}.
(Fourth Panel) Same as second panel for Ly$\alpha$.
(Bottom Panel) Same as second panel for \ion{Si}{4}.
}\label{fig:broad_abs_lcs}
\end{figure}

Table \ref{tab:absorp_lc_sample}\footnote{Table \ref{tab:absorp_lc_sample} appears at the end of the paper to facilitate the formatting.} shows
sample portions of the light curves for the broad absorption in \ion{C}{4}
and the intrinsic narrow absorption associated with \ion{C}{2} $\lambda 1334$.
Full light curves for all features listed in
Table \ref{tab:absorption_intervals} are published in the on-line version
of this paper.
Figure \ref{fig:sample_ew_lcs}
shows sample light curves for the equivalent widths of the
narrow absorption features associated with Component \#1 for
\ion{C}{2} $\lambda 1334$, \ion{Si}{3} $\lambda 1206$, and
\ion{Si}{4} $\lambda 1393$, plus Component \#3 for \ion{Si}{4} $\lambda 1393$,
all compared to the UV continuum flux at 1367 \AA.
Light curves for the broad absorption in \ion{C}{4}, \ion{N}{5}, Ly$\alpha$,
and \ion{Si}{4} are shown in Figure \ref{fig:broad_abs_lcs}.

\section{Analyzing Results from the Models}
\label{sec:results}

\subsection{Velocity Resolved Light Curves for Deblended Emission Lines}
\label{sec:lags}
Our absorption-corrected, deblended emission line profiles described in
\S\ref{sec:deblended_fluxes} allow us to remove the uncertainties in
emission-line lags that may have been introduced by the variable intrinsic
absorption in NGC 5548, as well as to separate the behaviors of
adjacent blended lines.
In addition, for the brightest two lines, Ly$\alpha$ and \ion{C}{4}, we can
determine velocity-binned lags for each line uncontaminated by absorption or
blended contributions from other lines.

Following \cite{DeRosa15} we measured the emission-line lags for the species
tabulated in Table \ref{tab:light_curves_int} by cross-correlating the time
series with the continuum light curve using the
interpolation cross-correlation method (ICCF) as implemented by
\cite{Peterson04}.
The procedure and the resulting associated uncertainties are described in detail
by \cite{DeRosa15}. Briefly, the technique uses a Monte-Carlo method
of ``flux randomization and random subset selection" to generate a large set
of realizations of the light curves.
For each realization, we determine the cross correlation function,
its maximum correlation coefficient $r_{max}$,
and associated peak lag, $\tau_{peak}$.
We also use the region surrounding the peak with $r(\tau) > 0.8$ to
calculate the centroid of the cross-correlation function, $\tau_{cent}$.
A few thousand realizations of each cross-correlation function then gives
distribution functions for $\tau_{peak}$ and $\tau_{cent}$ from which we
measure the median values to give the lags for each emission line as
presented in Table \ref{tab:lags}. The associated uncertainties
represent the 68\% confidence intervals of each Monte-Carlo distribution.

While \cite{DeRosa15} arbitrarily split the data set in two midway through the
campaign, we now know that a more logical breakpoint for examining any changes
is at day 75 in the campaign, which is the beginning of the period when the
broad emission line fluxes become decorrelated from the continuum variations
\citep{Goad16}, also known as the BLR holiday. We therefore quote lags
not only for the full campaign, but also for the first
75 days, the pre-holiday period, when the emission line and continuum fluxes
correlated normally; for the holiday period, days 76--129;
and for the post-holiday period at the end of the campaign.
Comparing the lags in Table \ref{tab:lags} to \cite{DeRosa15},
we see that Ly$\alpha$ is slightly shorter, \ion{Si}{4} is longer, and
\ion{C}{4} and \ion{He}{2} are about the same.
Within the error bars, the lags for the UV model data set are
consistent with the prior results using the original data.

\setcounter{table}{5}
\begin{deluxetable*}{lccccc}
\tablecaption{Emission-Line Lags from the Modeled NGC 5548 Spectra\label{tab:lags}}
\tablehead{\colhead{Emission Line} & \colhead{Ly$\alpha$} & \colhead{\ion{N}{5}} & \colhead{\ion{Si}{4}} & \colhead{\ion{C}{4}} & \colhead{\ion{He}{2}} }
\startdata
\multicolumn{6}{c}{Whole Campaign--Modeled Data} \\
\hline
$\tau_{\rm cent}$ & 5.1$\pm$0.3   & 7$\pm$8       & 8.1$\pm$0.7   & 5.8$\pm$0.5   & 2.2$\pm$0.3   \\
$\tau_{\rm peak}$ & 5.1$\pm$0.6   & 6$\pm$7       & 8.1$\pm$1.0   & 5.7$\pm$0.6   & 1.9$\pm$0.4   \\
$r_{\rm peak}$   & 0.71$\pm$0.03 & 0.17$\pm$0.06 & 0.16$\pm$0.04 & 0.39$\pm$0.04 & 0.56$\pm$0.03 \\
\hline
\multicolumn{6}{c}{Pre-BLR Holiday, THJD=56691--56765} \\
\hline
$\tau_{\rm cent}$ & 4.8$\pm$0.3  & ...   & 8.0$\pm$0.5   & 4.4$\pm$0.3   & 2.4$\pm$0.4   \\
$\tau_{\rm peak}$ & 4.8$\pm$0.4  & ...   & 8.1$\pm$0.6   & 4.5$\pm$0.5   & 2.2$\pm$0.4   \\
$r_{\rm peak}$    & 0.94$\pm$0.01 & ...  & 0.7$\pm$0.04  & 0.91$\pm$0.02 & 0.91$\pm$0.02 \\
\hline
\multicolumn{6}{c}{BLR Holiday, THJD=56766--56829} \\
\hline
$\tau_{\rm cent}$ & 5$\pm$1       & 4.5$\pm$0.4     & 7.3$\pm$1     & 7.1$\pm$0.6   & 2.1$\pm$0.4   \\
$\tau_{\rm peak}$ & 5$\pm$1       & 4.4$\pm$0.6     & 7.3$\pm$1     & 7.3$\pm$0.7   & 2.1$\pm$0.5   \\
$r_{\rm peak}$    & 0.74$\pm$0.06 & 0.79$\pm$0.04   & 0.66$\pm$0.06 & 0.73$\pm$0.07 & 0.75$\pm$0.05 \\
\hline
\multicolumn{6}{c}{Post-BLR Holiday, THJD=56830--56866} \\
\hline
$\tau_{\rm cent}$ & 6$\pm$1       & 5$\pm$3             & 10$\pm$2      & 8$\pm$1       & 7$\pm$5   \\
$\tau_{\rm peak}$ & 7$\pm$1       & 2$^{+1}_{-4}$      & 10$\pm$3      & 8.5$\pm$2.2   & 10$^{+1}_{-8}$   \\
$r_{\rm peak}$    & 0.85$\pm$0.04 & 0.72$\pm$0.06       & 0.77$\pm$0.19 & 0.80$\pm$0.13 & 0.63$\pm$0.09 \\
\hline
\hline
\multicolumn{6}{c}{Whole Campaign--Original Data} \\
\hline
$\tau_{\rm cent}$ & 6.2$\pm$0.3   & ...       & 5.3$\pm$0.7   & 5.3$\pm$0.5   & 2.5$\pm$0.3   \\
$\tau_{\rm peak}$ & 6.1$\pm$0.4   & ...       & 5.4$\pm$1.1   & 5.2$\pm$0.7   & 2.4$\pm$0.6   \\
$r_{\rm peak}$ & 0.77$\pm$0.02 & ... 	& 0.46$\pm$0.06 & 0.36$\pm$0.04 & 0.65$\pm$0.03 \\
\hline
\multicolumn{6}{c}{Pre-BLR Holiday, THJD=56691--56765} \\
\hline
$\tau_{\rm cent}$ & 5.8$\pm$0.3   & ...   & 5.2$\pm$0.8   & 4.3$\pm$0.3   & 2.4$\pm$0.4   \\
$\tau_{\rm peak}$ & 5.9$\pm$0.4   & ...   & 5$\pm$1	    & 4.4$\pm$0.5   & 1.4$^{+0.9}_{-0.1}$ \\
$r_{\rm peak}$ & 0.94$\pm$0.01 & ...   & 0.7$\pm$0.04  & 0.93$\pm$0.02 & 0.90$\pm$0.02 \\
\hline
\multicolumn{6}{c}{BLR Holiday, THJD=56766--56829} \\
\hline
$\tau_{\rm cent}$ & 5.2$\pm$0.6  	  & ...   & 6$\pm$2       & 6.9$\pm$0.9   & 3.3$\pm$0.6   \\
$\tau_{\rm peak}$ & 5$^{+1}_{-0.4}$  	  & ...   & 6$\pm$2       & 7$\pm$1       & 3.3$\pm$0.8   \\
$r_{\rm peak}$ & 0.84$\pm$0.04 	  & ...   & 0.52$\pm$0.09 & 0.64$\pm$0.09 & 0.74$\pm$0.06 \\
\hline
\multicolumn{6}{c}{Post-BLR Holiday, THJD=56830--56866} \\
\hline
$\tau_{\rm cent}$ & 7$\pm$1  	  & ...      & 7$\pm$1       & 8$\pm$1       & 3$\pm$1   \\
$\tau_{\rm peak}$ & 8$\pm$2  	  & ...      & 7$\pm$2       & 8$\pm$2       & 3$\pm$1   \\
$r_{\rm peak}$ & 0.83$\pm$0.08 & ...      & 0.75$\pm$0.11 & 0.82$\pm$0.12 & 0.81$\pm$0.05 \\
\enddata
\tablecomments{Delays measured in days in the rest frame of NGC 5548.}
\tablenotetext{a}{Centroid of the ICCF lag distribution for $r > 0.8 r_{peak}$}
\tablenotetext{b}{Peak lag}
\tablenotetext{c}{Peak correlation coefficient}
\end{deluxetable*}

Comparing results for the different time intervals within the campaign
reveals an interesting evolution in the emission-line lags.
As expected, during the pre-holiday period when the emission-line fluxes
correlated
well with the continuum fluctuations, correlation coefficients are high,
exceeding $r = 0.9$ for Ly$\alpha$, \ion{C}{4} and \ion{He}{2}.
For \ion{N}{5}, however, the correlation is so poor that we cannot determine
a lag in the pre-holiday period, as expected from the lack of any strong
features in this region of the light curve in Figure \ref{fig:light_curves_int}.
During the holiday period, correlation coefficients are lower, but they are
still very good, with $r>0.65$ for all lines.
Lags for all lines during the holiday are about the same as for the pre-holiday
period except for \ion{C}{4}. The \ion{C}{4} lag during the holiday is
significantly longer than for the pre-holiday period by almost three days.
In the post-holiday period, correlation coefficients are again very good,
and \ion{C}{4} shows significantly longer lags compared to the pre-holiday
period. Other lines hint at such a difference, but not significantly.
These changing lags with time explain why the correlation coefficients for
the overall campaign are low despite the much longer data set.
Together with changes in the velocity-resolved lags discussed below,
this may indicate that significant changes in the structure
(or at least our viewpoint) or the illumination, or both, of the BLR are
occurring on the timescale of our campaign.
Given that the orbital timescale at a radius of 1 light day in NGC 5548
is 115 days, such changes seem plausible.

To ensure that this apparent increase in the  emission-line lags over the
course of the campaign is not an artifact of our modeling of the data, we
reanalyzed the original data of \cite{DeRosa15} by splitting it into the same
time intervals. The results are given in the bottom half of
Table \ref{tab:lags}.
The lags for the whole campaign replicate the original results of
\cite{DeRosa15}, and we see the same lengthening of lags toward the end of the
campaign with similar values to those found using the modeled data.

\begin{figure*}
\centering
\resizebox{\hsize}{!}{\includegraphics[angle=0, width=0.9\textwidth]{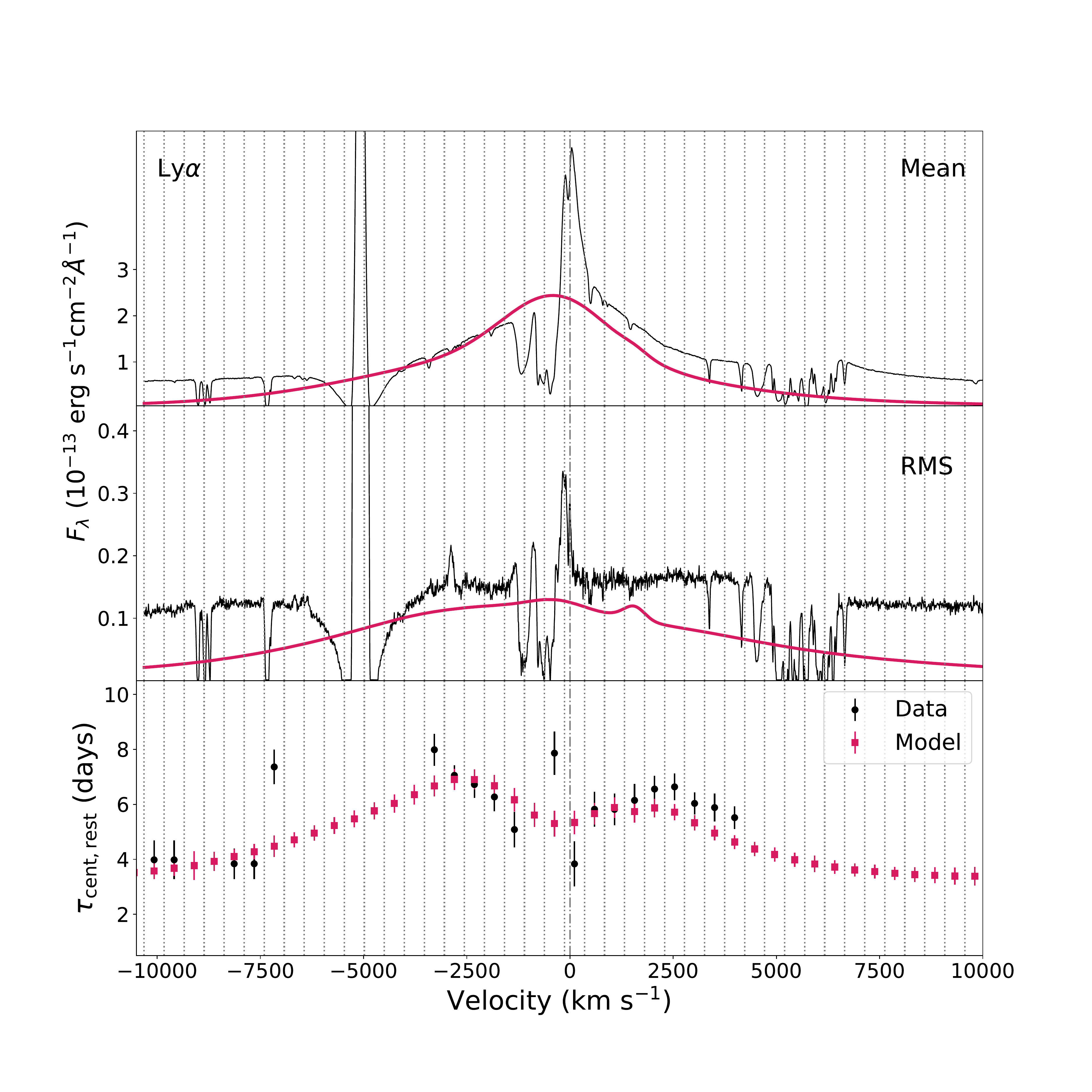}}
\caption{
(Top panel) Comparison of the mean spectrum in the Ly$\alpha$
region of NGC 5548 (black data curve) to the best-fit model spectrum corrected
for absorption with the continuum subtracted and
narrow and intermediate emission-line components removed (smooth red curve).
(Middle panel) Same as top panel for the RMS spectrum.
(Bottom Panel) We compare velocity-binned centroids from the interpolated
cross-correlation function across the Ly$\alpha$ profile obtained using the
original data (black dots) to the absorption-corrected model with narrow and
intermediate emission-line components removed (red squares).
Lags are in the rest frame of NGC 5548.
}\label{fig:lya_velocity_lags}
\end{figure*}

\begin{figure*}
\centering
\resizebox{\hsize}{!}{\includegraphics[angle=0, width=0.9\textwidth]{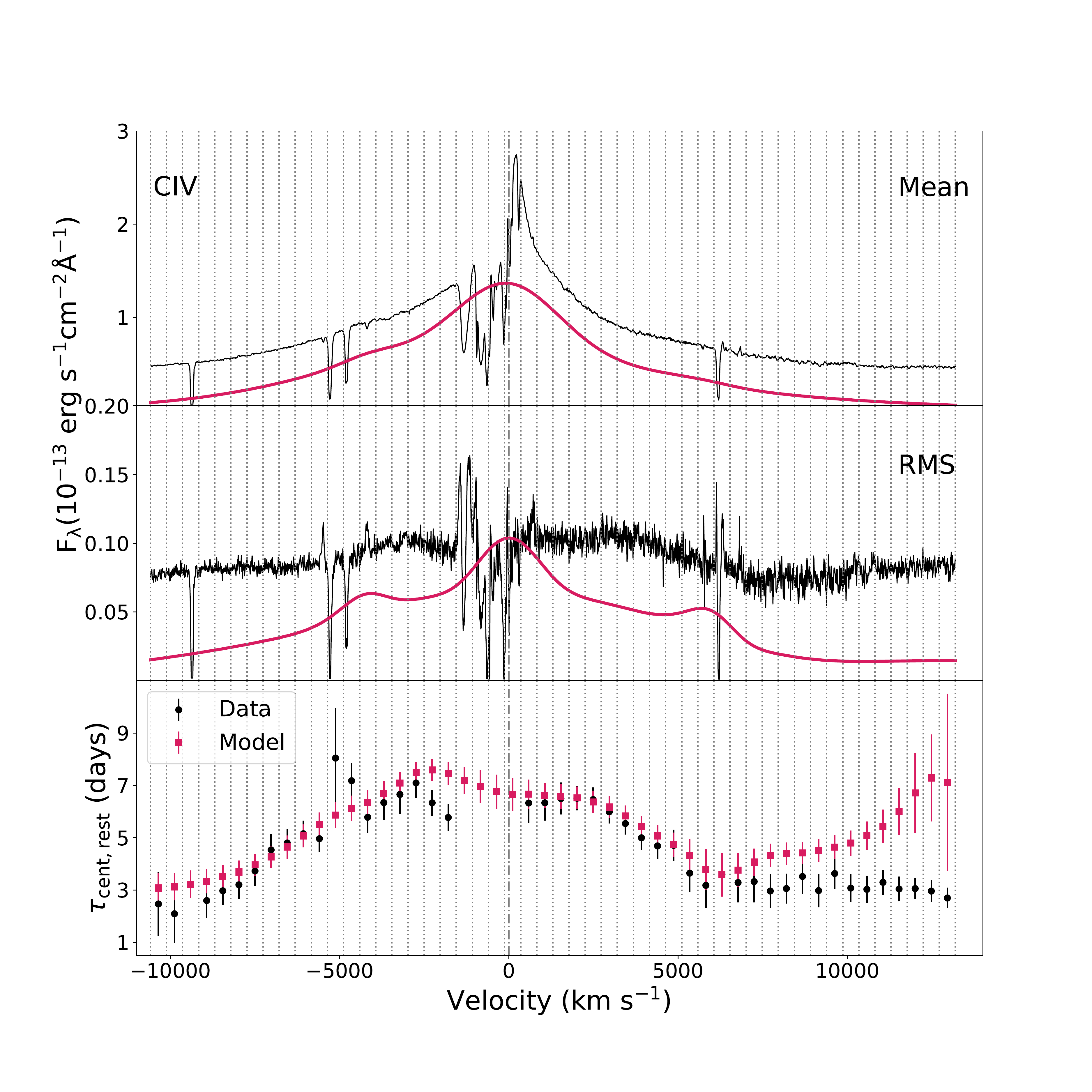}}
\caption{
Same as \ref{fig:lya_velocity_lags}, but for \ion{C}{4}.
}\label{fig:c4_velocity_lags}
\end{figure*}

\begin{figure*}
\centering
\resizebox{\hsize}{!}{\includegraphics[angle=0, width=0.9\textwidth]{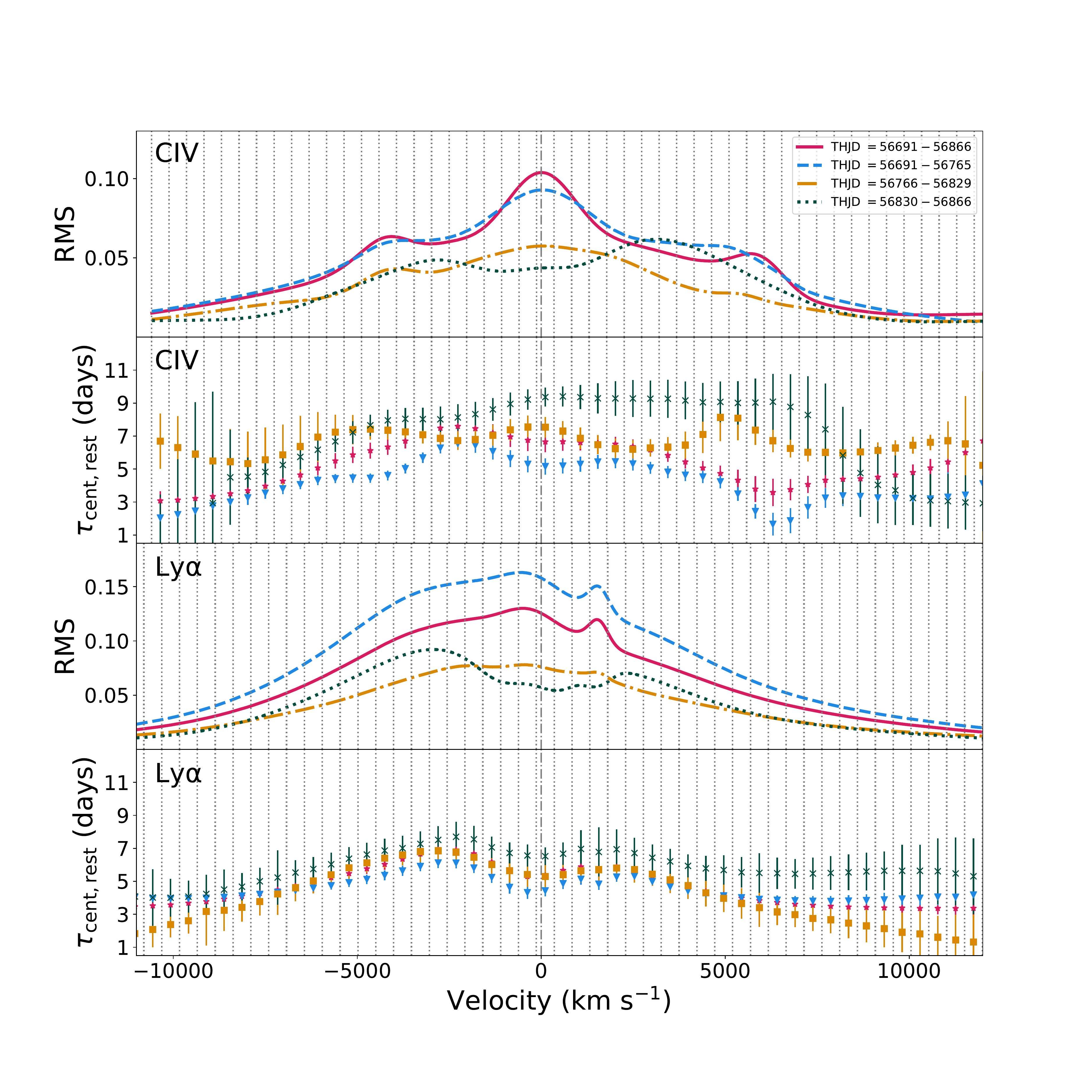}}
\caption{
(Top panel) Comparison of the RMS spectra in the \ion{C}{4} emission-line
region of NGC 5548 for four different time intervals.
All curves refer to the modeled UV spectrum as corrected for absorption with
the continuum subtracted and narrow and intermediate emission-line components
removed.
The smooth red curve is for the full campaign, THJD = 56691--56866.
The dotted blue curve is for the first 75 days of the campaign, the pre-holiday
period, THJD = 56691--56765.
The dash-dot gold curve is for the period of the BLR holiday,
THJD = 56766--56829.
The dotted green curve is for the post-holiday period,
THJD = 56830--56866.
(Second panel) We compare velocity-binned centroids from the interpolated
cross-correlation function across the absorption-corrected model for \ion{C}{4}
with narrow and intermediate emission-line components removed.
Lags are in the rest frame of NGC 5548.
Red triangles show results for the full campaign, THJD = 56691--56860.
Blue triangles are for the first 75 days of the campaign, the pre-holiday
period, THJD = 56691--56765.
Gold squares are for the period of the BLR holiday, THJD = 56765--56829.
Green crosses are for the  post-holiday period, THJD = 56830--56866.
(Third panel) Same as top panel, but for the Ly$\alpha$ emission line.
(Bottom Panel) Same as second panel, but for the Ly$\alpha$ emission line.
}\label{fig:lyac4_velocity_lags}
\end{figure*}

\begin{figure}
\centering
\resizebox{\hsize}{!}{\includegraphics[angle=0, width=0.9\textwidth]{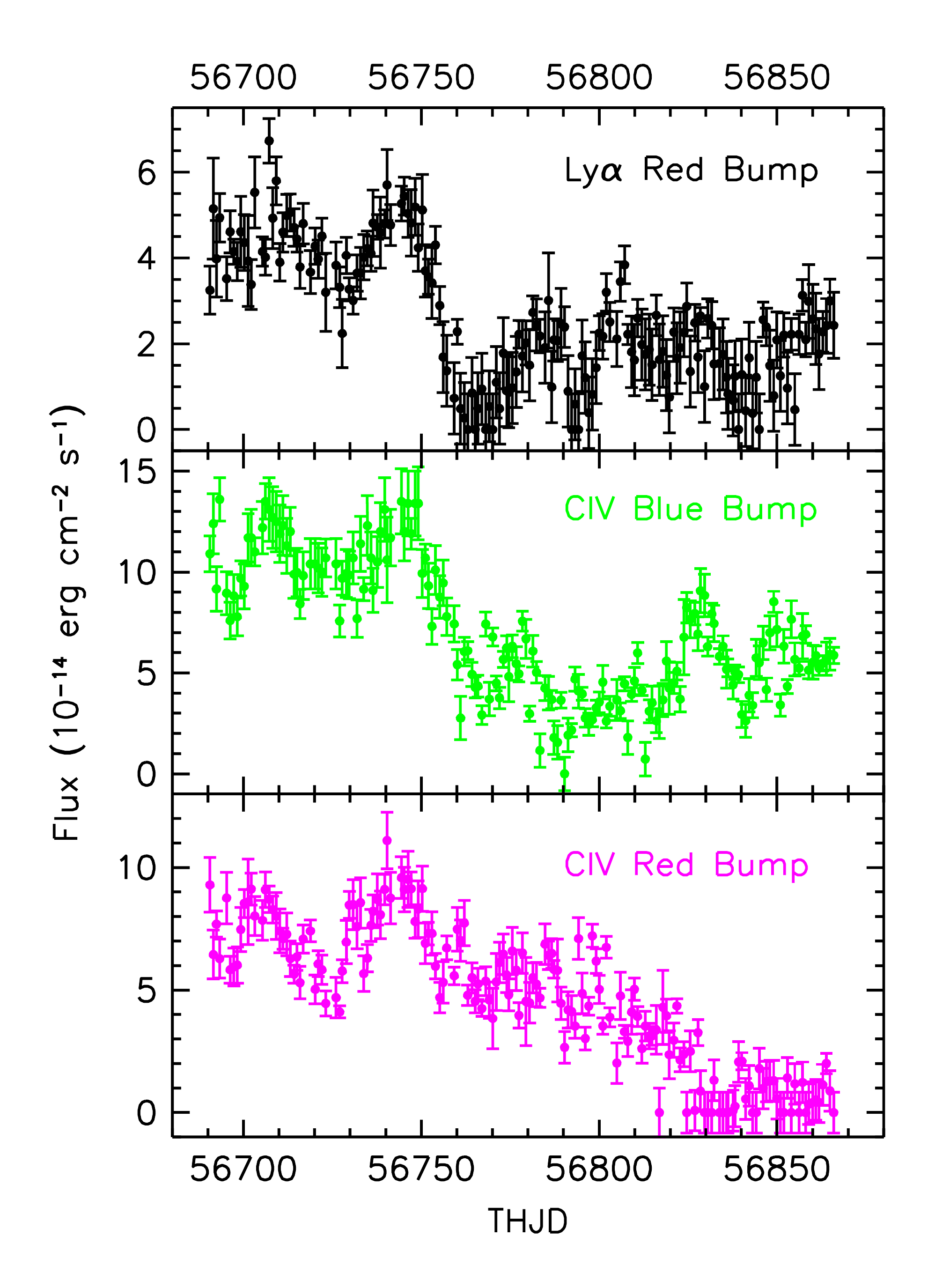}}
\caption{(Top Panel) Light curve (points in black with 1-$\sigma$ error bars)
for the flux in the emission bump on the
red wing of Ly$\alpha$ at $\sim +1500~{\rm km~s^{-1}}$.
(Middle Panel) Same for the emission bump on the
blue wing of \ion{C}{4} at $\sim -4300~{\rm km~s^{-1}}$ (green).
(Bottom Panel) Same for the emission bump on the 
red wing of \ion{C}{4} at $\sim +5700~{\rm km~s^{-1}}$ (magenta).
Fluxes are in units of of $10^{-14}~\rm erg~cm^{-2}~s^{-1}$.
}\label{fig:bump_lcs}
\end{figure}

Also interesting are the velocity-binned results for Ly$\alpha$ and \ion{C}{4}.
Absorption in NGC 5548 largely obscured the inner few thousand $\rm km~s^{-1}$
of the blue side of the profile of each emission line,
and \ion{N}{5} or \ion{He}{2} emission contaminated the far red wings.
Following \cite{DeRosa15}, we use bins of $500~\rm  km~s^{-1}$ spanning each
profile.
Figure \ref{fig:lya_velocity_lags} compares the mean spectrum for the modeled
broad component of Ly$\alpha$, its corresponding root-mean-square (RMS)
spectrum, and finally the velocity-binned profile to the original data from
\cite{DeRosa15}.
All data are for the full campaign.
Our modeled profile provides full velocity coverage across the Ly$\alpha$
emission line.
The most noticeable characteristic of the lag profile is its distinct ``M"
shape, with a local minimum in the lag near zero velocity, and maxima on the
red and blue sides at $\pm2500~\rm km~s^{-1}$.
A prominent feature in the RMS spectrum is the ``red bump" on the
Ly$\alpha$ emission-line profile at $+1500~\rm km~s^{-1}$, which loosely
corresponds to the local peak in the the velocity-dependent lag profile on the
red wing of Ly$\alpha$.

Figure \ref{fig:c4_velocity_lags} shows the corresponding set of results
for the \ion{C}{4} emission line.
For \ion{C}{4} there are emission bumps on both the red and blue wings of the
profile in the RMS spectrum. These bumps are at higher velocity than the red
bump in Ly$\alpha$, at roughly $\pm5000~\rm km~s^{-1}$, and they appear to
correspond to local minima in the lag profile,
as opposed to the maxima seen in Ly$\alpha$.
As with Ly$\alpha$, \ion{C}{4} shows a slight hint of an ``M" shape to its
profile with a shorter lag near the center and local peaks at
$\pm2500~\rm km~s^{-1}$.
However, the contrast is not as distinctive as in Ly$\alpha$.
The central dip in \ion{C}{4} has a confidence level of only $\sim 90$\%.

Examining the velocity-binned profiles for the separate, distinct time intervals
of the campaign reveals even more complex behavior.
Figure \ref{fig:lyac4_velocity_lags} compares the RMS spectra and the
velocity-dependent lags for \ion{C}{4} and Ly$\alpha$ 
from the full campaign to the first 75 days of the campaign, the pre-holiday
period, the period of the BLR holiday, and the post-holiday period
concluding the campaign.
The emission bumps on the red and blue wings of \ion{C}{4} and on the red
wing of Ly$\alpha$ are most prominent early in the campaign, and diminish in
flux (or, disappear in the case of \ion{C}{4} red) by the end of the campaign.
We show light curves for these features in Figure \ref{fig:bump_lcs}.
The red emission bump in \ion{C}{4} also has associated features in the lag
profiles that evolve from a local minimum on the red side of the bump during
the pre-holiday period to a local maximum in the lag on the blue side of the
bump.
These more detailed changes in the emissivity profile and the lag profile
again suggest that we are seeing changes in the structure of the BLR
over the course of the campaign.
This may be due to the presence of some outflowing components, as discussed
in \S5, but this is speculative, and more easily investigated with
two-dimensional reverberation maps and models.

\subsection{Physical Characteristics of the Narrow and Broad Absorbers based on the Mean Spectrum}

The very high S/N of the mean spectrum makes accurate measures of weak
features possible. These are particularly useful since they are often
unsaturated, and can therefore provide better diagnostic information on
physical conditions in the absorbing gas.
The last four columns of Table \ref{tab:absorption_intervals} give the
EW, the velocity relative to the systemic velocity of the host galaxy,
the covering factor, and the inferred column density for the narrow
absorption lines in the mean spectrum.
For Galactic ISM features, the inferred column densities are at best lower
limits since the lines are saturated, and since the profiles have not been
corrected for the COS line spread function.
For absorption lines intrinsic to NGC 5548, the line widths are broad enough
(FWHM typically $> 80~\rm km~s^{-1}$) that the COS line spread function
has little effect.
To measure column densities, we integrate the apparent optical depth across the
absorption line profile between the wavelength limits given in
Table \ref{tab:absorption_intervals},
assuming a uniform covering factor, as given in the next-to-last column of the
table \citep[see Appendix A of][for a description of the technique]{Arav15}.
For doublets and for absorption lines with multiple transitions
(e.g., \ion{P}{5}, or \ion{Si}{2}), we can determine the covering factor such
that integration of each line profile gives consistent column densities.
For other absorption lines, particularly the deep, heavily blended features
associated with Ly$\alpha$, \ion{N}{5}, and \ion{C}{4}, we use a covering factor
determined by the deepest point of the absorption-line profile and assume the
line is saturated. Like \cite{Arav15}, this gives a lower limit on the
column density.

Examining the covering factors in Table \ref{tab:absorption_intervals}
is instructive. Lines far from the centers of the bright emission lines
(e.g., \ion{P}{5}, \ion{Si}{2} $\lambda 1302$, \ion{C}{2} $\lambda 1335$)
have covering factors near unity, indicating that the absorbing gas fully
covers, or nearly fully covers, the continuum emission region.
These absorption lines are also multiplets, so their covering factors are well
determined.
Other absorption lines embedded in the profiles of bright emission lines
such as (\ion{C}{3}* $\lambda 1176$, \ion{Si}{2} $\lambda 1193$,
and \ion{Si}{2} $\lambda 1260$),
have well determined covering factors that are significantly less than unity,
and vary depending on their distance from the center of the emission line.
The sense of this
variation is such that covering factors are lower for lines in the brighter
portions of the emission line profile. From this we conclude that, at least for
Component \#1, the absorbing gas nearly fully covers the continuum-emitting
region, but only covers less than half of the BLR.

The column densities in Table \ref{tab:absorption_intervals} for the
STORM campaign mean spectrum are factors of several lower than those
observed by \cite{Arav15} during the {\it XMM-Newton} campaign.
This is consistent with the higher brightness of NGC 5548 during the
STORM campaign---
$\rm \langle F_{\lambda 1367} \rangle = 4.30 \times 10^{-14}~erg~cm^{-2}~s^{-1}~\mbox{\AA}^{-1}$
vs.\,$\rm \langle F_{\lambda 1367} \rangle = 3.11 \times 10^{-14}~erg~cm^{-2}~s^{-1}~\mbox{\AA}^{-1}$.
Simple scaling of the continuum would imply an increase in the ionization
parameter of $\Delta \rm log$\,$U$ of 0.14. 
Although this seems small, it is sufficient to account for the observed
differences since the weak, low-ionization species in Component \#1 are formed
in a thin hydrogen ionization front, and their column densities are highly
non-linear with changes in ionizing flux.

We tabulate the properties of the broad absorption features separately
since these are most likely associated with the soft X-ray obscurer
discovered by \cite{Kaastra14}.
To obtain the equivalent width (EW), the mean transmission-weighted velocity,
the transmission-weighted velocity dispersion,
and the column density of each broad absorption trough, we use the normalized
mean spectrum of NGC 5548.
For each trough we use the highest blue-shifted velocity at which the trough
drops by more than $1 \sigma$ below the normalized spectrum.
All troughs are integrated up to zero velocity.
Since these broad troughs are
well resolved, we use the apparent optical depth method of \cite{Savage91}
to calculate the column densities of each trough.
The deepest absorption troughs in \ion{N}{5}, \ion{Si}{4}, and
\ion{C}{4} appear to be saturated since they have similar depths at the
velocities of the red and blue members of their respective doublets.
Since Ly$\alpha$ is of similar depth, we also assume it is saturated.
We therefore measure a covering factor $\rm C_f$ at the deepest point of the
trough and use this in our apparent optical depth calculation
\citep{Arav02}. This column density is only a lower
limit to the actual column density.
The shallower absorption troughs are significantly less deep. If they were
saturated and had similar covering factors,
they would likely have similar depths to those of the stronger ions.
We therefore assume that they lie on the linear portion of the curve of growth
and use their apparent optical depths to obtain a direct measure of the
column density assuming covering factors of unity.
Table \ref{tab:broadabscolumns}
summarizes the broad absorption trough properties in detail.

\begin{deluxetable*}{l c c c c c c c c}
  \tablecaption{Properties of the Broad Absorption Troughs in the NGC~5548 Mean Spectrum\label{tab:broadabscolumns}}
\tablehead{ \colhead{Line} & \colhead{$\lambda_o$\tablenotemark{a}} & \colhead{$v_1$\tablenotemark{b}} & \colhead{$v_2$\tablenotemark{c}} & \colhead{$v_o$\tablenotemark{d}} & \colhead{$\sigma_v$\tablenotemark{e}} & \colhead{EW\tablenotemark{f}} & \colhead{$\rm log(N_{ion})$\tablenotemark{g}} & \colhead{$\rm C_f$\tablenotemark{h}} \\
 \colhead{}  & \colhead{(\AA)} & \colhead{($\rm km~s^{-1}$)} & \colhead{($\rm km~s^{-1}$)} & \colhead{($\rm km~s^{-1}$)} & \colhead{($\rm km~s^{-1}$)} & \colhead{(\AA)} & \colhead{($\rm cm^{-2}$)} & \colhead{} }
\startdata
\ion{P}{5} & 1122.99  & $-1388$ & 0 & $-\phantom{0}751$ & $\phantom{0}726$ & $0.07 \pm 0.04$ & $13.11 \pm 0.03$ & 1.0 \\
\ion{C}{3}*  & 1175.8  & $-5323$ & 0 & $-1250$ & $1054$ & $0.06 \pm 0.01$ & $12.72 \pm 0.07$ & 1.0 \\
Ly$\alpha$   & 1215.67  & $-6344$ & 0 & $-1931$ & $1160$ & $2.95 \pm 0.02$ & $>14.80$ & 0.25 \\
\ion{N}{5}   & 1240.51  & $-4907$ & 0 & $-1697$ & $1054$ & $1.70 \pm 0.02$ & $>14.81$ & 0.17 \\
\ion{Si}{2}  & 1260.42 & $-7894$ & 0 & $-4592$ & $\phantom{0}986$ & $0.09 \pm 0.01$ & $12.80 \pm 0.05$ & 1.0 \\
\ion{C}{2}  & 1334.53  & $-2067$ & 0 & $-\phantom{0}953$ & $\phantom{0}439$ & $0.04 \pm 0.01$ & $12.96 \pm 0.09$ & 1.0 \\
\ion{Si}{4} & 1398.27  & $-4913$ & 0 & $-1878$ & $1005$ & $0.84 \pm 0.02$ & $>13.83$ & 0.06 \\
\ion{C}{4}  & 1549.48  & $-4627$ & 0 & $-1893$ & $1091$ & $0.63 \pm 0.01$ & $>14.02$ & 0.07 \\
\enddata
\tablenotetext{a}{Rest wavelength. For doublets, the quoted wavelength is the average.}
\tablenotetext{b}{Starting velocity of the absorption trough.}
\tablenotetext{c}{Ending velocity of the absorption trough.}
\tablenotetext{d}{Transmission-weighted velocity centroid of the absorption trough.}
\tablenotetext{e}{Transmission-weighted velocity dispersion of the absorption trough.}
\tablenotetext{f}{Equivalent width of the absorption trough.}
\tablenotetext{g}{Inferred ionic column density assuming the trough is saturated.}
\tablenotetext{h}{Covering factor at the deepest point in the absorption trough.}
\end{deluxetable*}

\begin{figure}[!tbp]
\centering
\resizebox{1.0\hsize}{!}{\includegraphics[angle=0]{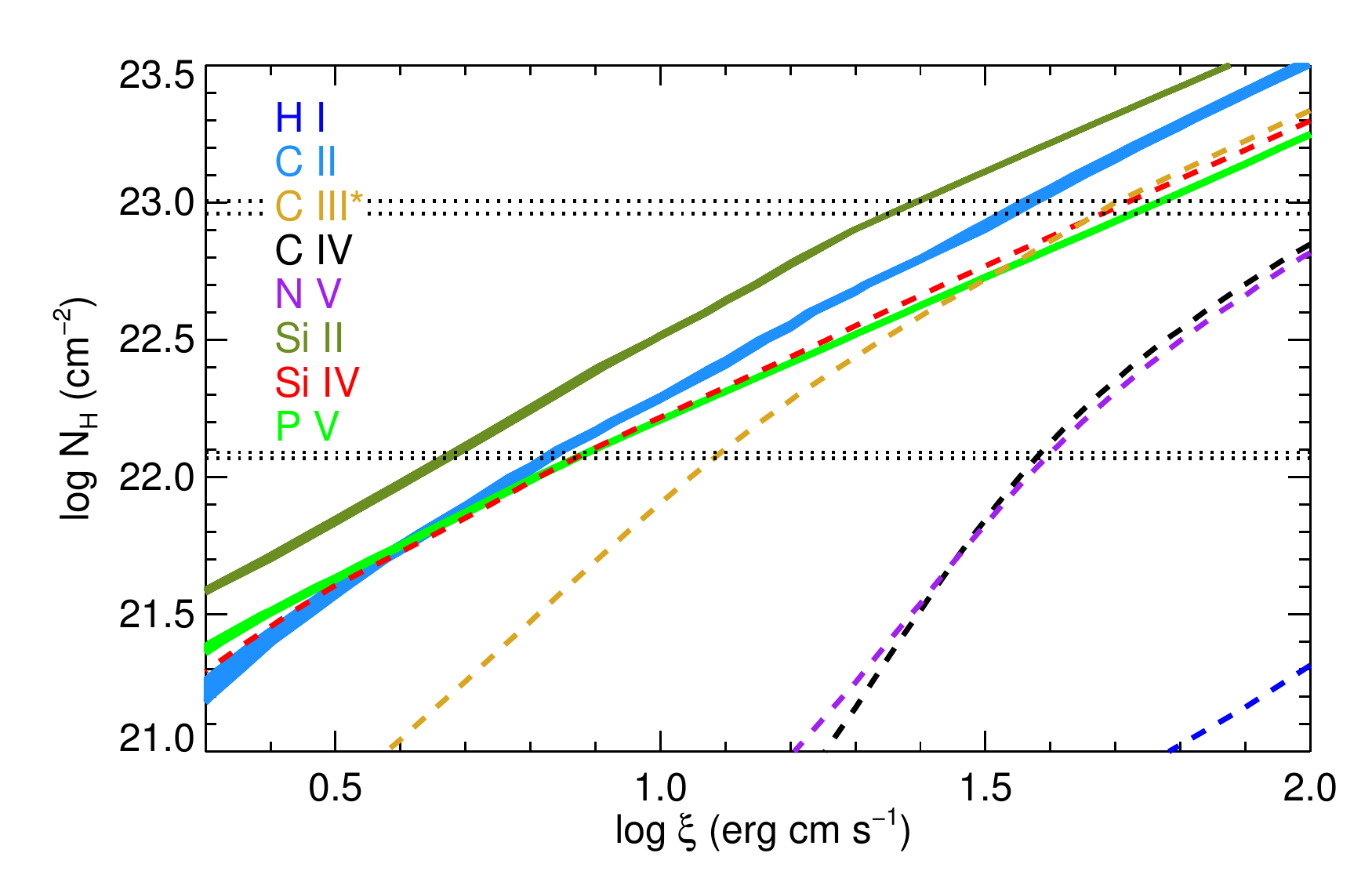}}
\caption{
Photoionization model constraints on the obscurer in NGC~5548.
Dotted black lines give the range of total column density for the two
components of the obscurer \citep{Kaastra14}.
Thick colored bands show the column densities with associated 1-$\sigma$
uncertainties for the weak,
low-ionization troughs in the mean UV spectrum of NGC 5548.
Nominally, photoionization solutions should lie along these lines.
For the strong, saturated features, we use dashed colored lines for
the lower limits on their column densities.
Allowed photoionization solutions lie above these lines.
}
\label{fig:nion_xi}
\end{figure}

We can now use these measures of ionic column densities in the UV together with
the X-ray opacity measurements from the {\it XMM-Newton} spectra of
\cite{Kaastra14} to determine the ionization state of the obscurer more
accurately.
Because of the low X-ray flux resulting from the heavy X-ray absorption,
the {\it XMM-Newton} spectra have no detectable spectral
features that we can use to determine the ionization properties of the obscurer.
However, the X-ray spectrum does provide a
good measure of the total column density.
As in the study of the obscurer in NGC 3783 \citep{Kriss19}, we examine
joint photoionization models that use the UV ionic column
densities along with the total column density determined from the X-ray
observations to more precisely determine the physical properties of the
obscurer. Since the obscuring gas is illuminated by the bare active nucleus in
NGC 5548, we use the unobscured spectral energy distribution (SED) shown
in Figure 2 of \cite{Arav15} for our photoionization models.
Using Cloudy v17.00 \citep{Ferland17}, we run a grid of models covering a range
of $-1.5$ to 2.0 in ionization parameter log $\xi$,
and total column densities from log $\rm N_H = 21.0$ to 23.5.
\footnote{The ionization parameter is defined as $\xi = L_{ion} / (n r^2)$,
where $L_{ion}$ ($\rm erg~s^{-1}$) is the ionizing luminosity obtained by
integrating from 1 to 1000 Ryd, $n$ is the density $(\rm cm^{-3})$,
and $r$ (cm) is the distance of the obscurer from the AGN.
We will also use the ionization parameter defined by
$U = {Q_H} / ({4 \pi r^2 n_H c})$, where
$Q_H$ is the rate of incident ionizing photons above the Lyman limit,
r is the distance to the absorbing gas from the nucleus,
$n_H$ is the total hydrogen number density, and
c is the speed of light.
For the SED of \cite{Arav15}, the conversion from $\xi$ to $U$ is
log $U$ = log $\xi - 1.6$.
}
Figure \ref{fig:nion_xi} shows the allowed space of photoionization solutions,
which are at the two points where the X-ray column densities of the two
obscurer components intersect the measured column densities of
\ion{C}{2}, \ion{C}{3}*, \ion{Si}{2},
and \ion{P}{5} in the grid of photoionization models.
Note that no single solution fits all measured ions, but some of this
incommensurability could be due to the unknown covering fraction of the
weak, low-ionization species.
In addition, all these low-ionization species are produced in very narrow
ionization fronts, so there are likely systematic errors in the photoionization
modeling \citep[see][]{Mehdipour16b}
that are larger than the statistical uncertainties we show in the figure.
\cite{Kaastra14} used only the {\it XMM-Newton} X-ray spectra to determine the
ionization parameter and column density of the obscurer,
Since there are no spectral features to constrain the X-ray models, a broad
range of ionization parameters produces acceptable fits.
As shown by \cite{Mehdipour17}, \cite{Kriss19} and \cite{Longinotti19},
including the UV absorption as an additional constraint indicates that the
obscuring gas likely has higher ionization, even in the observations of the
original {\it XMM-Newton} campaign.
From our new analysis that includes the broad UV absorption as part of the
solution, we see that the obscurer is much more highly ionized than
originally thought.
Component \#1, with $\rm log~N_H~cm^{-2} = 22.08$ actually has an ionization parameter
in the range $\rm log~\xi_1 = 0.8$--0.95, and Component \#2 is even more highly
ionized at $\rm log~\xi_2 = 1.5$--1.6.
These differ sufficiently from the original fits of \cite{Kaastra14}
($\rm log~\xi_1 = -0.8$ and $\rm log~\xi_2 = -4.5$)
that another look at the X-ray spectral analysis is warranted.

\subsection{Variability of the Narrow Absorption Features}
The narrow intrinsic absorption features in NGC 5548 lie at distances of
$\sim3$ to hundreds of parsecs \citep{Arav15}.
They vary on timescales of days
(as seen above in Figure \ref{fig:sample_ew_lcs})
to years \citep{Arav15}, and appear to be associated with the X-ray
warm absorber in NGC 5548 \citep{Mathur95,Crenshaw09,Kaastra14, Arav15}.
Using density sensitive transitions in the metastable excited states of
\ion{C}{3} and \ion{Si}{3}, \cite{Arav15} determined the density of the
gas producing the absorption features associated with Component \#1 as
$\rm log~n~cm^{-3} = 5.8 \pm 0.3$, placing it at a distance of
$3.5 \pm 1.0$ pc, consistent with the density and the 1--3 pc location of the
emission-line gas in the NLR \citep{Peterson13}.
Using the time variability of these absorption features, we can independently
measure the density by measuring the recombination time in the gas.
\cite{Krongold07} used the variability of the aggregate soft X-ray absorption
as it responded to continuum flux changes in NGC 4051 to estimate
recombination times, but here we have the opportunity to make such measurements
with distinct, resolved, absorption lines in the UV.

For gas in photoionization equilibrium, the population density $n_i$,
in a state $i$
depends on the balance between the ionizing photon flux causing ionizations to
more highly ionized states and recombinations from those states.
Following \cite{KK95}, this can be expressed as

\begin{equation}
dn_i/dt = -(F_{ion,i} \sigma_{ion,i} + n_e \alpha_{rec,i-1}) n_i 
\end{equation}
$$\phantom{0000000000} + n_e n_{i+1} \alpha_{rec,i} + F_{ion,i-1} \sigma_{ion,i-1} n_{i-1} .$$

For the ions we are measuring, generally $n_{i-1} \ll n_i \ll n_{i+1}$,
so we can simplify to
\begin{equation}
 dn_i/dt = -F_{ion,i} \sigma_{ion,i} n_i + n_e n_{i+1} \alpha_{rec,i} .
\end{equation}

\begin{figure}
\centering
\resizebox{\hsize}{!}{\includegraphics[angle=0, width=0.9\textwidth]{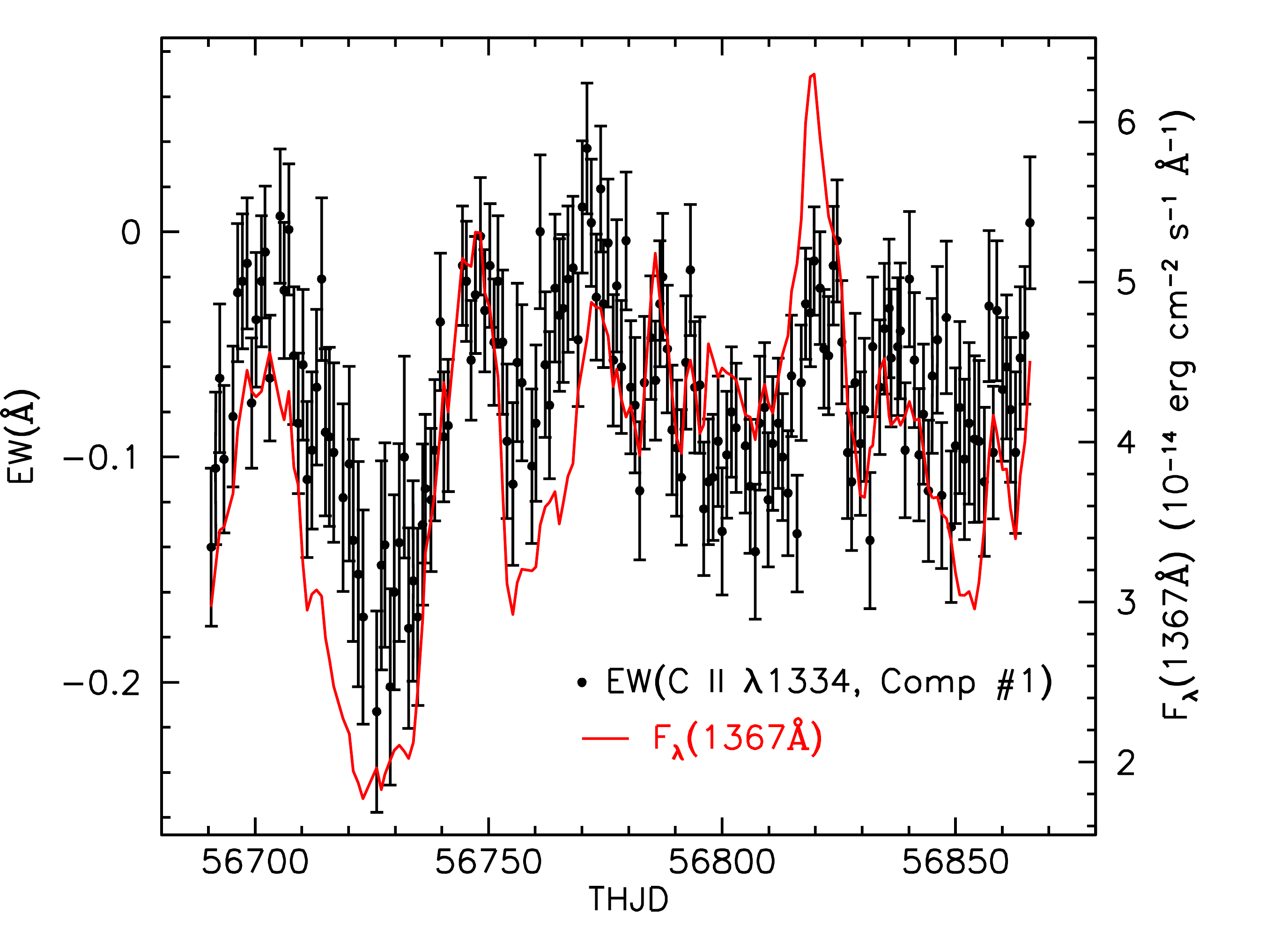}}
\caption{UV continuum light curve at 1367 \AA\ (red line) overlaid on the
variations in equivalent width for the absorption in \ion{C}{2} $\lambda 1334$
(black points)
associated with the intrinsic narrow absorption component \#1 in NGC 5548.}\label{fig:c2_ion_recom}
\end{figure}

In general, as long as there is a copious increase in the ionizing flux, the
$-F_{ion,i} \sigma_{ion,i} n_i$ term dominates,
and ions $n_i$ are destroyed instantly.
Conversely, when the flux decreases abruptly, $n_e n_{i+1} \alpha_{rec,i}$
dominates, and $n_i$ reappears more slowly, on the recombination timescale
\begin{equation}
 \tau_{rec} = (n_i / n_{i+1}) / (n_e \alpha_{rec,i}).
\end{equation}
Figure \ref{fig:c2_ion_recom} beautifully illustrates this simple behavior.
We can see the absolute magnitude of the EW of the
\ion{C}{2} $\lambda 1334$ absorption features decrease
in absolute value immediately when the continuum flux increases.
Conversely, when the continuum flux decreases, there is a noticeable delay
in the \ion{C}{2} response as it takes a measurable amount of time for
recombinations to repopulate the \ion{C}{2} ionization state.

Our high data quality and good sampling enable us to measure recombination
delays directly from our light curves.
To obtain an objective, empirical measure of the recombination time, we
cross-correlated the continuum light curve with the absorption line light
curves
\citep[using the interpolated cross-correlation function (ICCF) of][]{Peterson04},
but restricted our cross-correlation to time intervals when the
continuum flux had reached a peak and then fell to a minimum.
We obtained good cross-correlations only for a select number of absorption lines
in Components \#1 and \#3. We tabulate the measured recombination times in
Table \ref{tab:recombination_times}.
To convert these times into densities, we use Cloudy 17.00
\citep{Ferland17} to obtain the recombination rates for each of
the ions in Table \ref{tab:recombination_times}, assuming a fiducial density of log $\rm n = 4.8~cm^{-3}$.
The best-fit photoionization solutions from \cite{Arav15} give
log $\rm U = -1.5$ and log $\rm N_H = 21.5~cm^{-2}$ for Component \#1, and log $\rm U = -1.3$ and log $\rm N_H = 21.2~cm^{-2}$ for
Component \#3. We scale up the ionization parameters by the
ratio of the continuum fluxes at 1367 \AA\ for the mean spectrum
from the {\it XMM-Newton} campaign ($3.11 \times 10^{-14}~\rm erg~cm^{-2}~s^{-1}~\mbox{\AA}^{-1}$) to the value from the mean spectrum of the STORM campaign
($4.30 \times 10^{-14}~\rm erg~cm^{-2}~s^{-1}~\mbox{\AA}^{-1}$).
The {\tt print ionization rates} command in Cloudy then gives
total recombination rates for the relevant ionic states, and we
scale the fiducial density of log $\rm n = 4.8~cm^{-3}$
by the ratio of the inferred
recombination time from Cloudy to our measurements in
Table \ref{tab:recombination_times} to obtain the tabulated
inferred densities.

\begin{deluxetable}{lcc}
\tablecaption{Recombination Time Scales and Densities in NGC 5548 Absorption Components\label{tab:recombination_times}}
\tablehead{\colhead{Feature} & \colhead{$\tau_{rec}$} & \colhead{log $n_e$} \\
       & \colhead{(days)} & \colhead{($\rm cm^{-3}$)} \\}
\startdata
 \ion{C}{2}$\lambda1334$ \#1 & $2.84 \pm 0.69$ & $5.92 \pm 0.4$ \\
 \ion{Si}{3}$\lambda1206$ \#1 & $8.12 \pm 0.95$ & $4.99 \pm 0.1$ \\
 \ion{Si}{4}$\lambda1393$ \#1 & $3.81 \pm 0.71$ & $5.70 \pm 0.3$ \\
 \ion{Si}{3}$\lambda1206$ \#3 & $5.83 \pm 0.77$ & $5.13 \pm 0.1$ \\
 \ion{Si}{4}$\lambda1393$ \#3 & $5.54 \pm 0.71$ & $5.53 \pm 0.1$ \\
\enddata
\end{deluxetable}

These density measurements are reassuring, 
both for their internal consistency and for their
agreement with the independent determination of log $n_e = 4.8 \pm 0.3$
obtained by \cite{Arav15} using density sensitive absorption line ratios.
Despite our simplifying assumptions, it is gratifying that the atomic physics
of this gas produces such consistent results.
\cite{Silva19} discuss the limitations of our simplifying assumptions in
more detail, and produce a more rigorous time-dependent photoionization model
of the various light curves that verify these empirical results
with greater precision.

\begin{figure}
\centering
\resizebox{\hsize}{!}{\includegraphics[angle=0, width=0.9\textwidth]{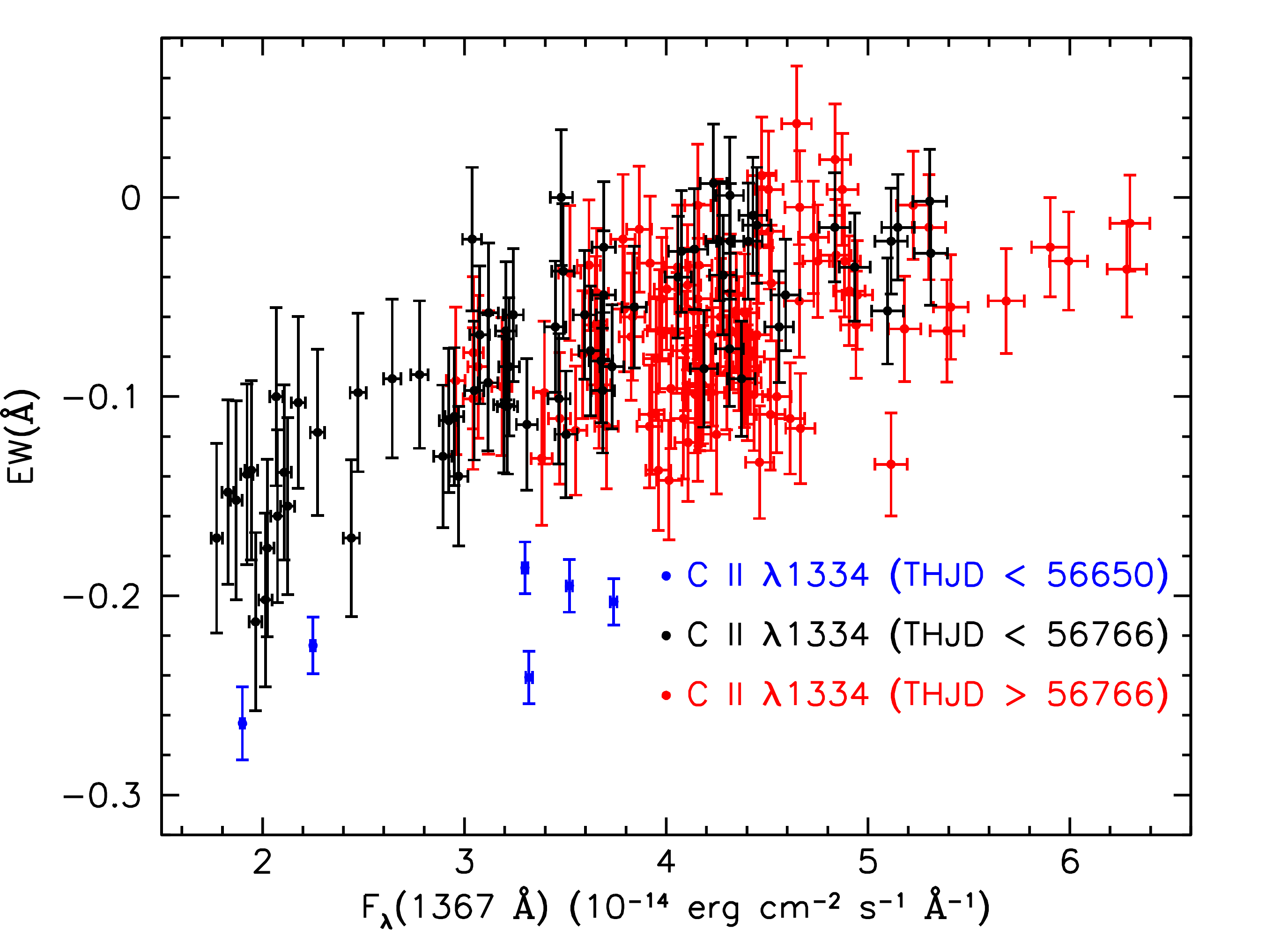}}
\caption{Correlation of the
equivalent width of the absorption line \ion{C}{2} $\lambda 1334$
associated with the intrinsic narrow absorption component \#1 in NGC 5548
with the UV continuum flux at 1367 \AA.
Blue points are for the six observations of the {\it XMM-Newton} campaign,
all prior to THJD=56650;
black points are for dates in the STORM campaign earlier than THJD=56766;
red points are for later times.
}\label{fig:c2_correlation}
\end{figure}

This classic ionization response describes the light curves of
the low-ionization transitions visible in Component \#1.
However, this idealized behavior only holds true for
approximately the first third of the campaign. At later times, the absorption
lines do not follow the continuum so closely, during either the ionization or
recombination phases.
The correlation plot comparing EW(\ion{C}{2}) to the UV continuum flux
$\rm F_\lambda(1367 \mbox{\AA})$ in Figure \ref{fig:c2_correlation}
shows a good linear correlation (linear correlation coefficient $r = 0.82$)
for the first 75 days of the campaign, or
THJD$<$56766,
but much more scatter at later times ($r = 0.39$).

\begin{figure}
\centering
\resizebox{\hsize}{!}{\includegraphics[angle=0, width=0.9\textwidth]{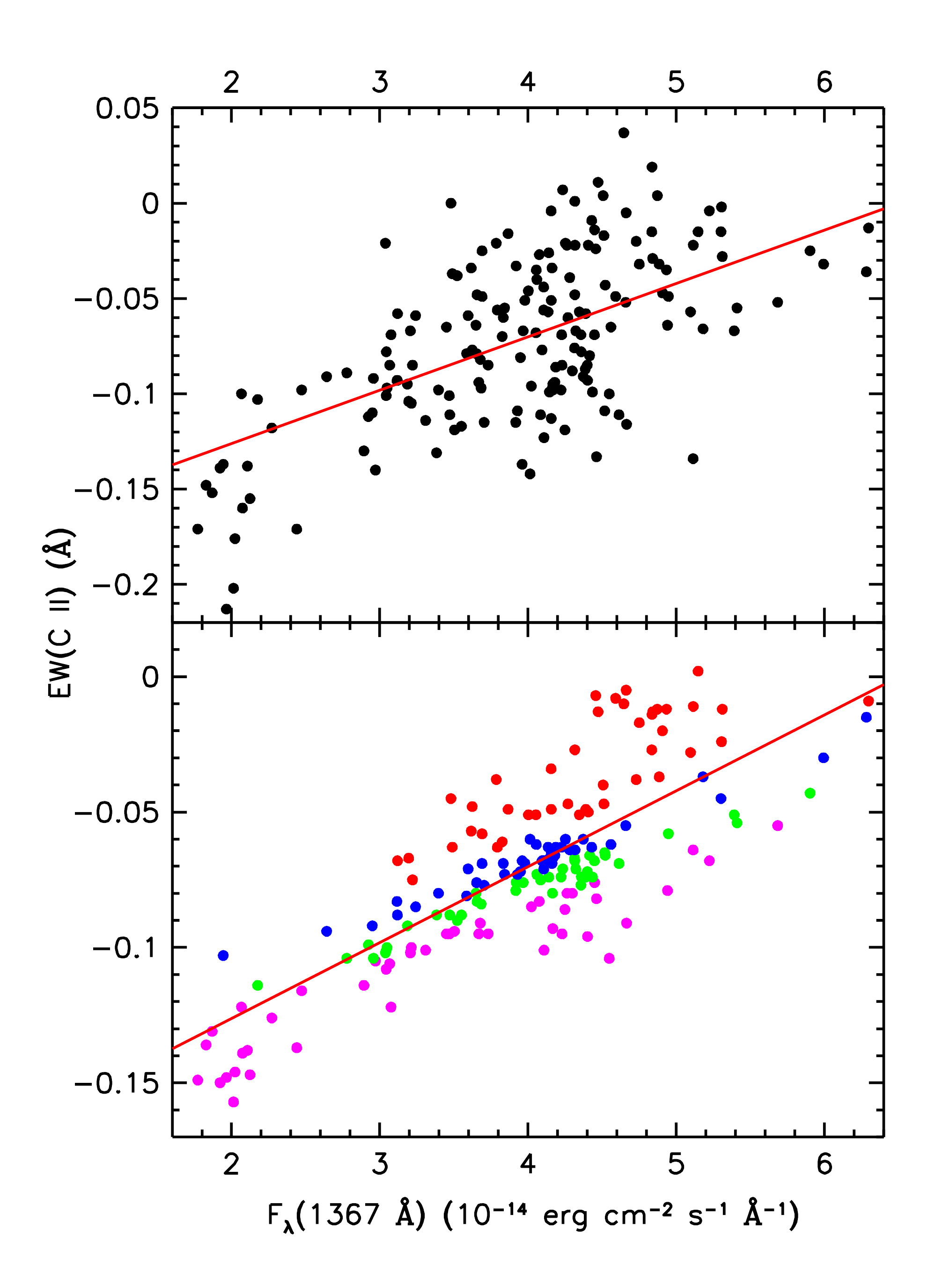}}
\caption{Fits to the correlation of the
equivalent width of the absorption line \ion{C}{2} $\lambda 1334$
associated with the intrinsic narrow absorption component \#1 in NGC 5548
with the UV continuum flux at 1367 \AA.
The top panel shows EW(\ion{C}{2}) (black points) for the entire STORM campaign.
The solid red line shows the best-fit linear correlation of EW(\ion{C}{2}) with
the UV continuum flux, $F_\lambda(1367 \mbox{\AA})$.
(Bottom Panel) Colored points show predicted equivalent widths at a given UV
flux for a joint correlation with the {\it Swift} hardness ratio, HR.
Red points have HR$<$0.768,
blue points have 0.768$\leq$HR$<$0.81,
green points have 0.81$\leq$HR$<$0.845, and
magenta points have HR$\geq$0.845.
}\label{fig:fitc2}
\end{figure}

A partial explanation for the inconsistent correlation of the absorption line
strength with the observed UV continuum is that the observed continuum is only
one determinant of the actual ionizing flux beyond the Lyman limit.
Historically, the extreme ultraviolet (EUV) continuum varies with greater
amplitude than the FUV \citep{Marshall97}.
We also know that the soft X-ray continuum is obscured by optically thick gas
that only partially covers the continuum source \citep{Kaastra14}.
This opaque partial coverage also shadows the ionizing ultraviolet, as shown by
the photoionization analysis of the UV absorption lines \citep{Arav15}.
The obscuration is also variable, with the variability predominantly
explained by variations in the covering fraction
\citep{DiGesu15, Cappi16, Mehdipour16}.
If the covering fraction of the obscurer is varying, then one would expect
variations in the ionizing flux that are independent of the strength of the
observed UV continuum.
To test this hypothesis, we did a joint correlation analysis of the variations
in EW(\ion{C}{2}) with the UV continuum and the hardness ratio as measured
with {\it Swift}.
The hardness ratio is defined as $\rm HR = (HX - SX) / (HX + SX)$, where
SX is the soft X-ray count rate in the 0.3--0.8 keV band, and HX is the
count rate in the hard X-ray band, 0.8--10.0 keV \citep{Edelson15}.
The hardness ratio is almost a direct measure of the covering fraction of the
obscurer \citep{Mehdipour16}.
A linear fit of the EW(\ion{C}{2}) to the UV continuum flux,
$F_\lambda (1367 \mbox{\AA})$ yields the red line shown in
Figure \ref{fig:fitc2}, with $\chi^2$ = 224.2 for 171 points and two degrees
of freedom. This is a good correlation, but the fit is not a statistically
acceptable predictor of the strength of the EW(\ion{C}{2}).
If we include the hardness ratio HR as an additional independent variable,
the goodness of fit improves dramatically to $\chi^2$ = 156.1, which is
statistically acceptable.
This strongly bolsters the interpretation that the ionizing continuum for
\ion{C}{2} is determined both by the observed intensity of the UV continuum
as well as the fraction of the ionizing continuum that is covered by the
X-ray obscurer.

The bottom panel of Figure \ref{fig:fitc2}
shows how including HR as a predictor of the
EW(\ion{C}{2}) improves the fit. The simple red line is the prediction based
solely on the UV continuum flux. The observed equivalent widths have a large
scatter about this line, as seen in the black points.
When the hardness ratio for each observation is taken into account, the
predicted equivalent widths then vary from the simple linear fit in a manner
consistent with variations in the covering fraction. For a given UV flux,
high hardness ratios, indicative of high covering fractions, mean more of the
ionizing continuum is obscured, so predicted equivalents widths are greater
in magnitude (more negative).
For low hardness ratios, covering fractions are lower, more ionizing flux leaks
past the obscurer, and equivalent widths decrease in magnitude as more
\ion{C}{2} is ionized.

An alternative possibility is that the shape of the ionizing continuum is
varying, and that the soft X-ray flux might be a good measure of this when
combined with the UV continuum flux.
We therefore tested a joint correlation of the EW(\ion{C}{2}) with the
UV continuum flux and the soft X-ray count rate, SX.
This also gives an improved fit, $\chi^2$ = 180.7, which is statistically
acceptable, but not as good as the fit using HR.
The improvement can readily be explained as a consequence of SX being
determined partially by intrinsic flux variations, but mostly by variations in
the covering fraction of the obscurer, as argued by \cite{Mehdipour16}.

\begin{figure}
\centering
\resizebox{\hsize}{!}{\includegraphics[angle=0, width=0.9\textwidth]{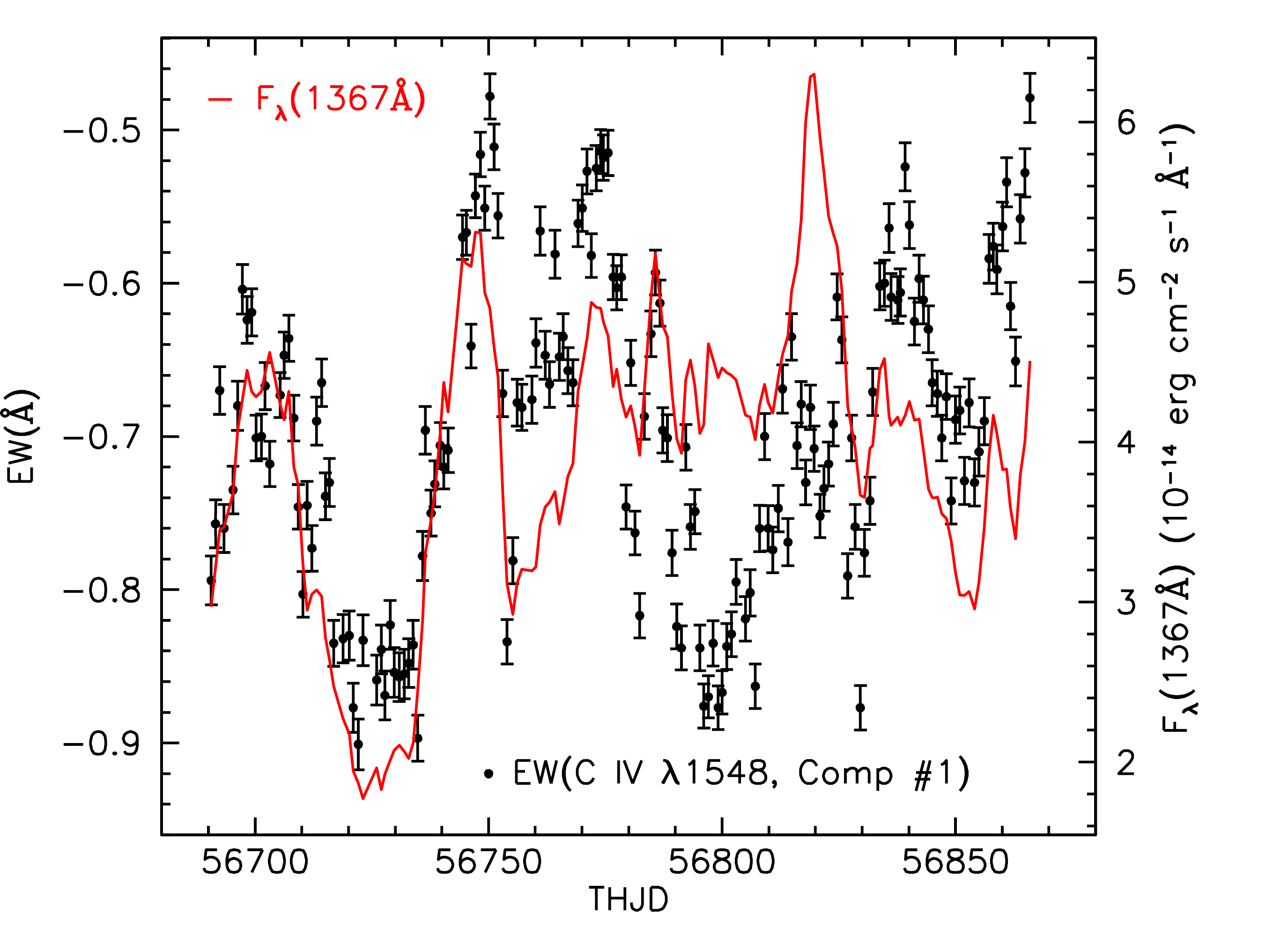}}
\caption{UV continuum light curve at 1367 \AA\ (red line) overlaid on the
variations in equivalent width for the absorption in \ion{C}{4} $\lambda 1548$
(black points)
associated with the intrinsic narrow absorption component \#1 in NGC 5548.}\label{fig:c4_ion_recom}
\end{figure}

\begin{figure}
\centering
\resizebox{\hsize}{!}{\includegraphics[angle=0, width=0.9\textwidth]{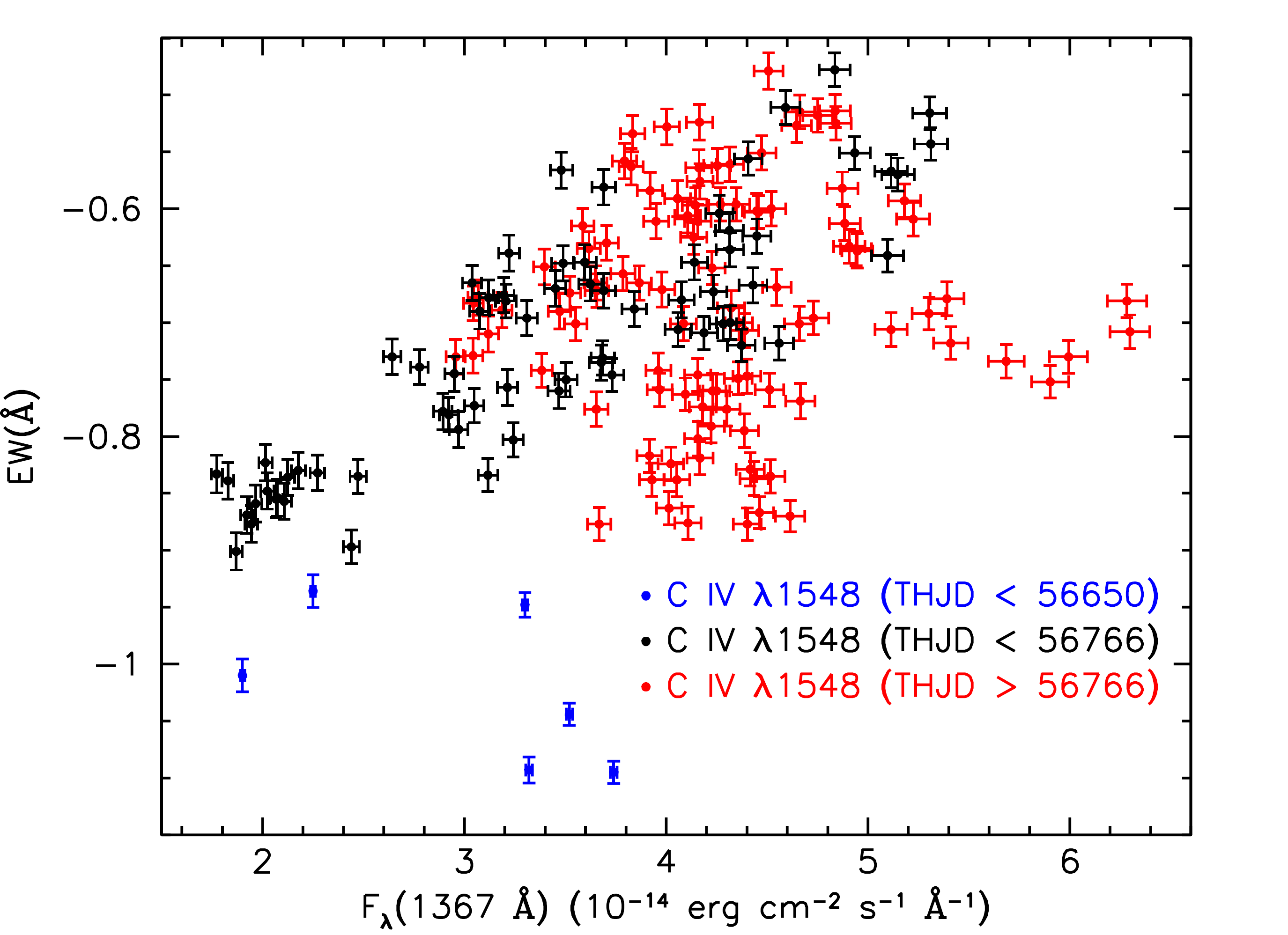}}
\caption{Correlation of the
equivalent width of the absorption line \ion{C}{4} $\lambda 1548$
associated with the intrinsic narrow absorption component \#1 in NGC 5548
with the UV continuum flux at 1367 \AA.
Blue points are for the six observations of the {\it XMM-Newton} campaign,
all prior to THJD=56650;
black points are for dates in the STORM campaign earlier than THJD=56766;
red points are for later times.
}\label{fig:c4_correlation}
\end{figure}

The lack of a direct, exclusive correlation with the observed UV continuum flux
is even more striking for
all the high-ionization absorption lines,
\ion{Si}{4}, \ion{C}{4} and \ion{N}{5}.
Figure \ref{fig:c4_ion_recom} illustrates these effects in the blue
component of \ion{C}{4} $\lambda$1548 associated with the absorbing gas in
Component \#1.
This transition has the highest blue-shift of any narrow intrinsic absorption
feature for \ion{C}{4},
and therefore it is not blended with any other components.
In Figure \ref{fig:c4_ion_recom} one can see that it tracks the UV continuum
variations very closely up to THJD=56766,
both during increases in flux and during decreases in flux.
For the first 75 days of the STORM campaign, before the BLR holiday,
\ion{C}{4} has a linear correlation coefficient of $r = 0.86$.
For the remainder of the campaign, however, the correlation drops
significantly, with $r = 0.06$.
In contrast to \ion{C}{2}, \ion{C}{4} does not show any recombination
delay when continuum flux levels fall, even during the first 75 days of the
campaign;
the absorption increases in strength almost immediately
when the flux levels rise (for THJD$<$56766).
The other high-ionization ions, \ion{Si}{4} and \ion{N}{5}, also
show this instantaneous response.
At later times, however, \ion{C}{4} becomes almost completely decorrelated.
The correlation plot comparing the EW(\ion{C}{4}) to the UV continuum flux
$\rm F_\lambda(1367 \mbox{\AA})$ in Figure \ref{fig:c4_correlation}
shows a tight correlation with the UV continuum during the first 75 days,
just like for \ion{C}{2}, but a large degree of scatter at later times.
During the period of the BLR holiday, after THJD=56776, there is no coherent
correlation between EW(\ion{C}{4}) and the UV continuum flux.
The particularly discordant points in the upper right corner of
Figure \ref{fig:c4_correlation} correspond to the FUV continuum flux peak
near the end of the holiday period at THJD=56820.

\begin{figure}
\centering
\resizebox{\hsize}{!}{\includegraphics[angle=0, width=0.9\textwidth]{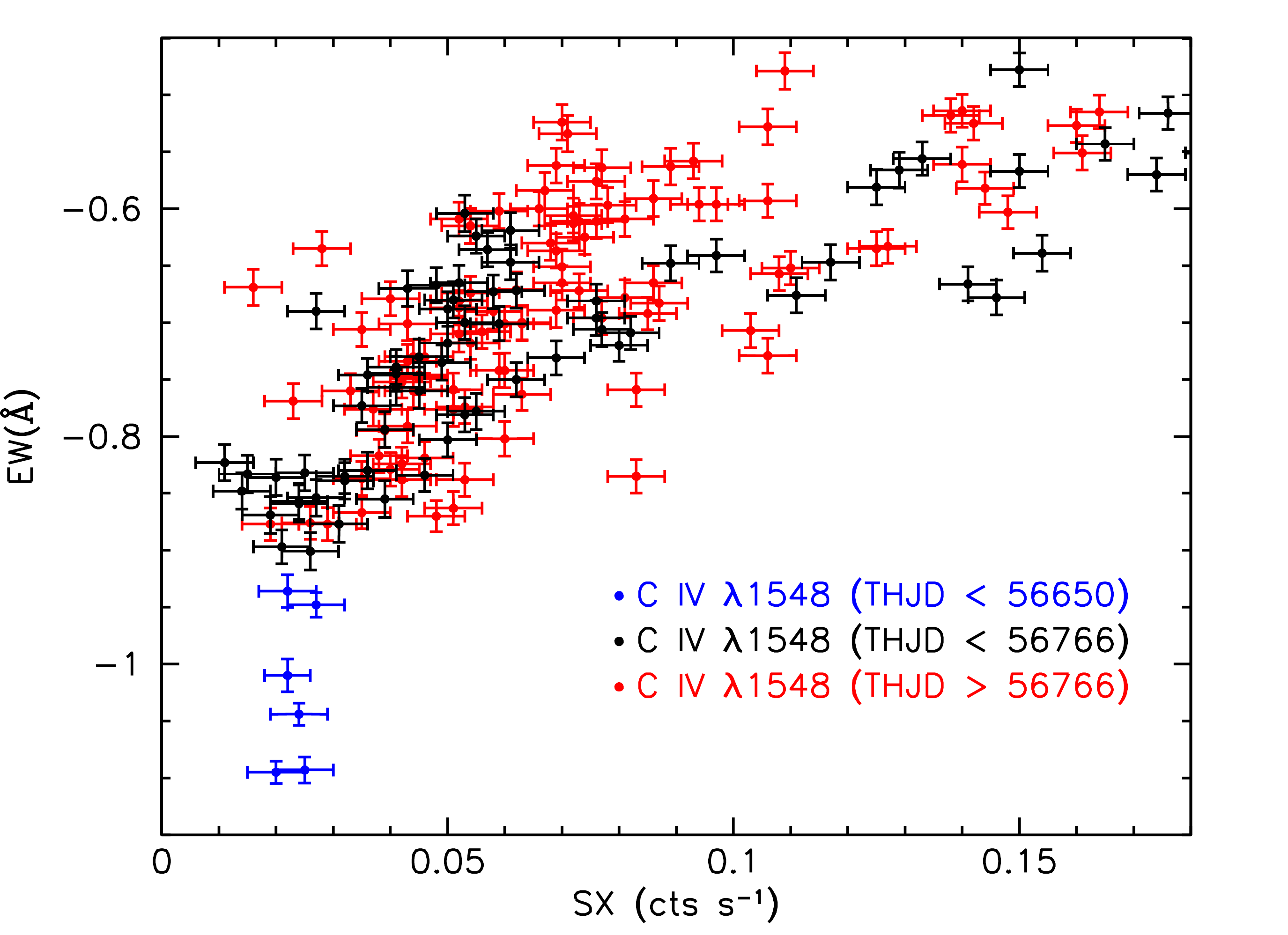}}
\caption{Correlation of the
equivalent width of the absorption line \ion{C}{4} $\lambda 1548$
associated with the intrinsic narrow absorption component \#1 in NGC 5548
with the {\it Swift} soft X-ray (SX) count rate.
Blue points are for the six observations of the {\it XMM-Newton} campaign,
all prior to THJD=56650;
black points are for dates in the STORM campaign earlier than THJD=56766;
red points are for later times.
}\label{fig:c4_sx_correlation}
\end{figure}

As for \ion{C}{2}, we investigated whether other variables might also have a
strong correlation with the ionizing flux that is controlling the population of
\ion{C}{4} ions.
In Figure \ref{fig:c4_sx_correlation} we compare EW(\ion{C}{4}) to the soft
X-ray flux as measured by {\it Swift} \citep{Edelson15}.
Although not perfect, the correlation improves significantly, suggesting
that the soft X-ray flux plays a more dominant role in controlling the
\ion{C}{4} ionic population than the FUV continuum.

\begin{figure}
\centering
\resizebox{\hsize}{!}{\includegraphics[angle=0, width=0.9\textwidth]{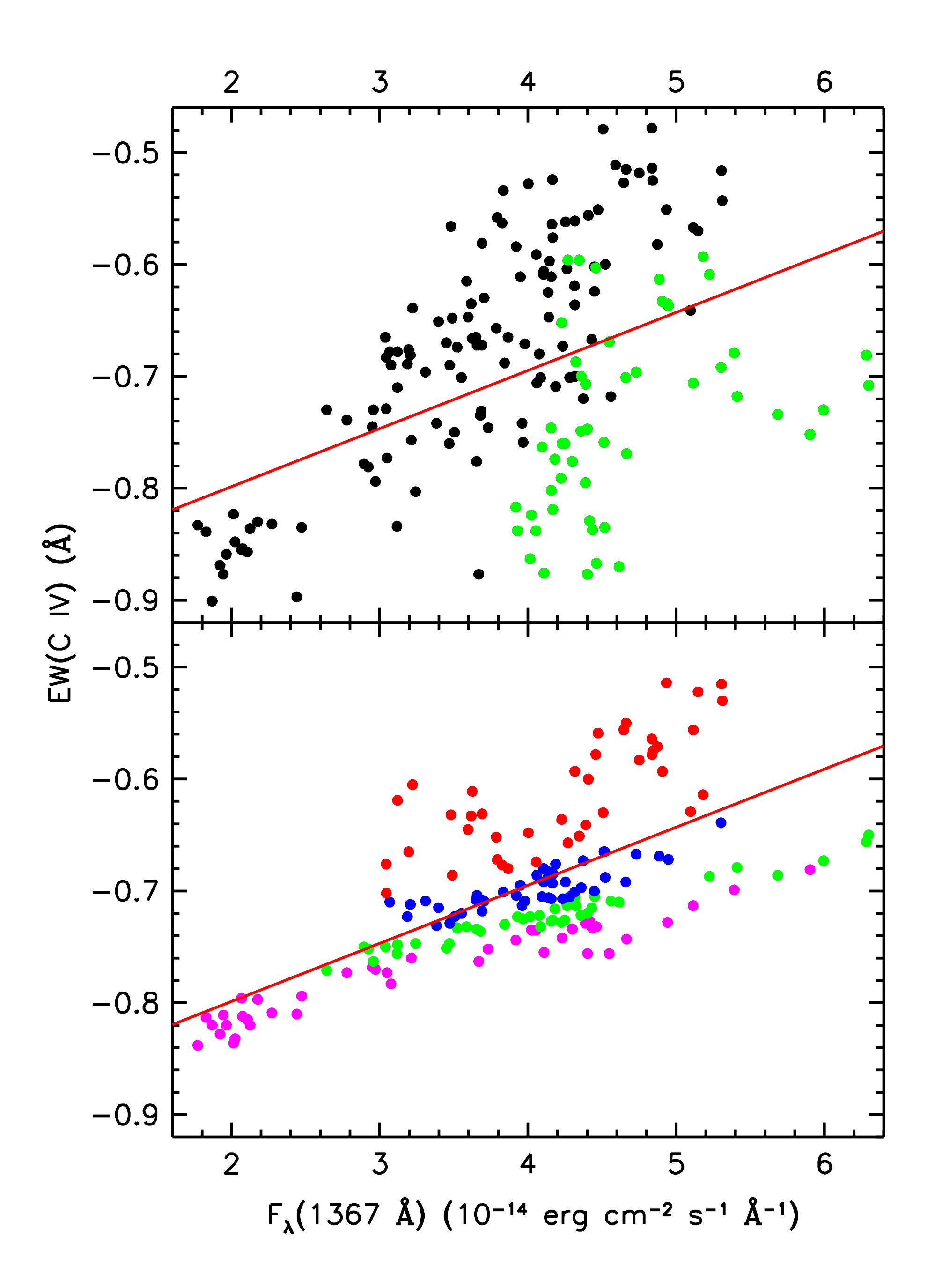}}
\caption{Fits to the correlation of the
equivalent width of the absorption line \ion{C}{4} $\lambda 1548$
associated with the intrinsic narrow absorption component \#1 in NGC 5548
with the UV continuum flux at 1367 \AA.
(Top Panel) Black and green points are for the entire STORM campaign.
Green points show values for EW(\ion{C}{4}) during the depths of the
BLR holiday, for times 56776$<$THJD$<$56827.
The solid red line shows the best-fit linear correlation of EW(\ion{C}{4}) with
the UV continuum flux, $F_\lambda(1367 \mbox{\AA})$.
(Bottom Panel)
Colored points show predicted equivalent widths at a given UV flux for a joint
correlation with the {\it Swift} soft X-ray flux from 0.3--0.8 keV, SX.
Red points have SX$>$0.085,
blue points have 0.085$\geq$SX$>$0.058,
green points have 0.058$\geq$SX$<$0.043, and
magenta points have SX$<$0.043.
}\label{fig:fitc4}
\end{figure}

Again, we investigated whether a multivariate correlation would be better than
either continuum measure on its own.
First, as expected, a simple linear fit of EW(\ion{C}{4}) vs.\,
$F_\lambda(1367 \mbox{\AA})$ is statistically unacceptable
with $\chi^2$=6271 for 171 points and 2 free parameters.
We then included the hardness ratio, HR, in the correlation, leading to
a dramatic improvement, but the fit is still poor with $\chi^2$=4314.
Finally, in contrast to \ion{C}{2}, using the soft X-ray flux in tandem
with $F_\lambda(1367 \mbox{\AA})$ gives the best result, $\chi^2$=3608.
This is still statistically unacceptable, but it does show the stronger
influence that the soft X-ray continuum is having on the \ion{C}{4} ionization
than the far-UV continuum.
As a final trial, assuming that the hardness ratio, HR, might serve as an
an additional measure of the covering fraction of the obscurer, we carried out
a trivariate analysis using SX, $F_\lambda(1367 \mbox{\AA})$, and HR.
This provided no additional improvement in $\chi^2$, with $\Delta \chi^2 < 1$.
Figure \ref{fig:fitc4} shows the correlation of the  EW(\ion{C}{4}) with
$F_\lambda(1367 \mbox{\AA})$, and how
including SX as an additional independent variable improves the fit.
Much of the scatter about the linear fit is explained by variations in the
soft X-ray flux.
The low equivalent width points in the upper right corner of the top panel of
Figure \ref{fig:fitc4} coincide with some of the lowest soft X-ray
fluxes observed as shown in the bimodal fit presented in the bottom panel.
The fact that it is the soft X-ray flux directly rather than
just a covering factor effect (as for \ion{C}{2}) may be telling us that the
extreme ultraviolet (EUV) and soft X-ray continuum shape is also varying,
since it is this portion of the continuum
that is responsible for ionizing \ion{C}{4}.
This is not the whole story, however. As shown in Figure \ref{fig:fitc4}, there
is a large set of points (shown in green) where variations in EW(\ion{C}{4}) are
not explained either by UV or soft X-ray continuum variations.
These green points encompass times from THJD=56776 to 56827.
This evidence, plus the complete decoupling from the continuum variations in
the light curve after THJD$\sim$56750, appears to be related to the
BLR holiday \citep{Goad16}, which sets in at the same time.
The green points in Figure \ref{fig:fitc4} lie in the deepest portion
of the BLR holiday.
In Paper VII \citep{Mathur17} we speculated that the BLR holiday was related
to changes in the EUV/soft X-ray continuum shape, and that these changes in
the higher-energy range of the spectral energy distribution (SED) were not
reflected in the visible FUV continuum.
The responses of the high-ionization intrinsic absorption lines enable us to
directly measure these changes in the SED.

While the responses of the broad emission lines also give insight into the
behavior of the continuum, their large-scale spatial distribution and differing
views of the continuum source complicate such an analysis.
Since the narrow intrinsic absorption lines in NGC 5548 are directly along our
line of sight, they are exposed to the same continuum that we see.
This allows us to take advantage of their unique point of view and to obtain
a direct measure of the continuum strength at a variety of energies.
To a good first approximation, the continuum flux at an ion's ionization
potential is the ionizing flux driving the population of that ion.
To use the strength of the absorption lines as a measure of this flux,
however, we must be certain that the absorption line variations are reflecting
changes in column density in response to changes in the ionizing flux rather
than being caused by either variations in covering factor, transverse motion,
or both.

\begin{figure}
\centering
\resizebox{\hsize}{!}{\includegraphics[angle=0, width=0.9\textwidth]{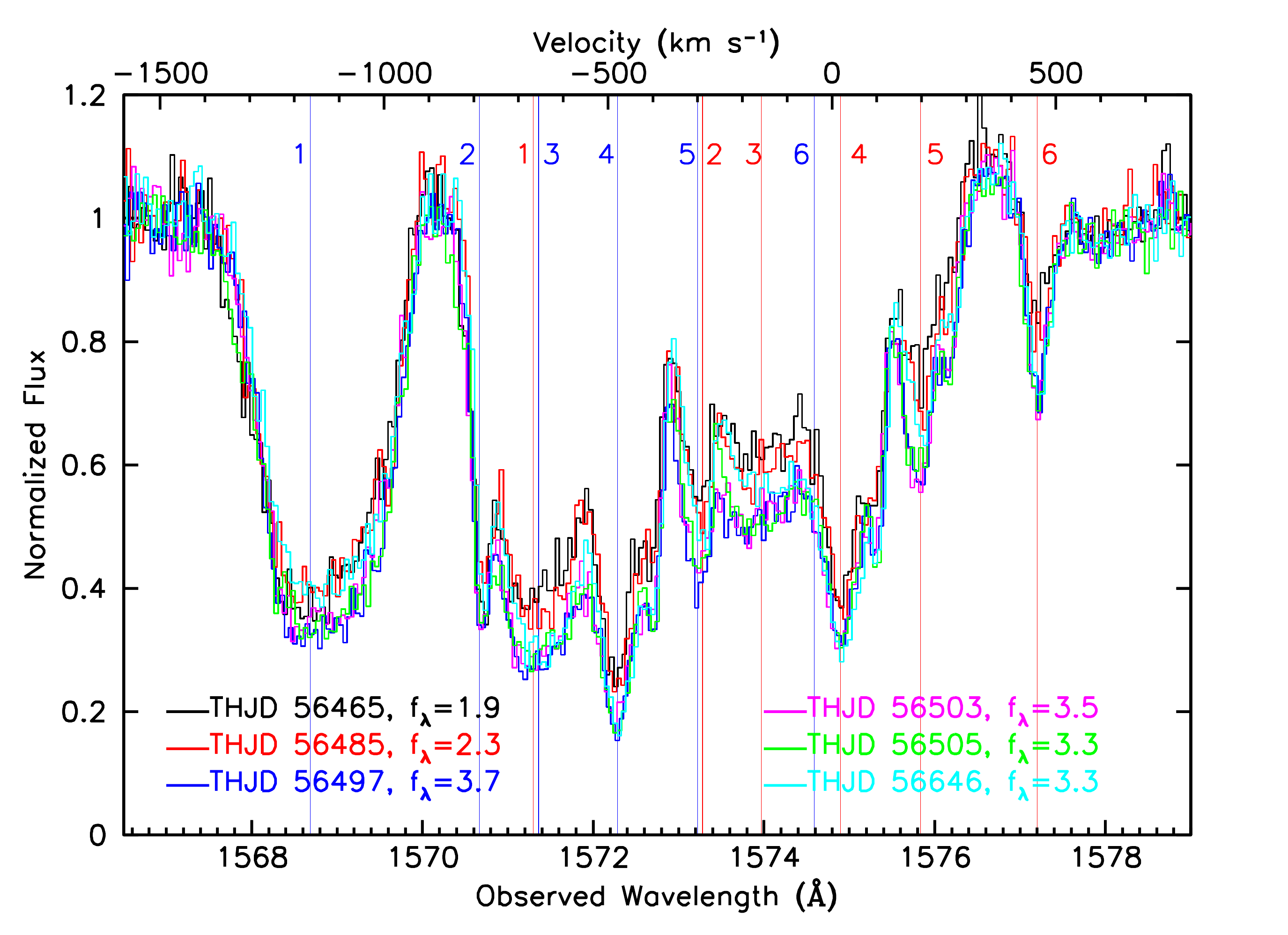}}
\caption{Illustration of variability in the narrow absorption components of
\ion{C}{4} in NGC 5548 during the {\it XMM-Newton} campaign \citep{Kaastra14}.
Normalized relative fluxes are plotted as a function of velocity relative to
the systemic redshift of $z=0.017175$ for \ion{C}{4} $\lambda1548$,
The colored curves are the spectra for the observation dates given in the key
along with the observed flux at 1367 \AA\ in units of
$10^{-14}~\rm erg~cm^{-2}~s^{-1}~\mbox{\AA}^{-1}$.
Thin vertical blue lines indicate the velocities of the blue components of
the doublets for the six intrinsic absorbers.
Vertical red lines show the locations of the corresponding red components.
Component \#1 shows little variability during the campaign, indicating high
saturation.
}\label{fig:c4xmm}
\end{figure}

\begin{figure}
\centering
\resizebox{\hsize}{!}{\includegraphics[angle=0, width=0.9\textwidth]{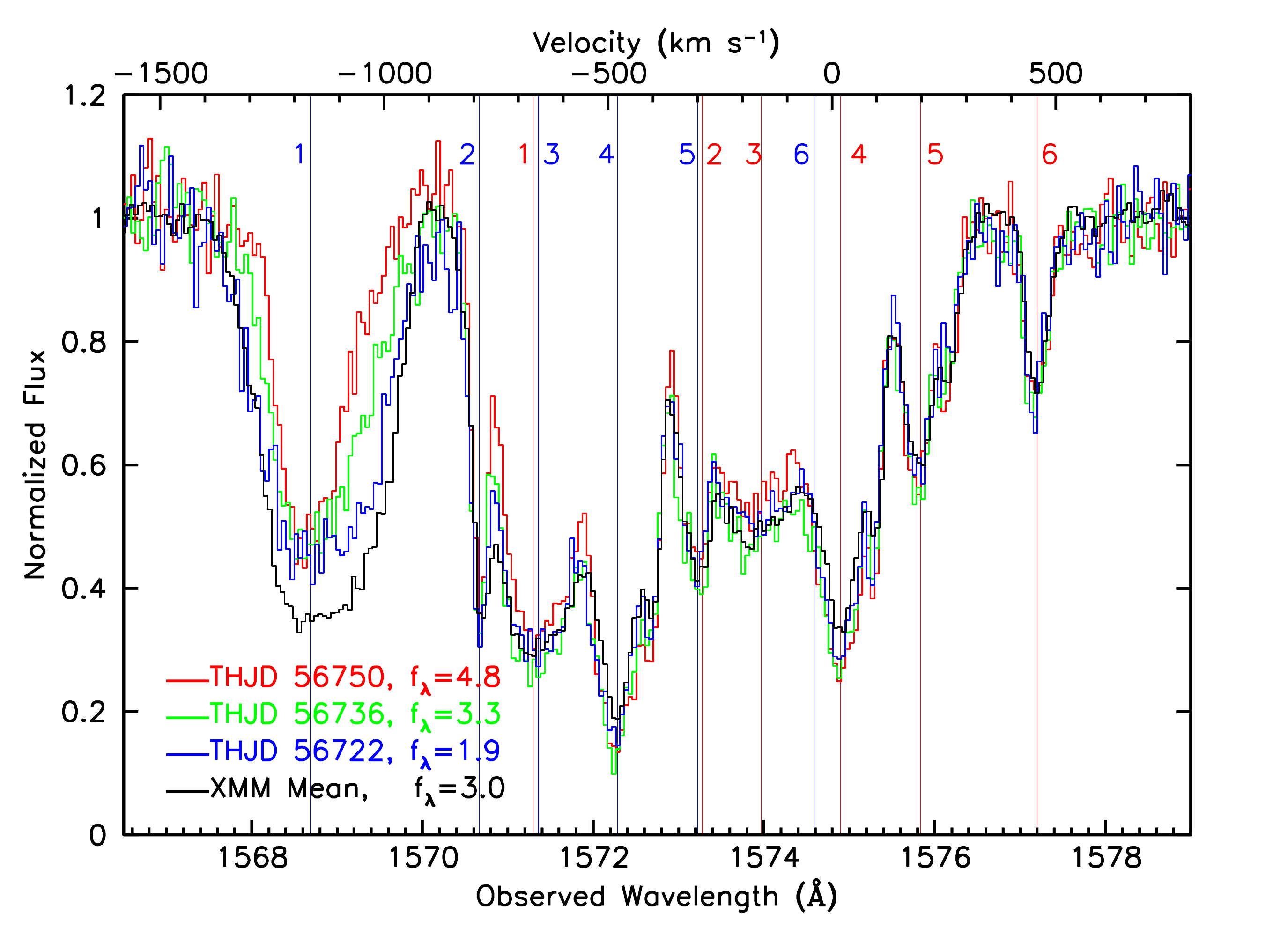}}
\caption{Illustration of variability in the narrow absorption components of
\ion{C}{4} in NGC 5548 campaign during the early part of the AGN STORM campaign
when the strength of \ion{C}{4} absorption in Component \#1 was
highly correlated with the FUV continuum flux.
Normalized relative fluxes are plotted as a function of velocity relative to
the systemic redshift of $z=0.017175$ for \ion{C}{4} $\lambda1548$,
The colored curves are the spectra for the observation dates given in the key
along with the observed flux at 1367 \AA\ in units of
$10^{-14}~\rm erg~cm^{-2}~s^{-1}~\mbox{\AA}^{-1}$.
The illustrated spectra show high (red), low (blue), and mean (green) flux
levels in comparison to the mean spectrum from the {\it XMM-Newton} campaign
(black).
Thin vertical blue lines indicate the velocities of the blue components of
the doublets for the six intrinsic absorbers.
Vertical red lines show the locations of the corresponding red components.
During AGN STORM, the absorption profile in Component \#1 varies as expected
for gas showing an ionization response.
}\label{fig:c4storm}
\end{figure}

The best evidence in favor of a dominant photoionization response is the good
correlation we see between the strength of the individual absorption lines
and either the FUV continuum (for low-ionization lines like \ion{C}{2}), or
the soft X-ray flux (for high-ionization lines like \ion{C}{4}).
Our biggest concern is for \ion{C}{4} because it was heavily saturated during
the {\it XMM-Newton} campaign. The photoionization models of \cite{Arav15}
show that it has an optical depth of 200 during the observations in 2014.
Figure \ref{fig:c4xmm} shows that the bottom of the absorption trough
in \ion{C}{4} $\lambda1548$ Component \#1 remained relatively constant in 2014,
in agreement with heavy saturation.
Likewise, Figure \ref{fig:c4_sx_correlation} shows that during the
{\it XMM-Newton} campaign, the strength of the absorption in \ion{C}{4}
$\lambda1548$ Component \#1 is independent of the soft X-ray flux, and
much stronger than during the AGN STORM campaign.
In contrast, Figure \ref{fig:c4_sx_correlation} shows a strong correlation
with the soft X-ray flux during the AGN STORM campaign, indicative
of a photoionization response.
In Figure \ref{fig:c4storm}, we show that the absorption line profiles
change as expected if the column density in \ion{C}{4} Component \#1
was varying in response to changes in the ionizing flux.

\begin{figure}
\centering
\resizebox{\hsize}{!}{\includegraphics[angle=0, width=0.9\textwidth]{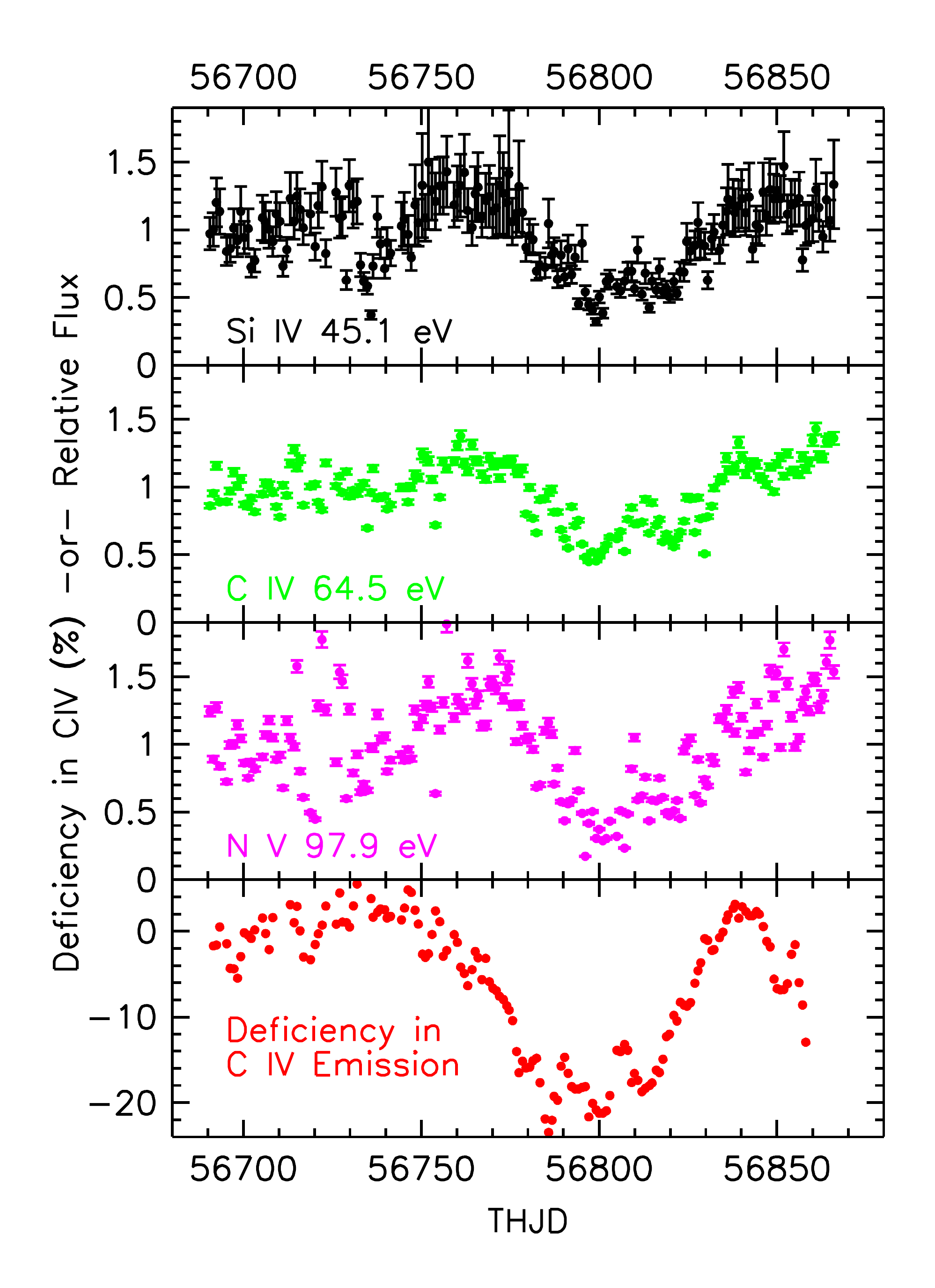}}
\caption{Relative EUV continuum fluxes at the ionization potentials of
\ion{Si}{4} (45.1 eV, black points), \ion{C}{4} (64.5 eV, green points) and
\ion{N}{5} (97.9 eV, magenta points)
as deduced from variations in the equivalent width of the absorption lines
\ion{Si}{4} $\lambda 1393$, \ion{C}{4} $\lambda 1548$ and
\ion{N}{5} $\lambda$1238
associated with the intrinsic narrow absorption component \#1 in NGC 5548.
Red points show the percentage deficiency in the flux of \ion{C}{4} broad
emission during the BLR holiday \citep{Goad16}.
}\label{fig:ip_lc}
\end{figure}

Assuming that the absorption lines in Component \#1 are indeed tracing a
response to change in the ionizing flux, we can use this behavior as a
diagnostic of the ionizing continuum.
For the first 75 days of the campaign (dates prior to THJD=56766)
both the absorption lines and the emission lines show a good correlation in
their variations with the FUV continuum flux at 1367 \AA.
We use this proportionality to derive a linear relationship between a line's EW
and the continuum flux of
\begin{equation}
\rm EW = a_0 + a_1 \times F_\lambda(1367) .
\end{equation}
During this early part of the campaign, when everything correlates well, we
assume that the continuum shape is that shown in Figure 4 of \cite{Mehdipour15}.
We designate points lying on this fiducial SED as $\rm F_0(\lambda)$, where
$\lambda$ may be either the observed wavelength 1367 \AA, or the
wavelength at the ionization potential of the relevant ion, $\lambda_{\rm IP}$.
We can then use variations in the EW of a given line to measure changes in the
flux at the ionization potential relative to this fiducial SED shape of
\begin{equation}
 \rm \frac{F(\lambda_{IP})}{F_0(\lambda_{IP})} = \frac{(EW - a_0) / a_1}{F_0(1367)} .
\end{equation}
Following this methodology, we derive relative ionizing continuum light curves
for each absorption line representative of the time-varying relative fluxes at
their ionization potentials.
Figure \ref{fig:ip_lc} shows these relative light curves for the blue
components of the \ion{Si}{4},
\ion{C}{4} and \ion{N}{5} doublets of the intrinsic absorption
line for Component \#1 in NGC 5548.
The figure also compares these relative light curves to the deficiency in the
broad \ion{C}{4} emission line flux during the BLR holiday \citep{Goad16},
defined as the percentage diminution in the \ion{C}{4} emission line flux
during the holiday compared to the expected flux based on the
correlation of \ion{C}{4} emission line flux with continuum flux observed
during the first 75 days of the AGN STORM campaign.
Note that the relative decrease in broad \ion{C}{4} emission line flux
occurs during the same time interval (56776$<$THJD$<$56827) as the inferred
decrease in EUV continuum flux at the ionization potentials of
\ion{Si}{4} (45.1 eV),
\ion{C}{4} (64.5 eV) and \ion{N}{5} (97.9 eV) deduced from the behavior
of the intrinsic absorption line.
All three inferred continua show significant correlations with the
deficit in the \ion{C}{4} emission.
However, there is a delay between the \ion{C}{4} deficit and the inferred
continua.
Using ICCF, we find delays and peak correlation coefficients of
$11.6 \pm 1$ days and $r = 0.66$ for \ion{Si}{4},
$10.5 \pm 1$ days and $r = 0.70$ for \ion{C}{4}, and
$12.3 \pm 1$ days and $r = 0.64$ for \ion{N}{5}.
These delays are actually with respect to the FUV continuum at 1158 \AA,
since \cite{Goad16} shifted the \ion{C}{4} light curve by $-5$ days to align it
with the continuum variations,
so the inferred EUV continua variations are delayed by 5--7 days relative to the
\ion{C}{4} deficit.
These delays are puzzling since one might expect the \ion{C}{4} deficit to lag
the continuum deficits rather than vice versa
if they were the direct cause of the BLR holiday. 
\cite{Dehghanian19a} investigate the complex responses of the intrinsic narrow
absorption lines to changes in the ionizing continuum and the intervening
obscurer in greater detail
although they also do not explain this puzzling behavior.

\subsection{Variability of the Broad Absorption Features}

As discussed in \S5.2, the broad absorption features that appear as extensive
blue-shifted troughs on all the resonant UV lines in the spectrum of NGC 5548
are the UV counterparts to the soft X-ray obscurer discovered by
\cite{Kaastra14}.
These broad UV absorption features also vary with time.
In fact, in the observations from that campaign, there was a significant
anticorrelation between the strength of the absorption troughs and the
soft X-ray flux as measured by {\it Swift}
(linear correlation coefficient $r = -0.37$).
As the soft X-ray intensity increased, the broad UV absorption strength
decreased. This is not due to a photoionization response, however.
The broad UV absorption lines are saturated, and so their strength is also
governed more by the covering factor than by the ionization state or column
density of the absorbing gas.

\begin{figure}
\centering
\resizebox{\hsize}{!}{\includegraphics[angle=0, width=0.9\textwidth]{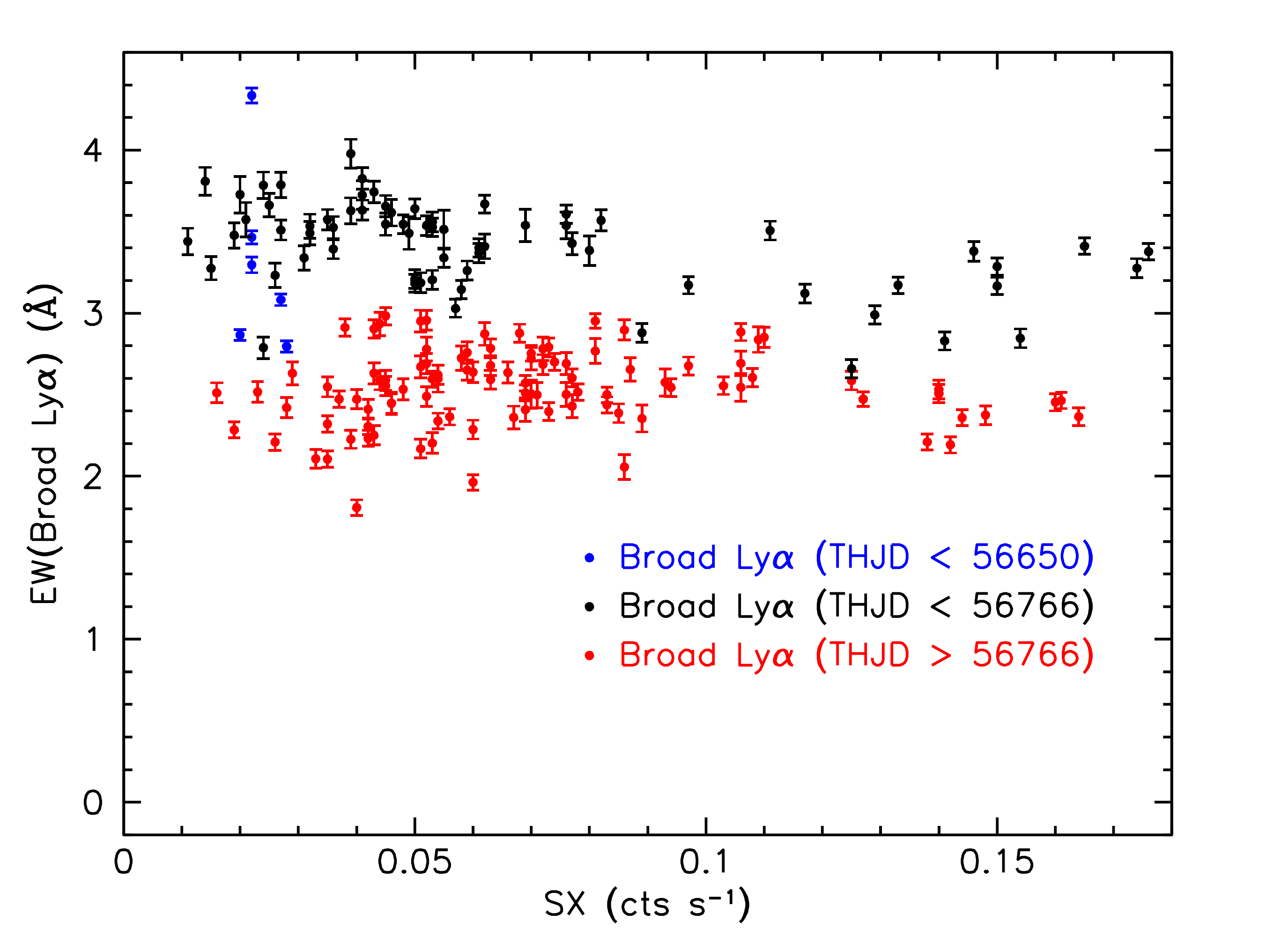}}
\caption{Correlation of the absolute value of the
equivalent width of the broad absorption in Ly$\alpha$
in NGC 5548 with the soft X-ray flux from {\it Swift}.
Blue points are for the six observations of the {\it XMM-Newton} campaign,
all prior to THJD=56650;
black points are for dates in the STORM campaign earlier than THJD=56766;
red points are for later times.
}\label{fig:broad_abs_vs_sx}
\end{figure}

The more extensive monitoring of the time variability of the broad absorption
features enabled by the STORM campaign reveals an even more complex picture.
As shown in Figure \ref{fig:broad_abs_vs_sx},
the anticorrelation of the EW of the broad absorption is not as clean as it
was in the six points available from the {\it XMM-Newton} campaign.

\begin{figure}
\centering
\resizebox{\hsize}{!}{\includegraphics[angle=-90, width=0.9\textwidth]{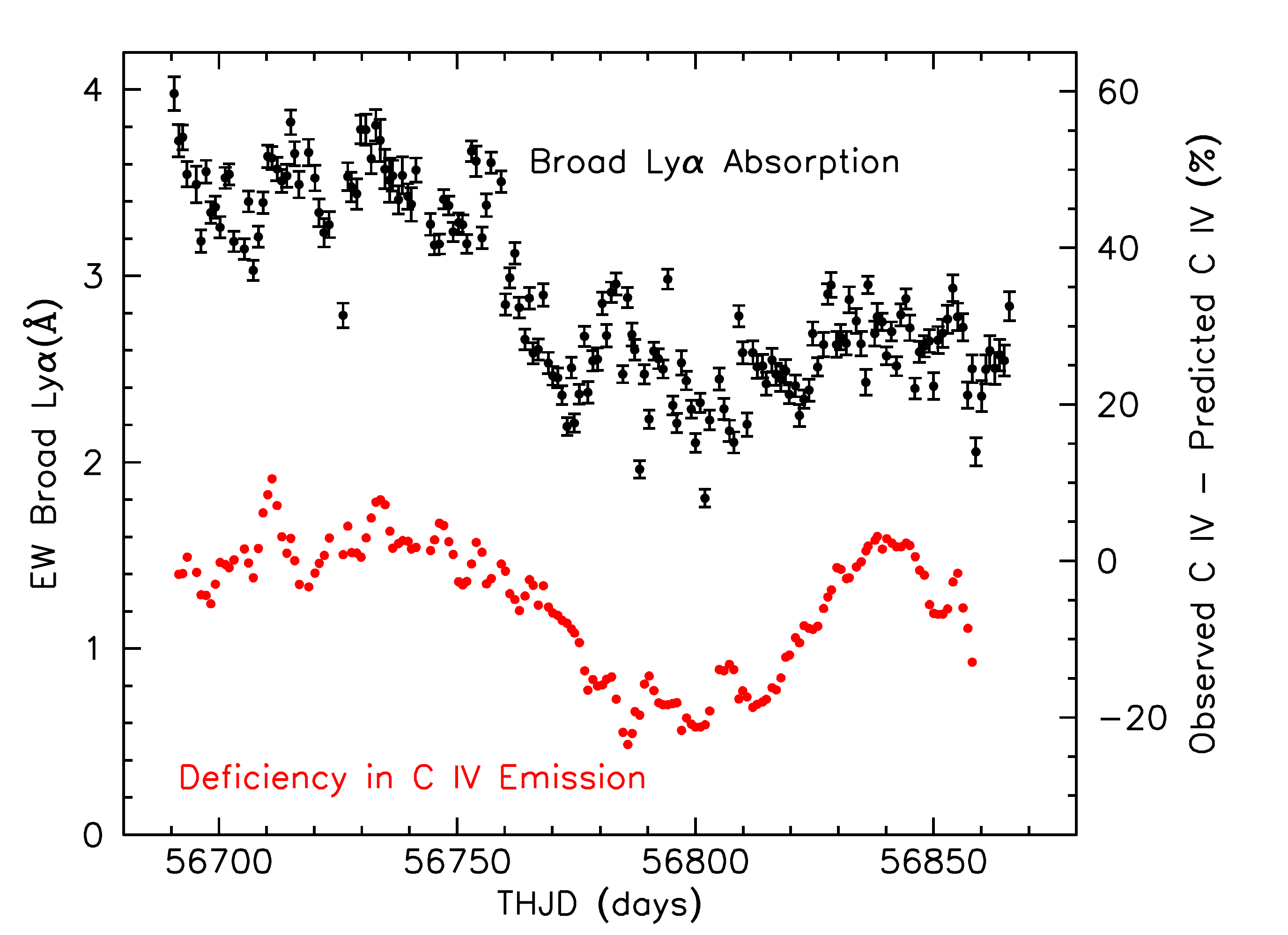}}
\caption{Light curve of the variations in the absolute value of the 
equivalent width of broad Ly$\alpha$
absorption in NGC 5548 (black points) compared to the percentage deficiency in
the flux of \ion{C}{4} broad emission (red points) during
the BLR holiday \citep{Goad16}.}\label{fig:broad_abs_vs_anomaly}
\end{figure}

While there is a rough anticorrelation ($r = -0.41$) early in the
campaign (THJD$<$56766), as with the behavior of the
narrow absorption lines, the correlation disappears after
THJD=56766 ($r = +0.02$),
approximately the time of the onset of the BLR holiday \citep{Goad16}.
Figure \ref{fig:broad_abs_vs_anomaly} shows this even more dramatically.
Light curves for the EWs of the broad absorption features are eerily similar
to the time variation of the anomalous deficit in the response of the
flux of the broad emission lines.
The connection between these two behaviors is puzzling.
On the one hand, the lack of response in the BLR during the holiday appears to
be due to a change in the SED, in which the soft X-ray excess decouples from
the visible UV continuum.
A decrease in the BLR flux during the holiday implies a decrease in the
ionizing UV driving the line emission.
On the other hand, a decrease in the broad absorption line EW implies a
decrease in the covering factor of the obscurer,
because these broad troughs appear to be saturated (\S4.2)
and we see no dramatic changes in their velocity widths..
This is also seen in the {\it Swift} monitoring of NGC 5548 \citep{Mehdipour16},
and it implies that more of the EUV and soft X-ray flux is being
transmitted past the obscurer.
Plus, since the BLR response represents a global view of the continuum source,
and our view of the obscurer is restricted purely to our line of sight,
this suggests that our view of the obscurer may not be a special
geometrical arrangement, such as a single cloud crossing our line of sight,
but rather a more axisymmetric arrangement affecting all sightlines that
illuminate the BLR.
\cite{Dehghanian19a} present further photoionization analysis of both the 
narrow and the broad absorbers that support this interpretation.

An approximate axisymmetric arrangement for the obscurer is even more natural
when one considers the timescales covered by our observations.
From the beginning of the {\it XMM-Newton} campaign in June 2013 to the end of
the reverberation mapping campaign in July 2014, the elapsed time was 415 days.
For a black-hole mass of $5.2 \times 10^7~\rm M_\sun$ \citep{Bentz15}, clouds at
a radius of 1 light day, at the innermost edge of the BLR.
would have an orbital timescale of 115 days.
Thus the obscurer and the innermost part of the BLR could have completed more
than one full revolution over the course of our observing programs.

\section{Discussion}
\label{sec:discussion}

Modeling the UV spectra of NGC 5548 from the STORM campaign accomplished our
primary goal of obtaining clean, absorption-corrected, deblended emission-line
profiles of all the principle UV emission lines.
We are also able to separate the non-varying contributions of the narrow-line
and intermediate line emission to the Ly$\alpha$ and \ion{C}{4} profiles and
produce uniform velocity-binned lag measurements across the full profiles of
Ly$\alpha$ and \ion{C}{4}.
These ``clean" line profiles are a key element for future detailed modeling of
the BLR to produce two-dimensional velocity delay maps using both MEMEcho
\citep{Horne19} and forward-modeling techniques \citep{Pancoast19}.
However, simple characteristics of the emission-line profiles and the
one-dimensional velocity delay profiles enable us to make some key inferences.

The Ly$\alpha$ velocity delay profile shows a distinctive
``M" shape, with short lags at high velocity,
peak lags of $\sim 6 - 8$ days at moderate velocities
($\pm 2500~\rm km~s^{-1}$), and a local minimum near zero velocity.
As we discuss in more detail later, this is very different from the
centrally peaked H$\beta$ profile observed in 2007 by \cite{Denney09}.
The \ion{C}{4} velocity delay profile is similar to Ly$\alpha$,
but it shows only ``shoulders" rather than a central dip.
This behavior was seen in the original raw data \citep{DeRosa15}, but we now
have a clean, uniform picture across both profiles.
This morphology is very similar to that seen in H$\beta$ during the
ground-based portion of the STORM campaign \citep{Pei17}
as well as during the 2015 campaign by \cite{Lu16} and \cite{Xiao18}.

The short lags at high velocity and longer lags at low velocity are a natural 
expectation for Keplerian profiles in a disk \citep{Welsh91, Horne04}.
In a simple disk geometry where the line-emitting portion of the disk is
truncated at both an inner and an outer radius, the lag profile across the
emission line 
extends from short lags at high velocities, corresponding to the inner radius of
the disk, and rises to peaks at lower velocities, corresponding to the outer 
radius of the disk \citep{Welsh91}. In the region near zero velocity, the 
near-zero lags from the near side of the disk produce a central dip. The sharp 
peaks of the lag profile produced by a thin disk in the models of \cite{Welsh91}
 is a direct consequence of their disk model having a monotonic, powerlaw 
emissivity and sharp edges. This leads to a ``cuspy" shape for the transfer 
function, $\Psi(\tau)$. The sharp peaks can be broadened into smoother bumps if 
the line-emitting portions of the disk do not have abrupt edges,
and if the disk is thickened, as in \cite{Pancoast14}, for example.
In the case of NGC~5548, the transfer function 
$\Psi(\tau)$ recovered from the data is smooth and bell-shaped \citep{Horne19}, 
which would result in smooth, round peaks in the lag profile.

Notably, the lag profile for \ion{C}{4} for the full campaign shows an upturn
to longer lags at high positive velocities
($> 10,000~\rm~km~s^{-1}$).
This feature was not evident in the original raw data, but may now appear
when the contamination of the red wing of \ion{C}{4} by the shorter lag of
\ion{He}{2} may have been biasing the result to lower values.
Longer lags at high positive velocities are indicative of outflowing gas
since redshifted, outflowing gas is on the far side
of the illuminating source relative to the observer.
See the models of \cite{Welsh91} for examples.
These long lags at high positive velocities are absent when we restrict the data
to the first 75 days of the campaign before the BLR holiday.

The \ion{C}{4} lag profile shows additional changes in structure for
different time intervals during the campaign, as discussed in \S\ref{sec:lags}.
The biggest change is the apparent change from a local minimum in the lag on
the red side of the red emission bump in the pre-holiday period
to a local maximum on the blue side of the bump during the holiday period.
These changes are in contrast to smoother, more uniform changes for
Ly$\alpha$, which mostly track the evolution to longer lags of the mean profile
at all velocities across the profile.
Although we cannot unambiguously assign physical explanations to
one-dimensional emission-line and lag profiles,
the difference in behavior between \ion{C}{4} and Ly$\alpha$
may reflect the differences in radiative transfer through the BLR for these
two emission lines.
Both lines are generally considered to be optically thick, with preferential
emission originating from the illuminated side of the BLR gas
\citep{KK79, KK81, Ferland92},
but the smoother evolution of the lag profile during the course of the campaign
may suggest that Ly$\alpha$ emission is less orientation dependent than that
of \ion{C}{4}.

The lag profiles across the emission lines in NGC 5548 have also
evolved on longer timescales over the past decade.
In the 2007 observations of the H$\beta$ profile by \cite{Denney09}, the
lag profile is centrally peaked with no dip, and it extends to velocities of
only $\pm6000~\rm km~s^{-1}$.
For a black hole mass of $\rm M_{BH} = 5.2 \times 10^7~\Msun$ \citep{Bentz15}
and an inclination of $i = 32^{\circ}$ \citep{Pancoast14}, this corresponds to
Keplerian velocities at an inner disk radius of 2.3 light days.
The innermost bins span a range of $\pm1000~\rm km~s^{-1}$/sin $i$,
which correspond to Keplerian velocities
at an outer disk radius of $\sim80$ light days.

The AGN12 campaign \citep{DeRosa18} showed that by 2012 NGC 5548 had
developed a double peaked, ``M"-shaped lag profile.
By the time of the STORM campaign in 2014, this shape was well developed in
both H$\beta$ and in Ly$\alpha$.
For the STORM data, profiles reaching velocities of $\pm10\,000~\rm km~s^{-1}$
imply an inner disk radius of $\sim 0.8$ light day.
The double peaks at $\pm 2500~\rm km~s^{-1}$ lead to an outer disk radius of
13 light days.

\begin{figure}
\centering
\resizebox{\hsize}{!}{\includegraphics[angle=0, width=0.9\textwidth]{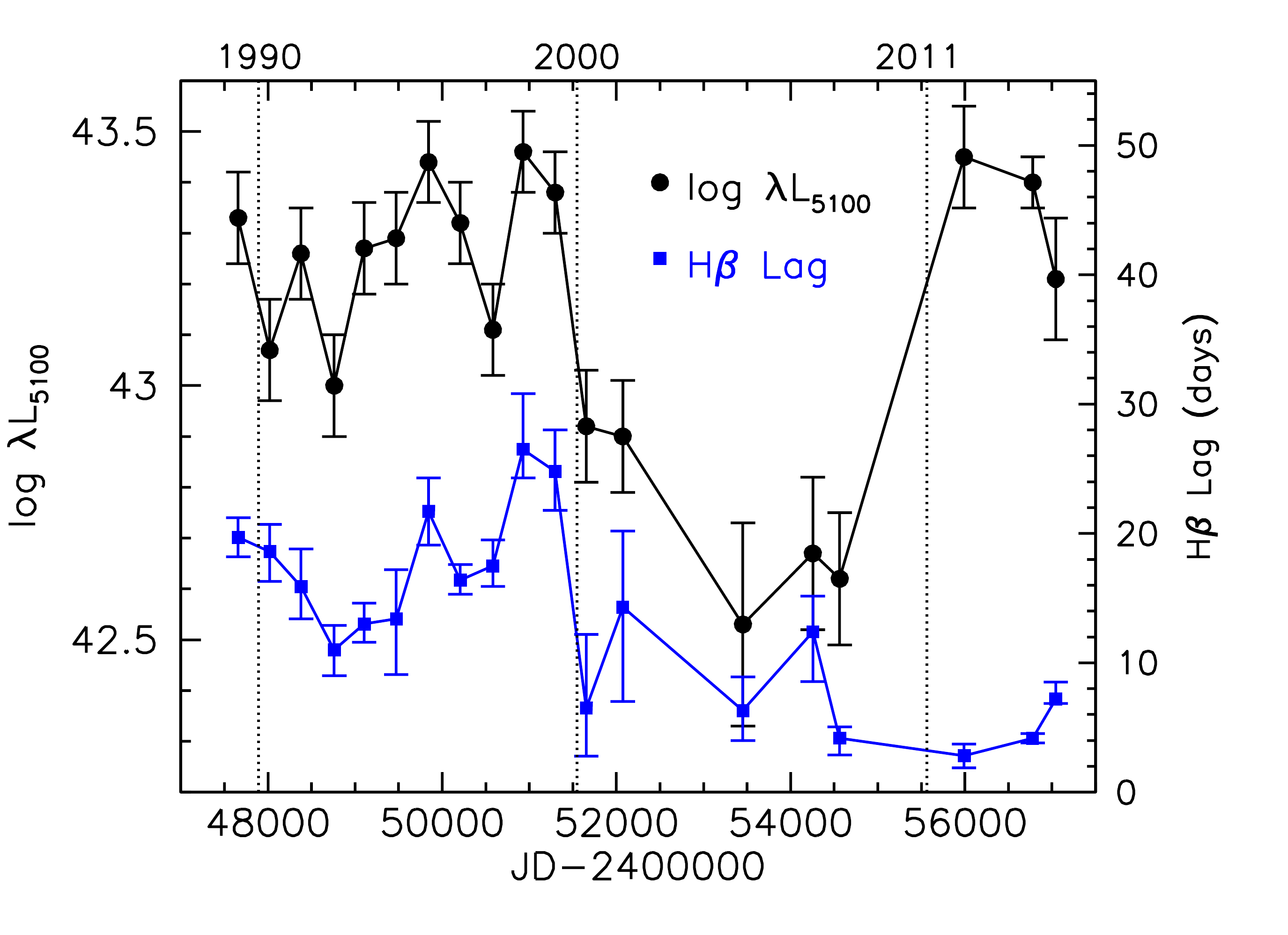}}
\caption{Time series showing the history of the optical luminosity of
NGC 5548 ($\lambda \rm L_{5100}$, black points) and
the reverberation lag size for the H$\beta$ emission line (blue points).
Vertical dotted lines show the dates 1990 January 1, 2000 January 1, and
2011 January 1.
Data prior to 2011 are from \cite{Kilerci15}, which are taken
from these original sources: \cite{Peterson02}, \cite{Bentz07}, \cite{Bentz09a},
\cite{Denney09}, and \cite{Peterson93}.
Data in 2012 are from \cite{DeRosa18}.
Data in 2014 are from the STORM campaign \citep{Pei17}.
Data in 2015 are from \cite{Lu16}.
\label{fig:rhb_lum}
}
\end{figure}

These changes in the implied overall structure of the BLR may explain the
departure of its measured size from the previously well established
BLR size/luminosity relation \citep[see Fig. 13 of][]{Pei17} as
the luminosity of NGC 5548 has evolved in time over the past two decades.
Figure \ref{fig:rhb_lum} compares the evolution in the
optical luminosity ($\lambda \rm L_{5100}$) of NGC 5548 to the size of the BLR
as measured by the reverberation lag of the H$\beta$ emission line.
During the 2007 campaign of \cite{Denney09}, NGC 5548 was near a minimum in its
luminosity. The integrated H$\beta$ lag was 5 days, and it fit well on the
BLR size/luminosity relation.
In 2012, NGC 5548 increased to normal brightness levels, but the overall
H$\beta$ lag remained low at 4 days \citep{DeRosa18}.
One must be careful in assessing the absolute values of these changes in size,
however. \cite{Goad14} note that emission line lags can be biased low by
factors of 2--3 for short continuum variability timescales
(as we observed in the STORM campaign),
and for campaigns of short duration (as were those in the 2000s).
During the lengthy STORM campaign in 2014, NGC 5548 remained bright, yet the
H$\beta$ lag remained small, at 4 days, factors of 5--6
smaller than historical values in the 1990s.
Notably, these changes coincide with the appearance of the X-ray obscurer in
February 2012, as shown in data from {\it Swift} monitoring \citep{Mehdipour16}.

Similar behavior has been seen in NGC 3783.
This Seyfert 1 also exhibited an obscuration event in 2016
with heavy, soft X-ray and broad UV absorption \citep{Mehdipour17}.
When the obscuration appeared, NGC 3783 was at a luminosity peak following a
low-luminosity state from 2011 through 2014 \citep{Kaastra18}.
Despite the high luminosity, its broad lines were significantly diminished in
flux, with moderate velocity regions ($\sim2800~\rm km~s^{-1}$) of the profile
apparently gone \citep{Kriss19}.
\cite{Kriss19} suggest that the changes in the structure of the BLR and the
appearance of the obscuring outflow are related.
In the context of disk-driven wind models
\citep{Murray97, Proga00, Proga04, Proga15, Waters16, Czerny17, Baskin18},
the transition from a normal BLR configuration to an extended low state
requires several years for the BLR to dynamically adjust to the reduced
radiation pressure from the accretion disk.
The BLR at large radii collapses back toward the disk plane, and the peak in
the BLR emissivity shifts to smaller radii.
When the continuum brightens, the excess material in the inner regions may be
blown away as a radiatively driven wind, observable as the obscurer.
The BLR reinflates, but this requires years, governed by the free-fall
timescale for the disk wind, which is 3.4 years at 10 light days in NGC 3783
for a black hole mass of $\rm M_{BH} = 2.3 \times 10^7~\Msun$ \citep{Bentz15}.

Timescales in NGC 5548 are similar, but shorter, given its slightly higher
black hole mass. At 10 light days in NGC 5548, Keplerian velocities are
$\sim5200~\rm km~s^{-1}$, and the free-fall time, defined as
\begin{equation}
\rm t_{ff} = \left(\frac{2 R^3}{G M_{BH}}\right)^{1/2} ,
\end{equation}
is 2.3 years.
The shrunken size of the BLR during the STORM campaign is also reflected in the
lack of stratification in the ionization structure that was evident in past
campaigns \citep{Clavel91, Korista95}.
Except for \ion{He}{2}, all the other broad emission lines have
indistinguishable lags.
After the STORM campaign, in 2015
NGC 5548 appears to be reverting back to the normal size/luminosity relation.
\cite{Lu16} observed an increase in the H$\beta$ lag to 7 days.
They suggest that this evolution in size was evidence for
radiation pressure affecting the size of the BLR, with a time delay likely
due to the dynamical timescale of the BLR, similar to our arguments above.

Although our modeling of the UV spectra from the NGC 5548 STORM campaign
have enabled us to understand the dynamics of the BLR better, and its
potential links to the appearance of X-ray and UV obscuration, we must still
work out the details of what produces the short timescale variations in the
obscuring gas, and how that affects the BLR emission.
These processes will be subjects of future papers.

\section{Summary}
\label{sec:summary}

We have modeled the HST UV spectra of NGC 5548 obtained during the 2014
Space Telescope and Optical Reverberation Mapping (STORM) campaign.
The model uses 97 emission and absorption components to construct an
emission model for all 171 spectra of NGC 5548 obtained during the campaign.
The model permits us to separate the broad absorption associated with the
soft X-ray obscuration discovered by \cite{Kaastra14}, and to correct for
narrow absorption produced by the intrinsic UV absorption lines of NGC 5548
as well as intervening interstellar features.
Using the model, we are able to produce absorption-corrected, deblended
emission-line profiles, velocity-binned light curves for Ly$\alpha$ and
\ion{C}{4}, and light curves for all narrow and broad absorption components.
The modeled emission-line profiles eliminate the non-varying flux of the
narrow-line and intermediate-line components of  Ly$\alpha$ and \ion{C}{4},
separate the contributions of \ion{N}{5} and \ion{He}{2} to their profiles,
and give a clear view of the emission-line time delay
across the full line profile.
The principle results of our study are:

\begin{enumerate}
\item
The time lags of 2--8 days
for the integrated emission lines are comparable to, or
slightly shorter than those in \cite{DeRosa15}, primarily due to the
elimination of the non-varying narrow-line and intermediate-line components
(\S4.1).
\item
The velocity-binned lag profiles have a distinct ``M" shape in Ly$\alpha$,
and less-prominent local maxima above a central minimum in \ion{C}{4} (\S4.1).
The local minima in the lag profiles are near zero velocity, and the local peaks
(Ly$\alpha$) or shoulders (\ion{C}{4}) are at $\pm2500~\rm km~s^{-1}$ (\S4.1).
This morphology is indicative of Keplerian motion in a disk-like configuration.
\item 
The narrow absorption lines vary in strength in response to changes in the
continuum intensity (\S4.3).
The lowest ionization absorption lines (Ly$\alpha$, \ion{C}{2}, and
\ion{Si}{2}) correlate with the continuum through the course of the whole
campaign. Their variations show a delayed and smoothed response to continuum
variations consistent with recombination of gas with densities of
$\rm log~n_e~cm^{-3} \sim 5$, similar to the density measured by \cite{Arav15}
using density-sensitive absorption lines of \ion{C}{3}* and \ion{Si}{3}*,
and similar to the density of the NLR inferred by \cite{Kraemer98} and
\cite{Peterson13}.
\item
High-ionization intrinsic narrow absorption lines (\ion{Si}{4}, \ion{C}{4}, and
\ion{N}{5}) correlate well with the continuum during the first 75 days of the
campaign (\S4.3).
Thereafter, they decorrelate, showing signs of the same ``holiday"
exhibited by the BLR \citep{Goad16}. During the holiday period, their strengths
correlate most closely with the soft X-ray flux as measured by {\it Swift},
which suggests that the soft X-ray flux is more closely related to the
ionizing extreme ultraviolet than the FUV continuum.
Using the response of these high-ionization absorption lines, we are able to
reconstruct the relative flux of the spectral energy distribution in the
extreme ultraviolet. We show that it is diminished in flux relative to the
FUV continuum during the period of the holiday.
\item
The broad absorption lines associated with the X-ray obscurer also vary during
the course of the campaign (\S4.4).
Early in the campaign, they show a rough
anticorrelation with the soft X-ray flux, as seen by \cite{Kaastra14}.
However, during the holiday period they decrease in strength, with behavior
mimicking the deficiencies in the flux of the broad emission lines.
This behavior is not understood.

\item
The departure of NGC 5548's BLR from the radius/lum\-i\-nos\-i\-ty
relationship of \cite{Peterson02} during the
AGN12 campaign in 2012 \citep{DeRosa18} coincides with the appearance of
the X-ray obscurer at about the same time \citep{Mehdipour16}.
We suggest the two events are related, and that the
obscurer is a manifestation of a disk wind launched by the brightening of
NGC 5548 in 2012 after a prolonged low-luminosity state which had led to the
collapse of the BLR (\S5).
\end{enumerate}

Modeling the UV spectra of NGC 5548 from the STORM campaign has yielded
crucial insights into the behavior of the emission-line and
absorption-line gas as well as properties of the ionizing continuum
that were not easily seen in the raw data.
These new insights may help us to resolve the structure of the BLR and
successfully interpret our measures of its two-dimensional reverberation maps.

\acknowledgments
We thank R. Plesha for formatting and posting the high-level data products in
MAST.
Support for {\it HST} program number GO-13330 was provided by NASA through a
grant from the Space Telescope Science Institute, which is operated by the
Association of Universities for Research in Astronomy, Inc., under NASA
contract NAS5-26555.
M.M.F., G.D.R., B.M.P., C.J.G., and R.W.P. are grateful for the support of the
National Science Foundation (NSF) through grant AST-1008882 to
The Ohio State University.
A.J.B. and L.P. have been supported by NSF grant AST-1412693.
E.M.C., E.D.B., L.M., and A.P. acknowledge support from Padua University through grants DOR1699945/16, DOR1715817/17, DOR1885254/18, and BIRD164402/16.
M.C. Bentz gratefully acknowledges support through NSF CAREER grant AST-1253702
to Georgia State University.
S.B. was supported by NASA through Chandra award no. AR7-18013X issued by the
Chandra X-ray Observatory Center, operated by the Smithsonian Astrophysical
Observatory for NASA under contract NAS8-03060.
S.B. was also partially supported by grant HST-AR-13240.009.
M.C. Bottorff acknowledges HHMI for support through an undergraduate science
education grant to Southwestern University.
K.D.D. is supported by an NSF Fellowship awarded under grant AST-1302093.
R.E. gratefully acknowledges support from NASA under awards NNX13AC26G,
NNX13AC63G, and NNX13AE99G, and ADAP award 80NSSC17K0126.
G.J.F. and M.D. thank NSF (1816537), NASA (ATP 17-0141),
and STScI (HST-AR-13914, HST-AR-15018) for their support,
and the Huffaker scholarship for funding related travel.
B.D.M. acknowledges support from the Polish National Science Center grant
Polonez 2016/21/P/ST9/04025.
J.M.G. gratefully acknowledges support from NASA under award NNH13CH61C.
P.B.H. is supported by NSERC.
K.H. acknowledges support from the UK Science and Technology Facilities Council
through grant ST/J001651/1.
M.I. acknowledges support from the National Research Foundation of Korea (NRF) grant, No. 2017R1A3A3001362.
M.K. was supported by the National Research Foundation of Korea (NRF) grant funded by the Korean government (MSIT) (No. 2017R1C1B2002879).
M.D.J. acknowledges NSF grant AST-0618209.
SRON is financially supported by NWO, the
Netherlands Organization for Scientific Research.
B.C.K. is partially supported by the UC Center for Galaxy Evolution.
C.S.K. acknowledges the support of NSF grants AST-1814440 and AST-1515876.
D.C.L. acknowledges support from NSF grants AST-1009571 and AST-1210311.
P.L. acknowledges support from Fondecyt grant \#1120328.
A.P. acknowledges support from a NSF graduate fellowship and a
UCSB Dean's Fellowship.
C.S. acknowledges support from NOVA, the Nederlandse Onderzoekschool voor Astronomie.
J.S.S. acknowledges CNPq, National Council for Scientific and Technological
Development (Brazil) for partial support and The Ohio State University
for warm hospitality.
T.T. has been supported by NSF grant AST-1412315.
T.T. and B.C.K. acknowledge support from the Packard Foundation in the form of
a Packard Research Fellowship to T.T.
Support for A.V.F.'s group at U.C. Berkeley is provided by the TABASGO
Foundation, the Christopher R. Redlich Fund, and the Miller Institute for
Basic Research in Science (U.C. Berkeley).
M.V. and J.J.J. gratefully acknowledge support from the
Danish Council for Independent Research via grant no. DFF 4002-00275.
J.-H.W. acknowledges support by the National Research Foundation of Korea (NRF)
grant funded by the Korean government (No. 2010-0027910).
This research has made use of the NASA/IPAC Extragalactic Database (NED),
which is operated by the
Jet Propulsion Laboratory, California Institute of Technology,
under contract with the National Aeronautics and Space Administration.

\bibliographystyle{apj}
\bibliography{agn}

\setcounter{figure}{0}
\begin{figure*}[h]
\centering
    \includegraphics[angle=0,width=0.77\textwidth]{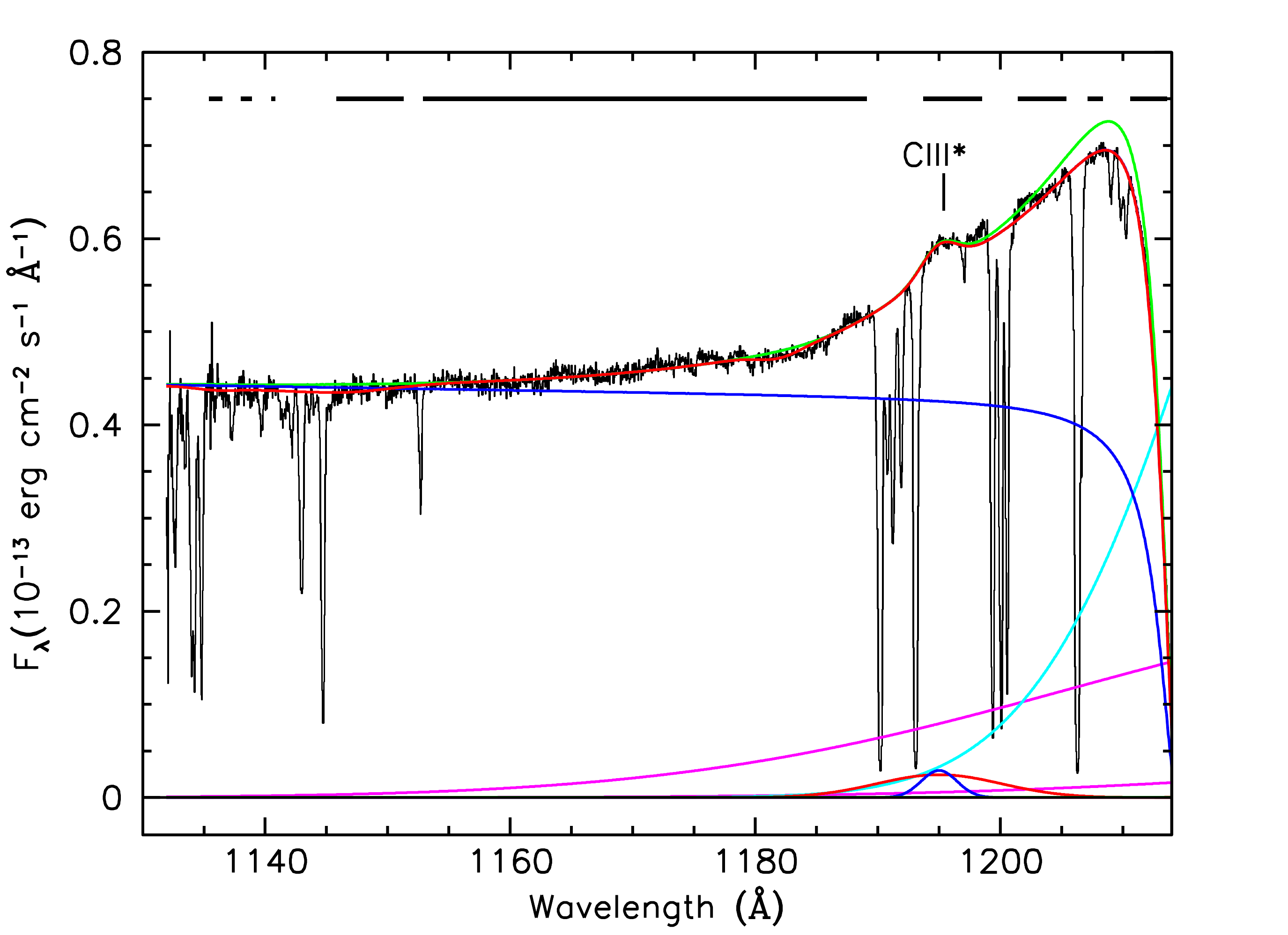}
    \caption{NGC 5548 Mean COS Spectrum, 1130--1214 \AA.
Best-fit model for the COS UV spectrum of NGC 5548 overlaid
        on the data (black).
        The total best-fit model is in red.
        The model corrected for absorption is in green.
	(Damped Galactic Ly$\alpha$ is not corrected.)
        The solid blue line is the continuum component absorbed by damped
        Galactic Ly$\alpha$.
	Emission lines from Table 1 are marked.
        Narrow emission-line region components are in green.
        Intermediate-line region components are in blue.
        Broad components are in red.
        Medium Broad components are in cyan.
        Very-broad components are in magenta.
	Thick horizontal bars across the top indicate wavelength intervals
	used for the fit.
	Fluxes and wavelengths are in the observed frame.
\label{fig:bestfit}}
\end{figure*}
\setcounter{figure}{0}
\begin{figure*}[h]
    \centering
    \includegraphics[angle=0,width=0.77\textwidth]{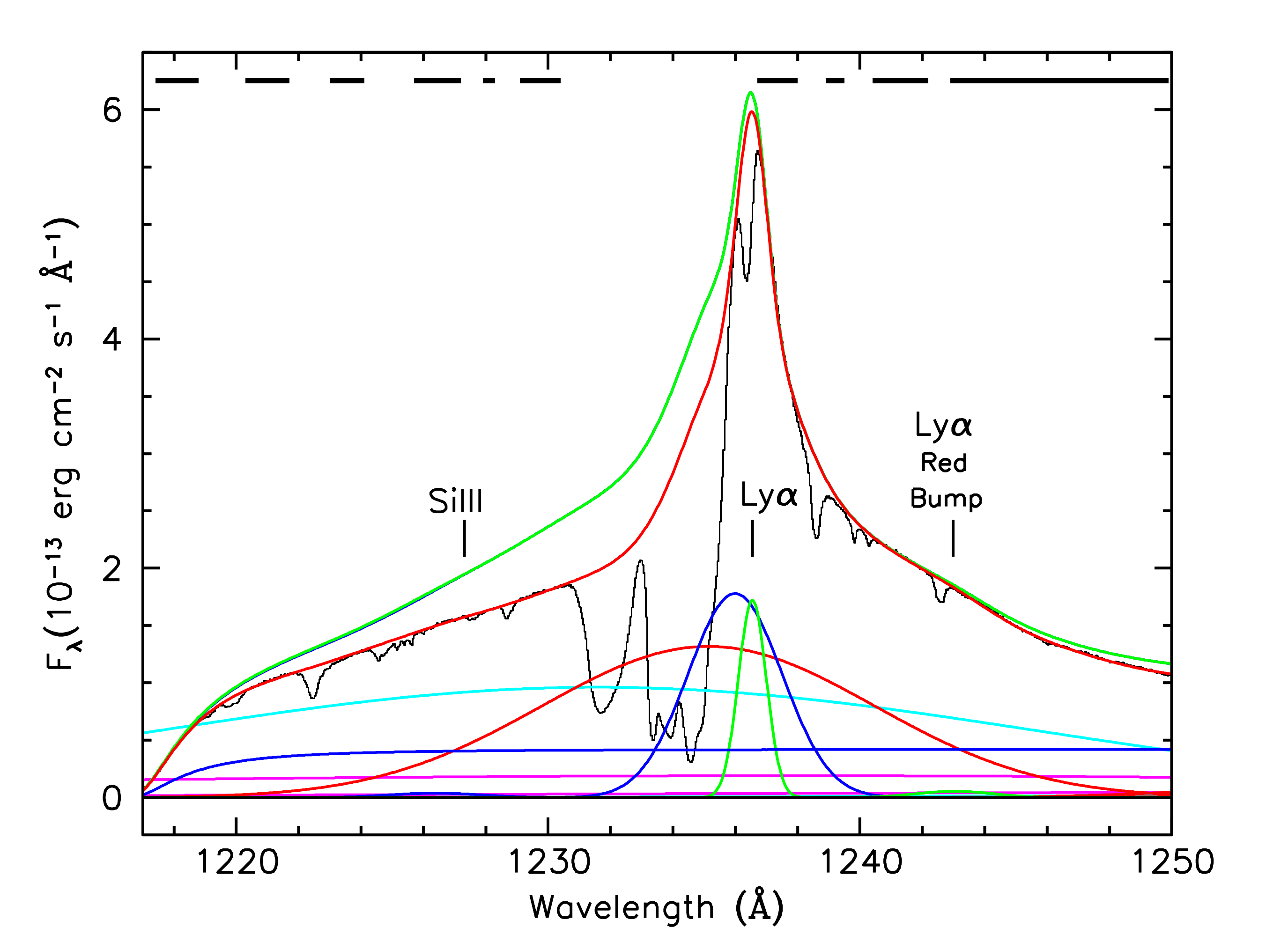}
    \caption{NGC 5548 Mean COS Spectrum, 1217--1250 \AA}\label{fig:1b}
\end{figure*}

\clearpage
\newpage
\setcounter{figure}{0}
\begin{figure*}[htb]
\centering
    \includegraphics[angle=0,width=0.77\textwidth]{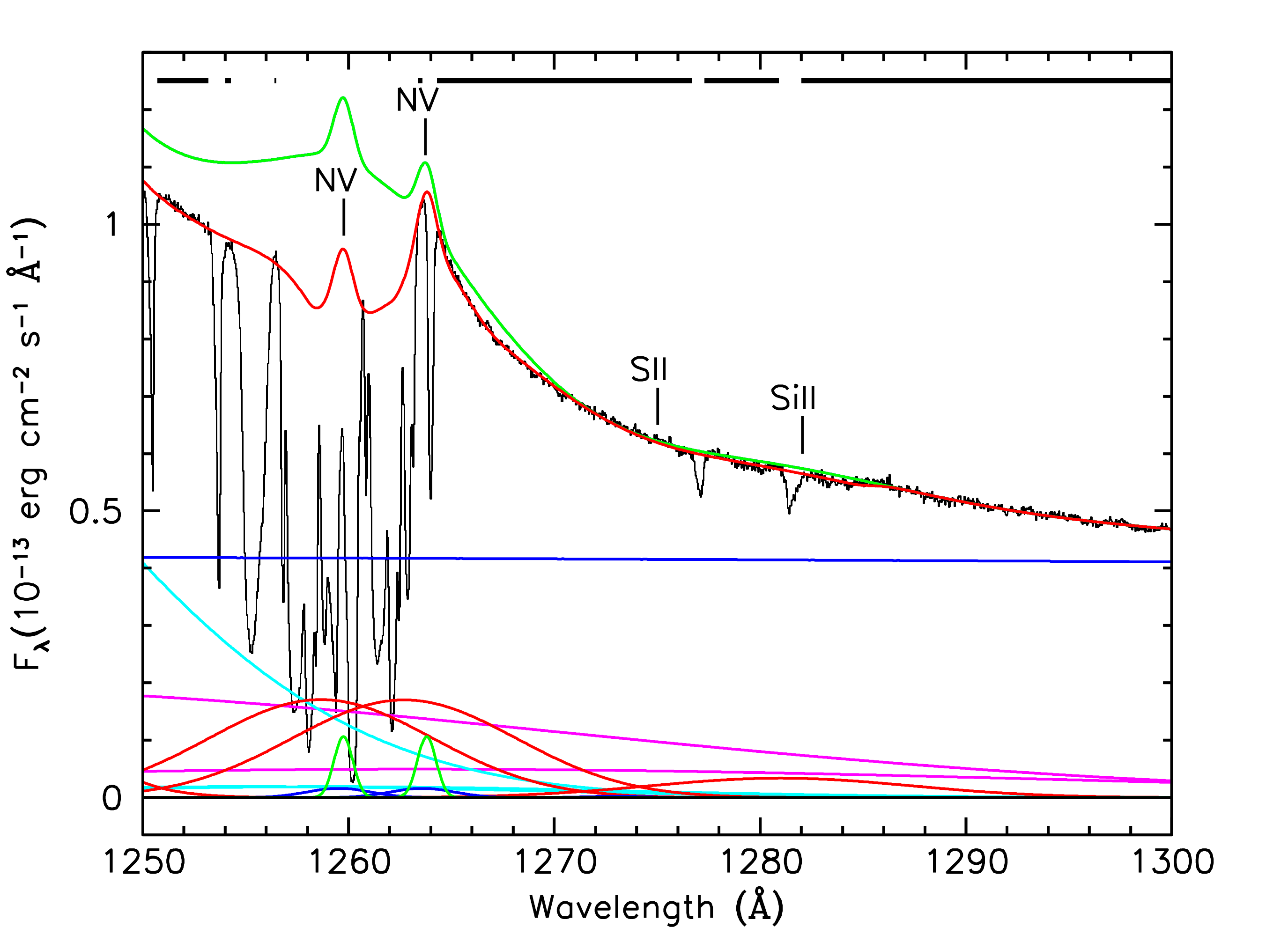}
    \caption{NGC 5548 Mean COS Spectrum, 1250--1300 \AA}\label{fig:1c}
\end{figure*}
\setcounter{figure}{0}
\begin{figure*}[h]
\centering
    \includegraphics[angle=0,width=0.77\textwidth]{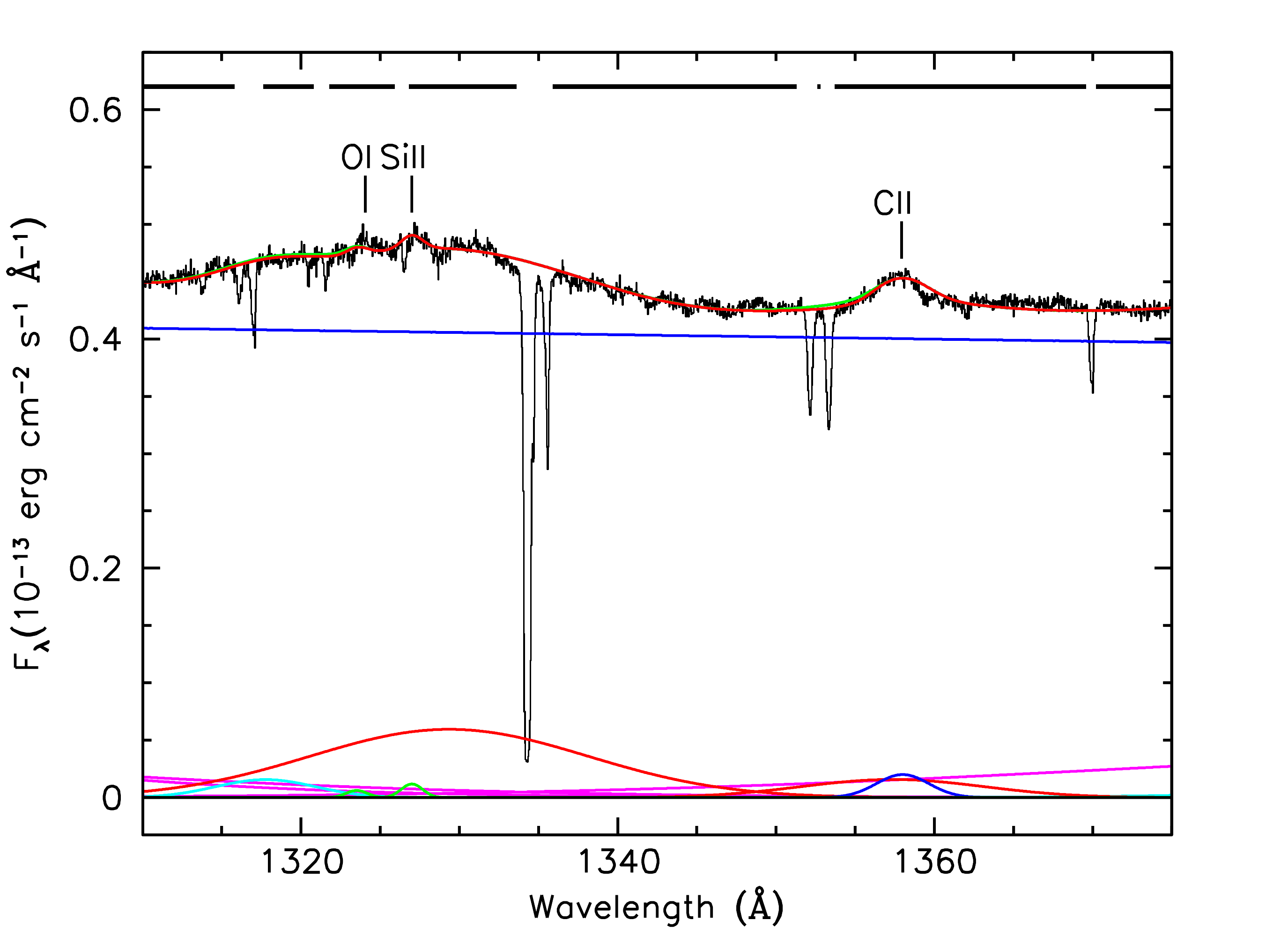}
    \caption{NGC 5548 Mean COS Spectrum, 1310-1375 \AA}\label{fig:1d}
\end{figure*}

\clearpage
\newpage
\setcounter{figure}{0}
\begin{figure*}[htb]
\centering
    \includegraphics[angle=0,width=0.77\textwidth]{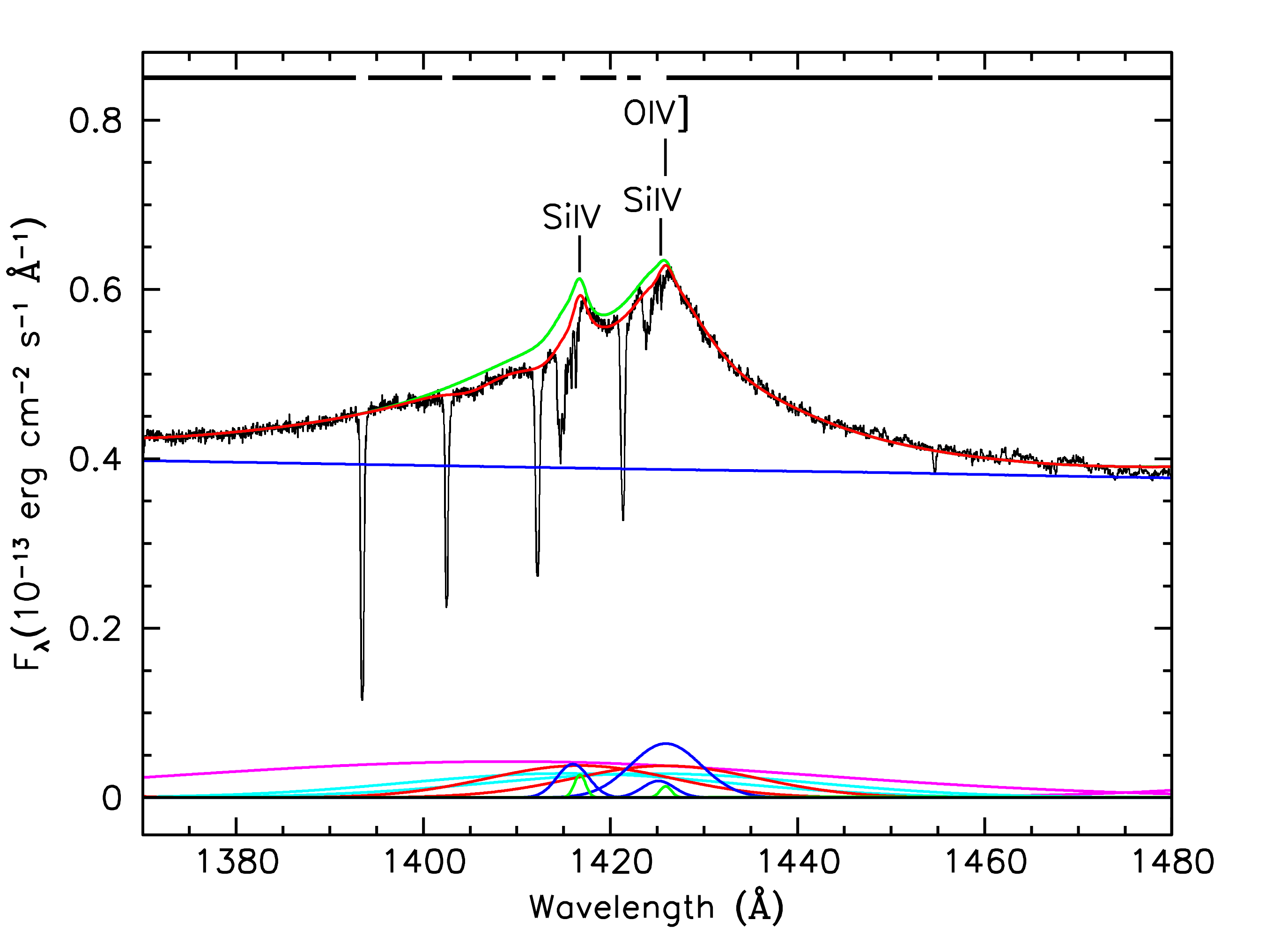}
    \caption{NGC 5548 Mean COS Spectrum, 1370--1480 \AA}\label{fig:1e}
\end{figure*}
\setcounter{figure}{0}
\begin{figure*}[h]
    \centering
    \includegraphics[angle=0,width=0.77\textwidth]{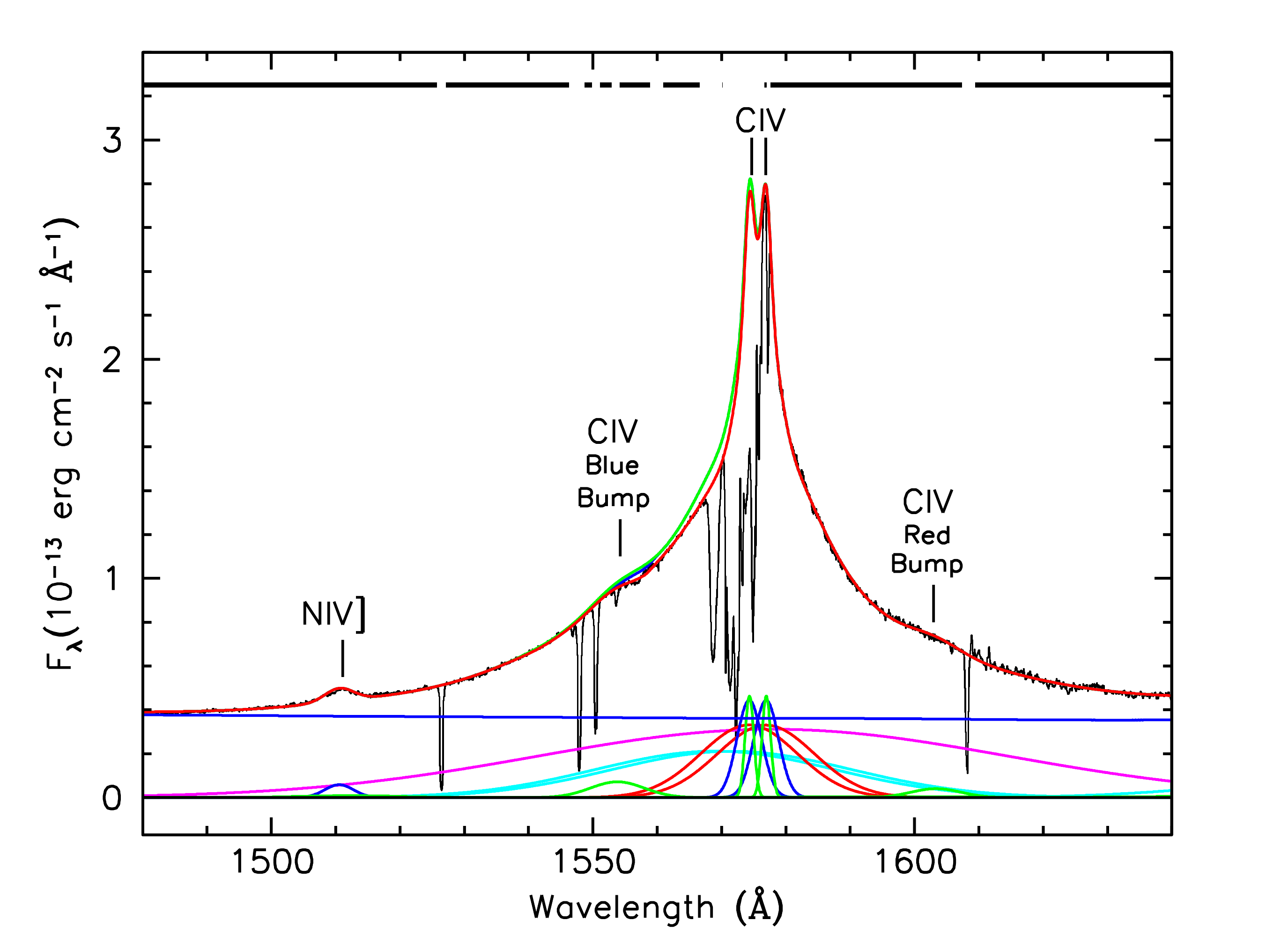}
    \caption{NGC 5548 Mean COS Spectrum, 1480--1640 \AA}\label{fig:1f}
\end{figure*}

\clearpage
\newpage
\setcounter{figure}{0}
\begin{figure*}[htb]
\centering
    \includegraphics[angle=0,width=0.77\textwidth]{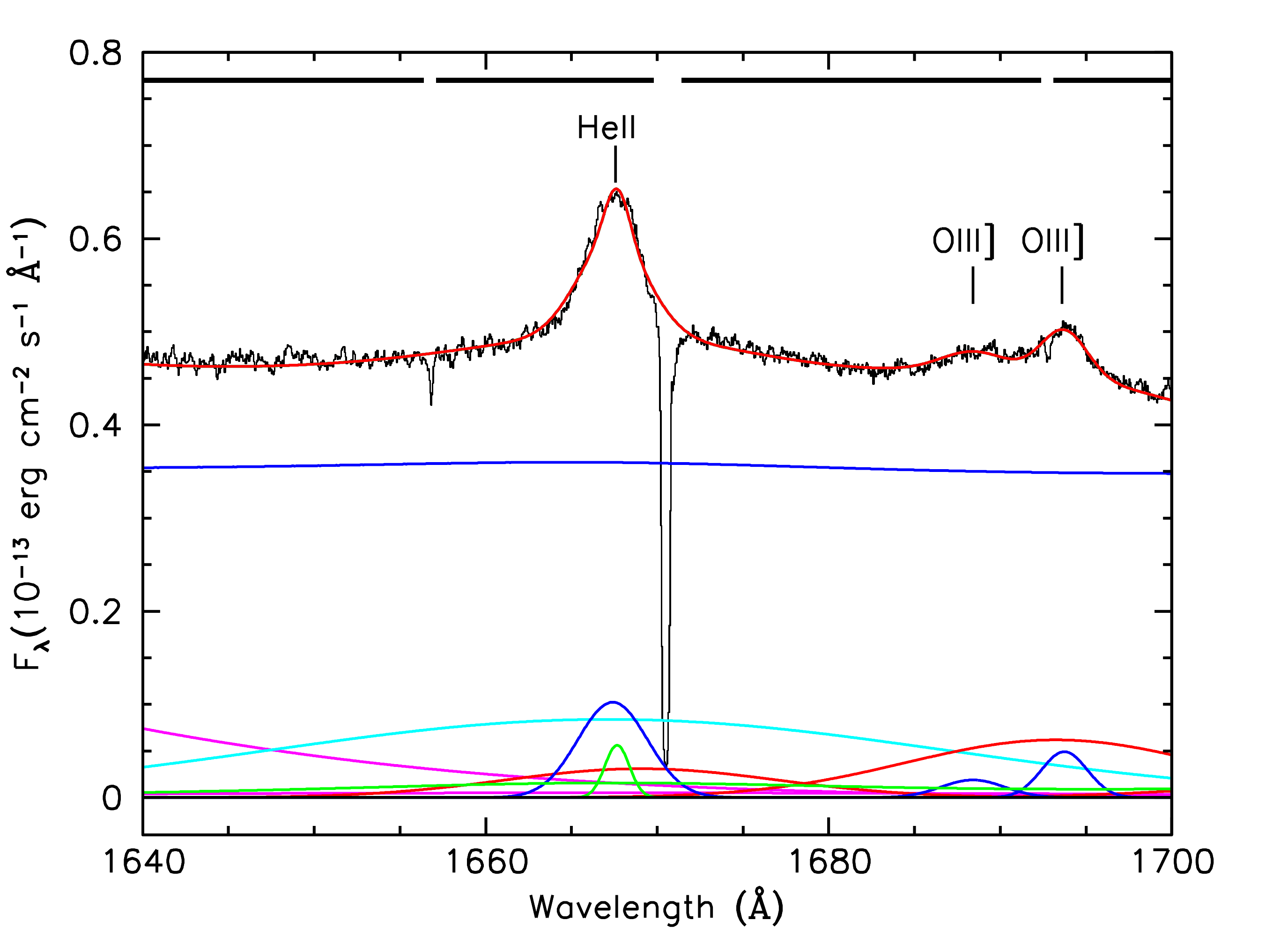}
    \caption{NGC 5548 Mean COS Spectrum, 1640--1700 \AA}\label{fig:1g}
\end{figure*}
\setcounter{figure}{0}
\begin{figure*}[h]
    \centering
    \includegraphics[angle=0,width=0.77\textwidth]{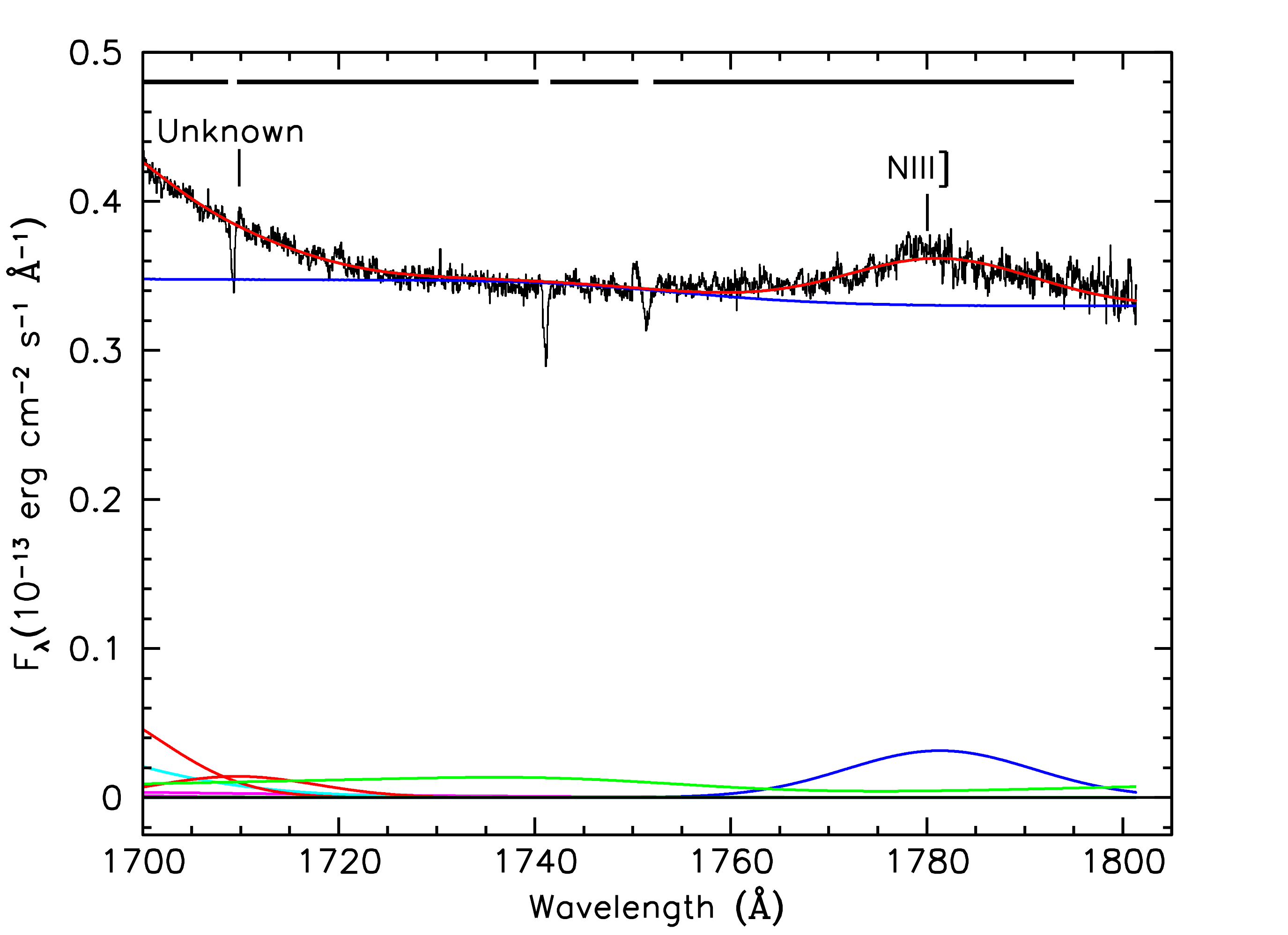}
    \caption{NGC 5548 Mean COS Spectrum, 1700--1805 \AA}\label{fig:1h}
\end{figure*}

\clearpage
\newpage
\setcounter{table}{0}
\startlongtable
\begin{deluxetable*}{l c c c c c c c}
\tablecaption{Model Parameters for the Mean Spectrum of NGC~5548\label{tab:parameters}}
\tablehead{\colhead{Feature} & \colhead{Component\tablenotemark{a}} & \colhead{$\rm \lambda_{vac}$} &  \colhead{$\rm \lambda_{obs}$} & \colhead{Flux\tablenotemark{b}} & \colhead{FWHM} & \colhead{Asymmetry\tablenotemark{c}} & \colhead{Freedom\tablenotemark{d}} \\
\colhead{} & \colhead{Number} & \colhead{(\AA)} & \colhead{(\AA)} &\colhead{} & \colhead{($\rm km~s^{-1}$)} &\colhead{ } & \colhead{Flags} \\
}
\startdata
Power Law $\rm F_{\lambda}$(1000) & 1 & 5.978\tablenotemark{e} & &  &  & & 0 \\
Power Law $\alpha$ & 1 &  0.840 &  & & & & 0 \\
E(B$-$V) & 2 & 0.017 &  & & & & $-$1 \\
$\rm R_V$ & 2 & 3.1 &  & & & & $-$1 \\
Galactic Ly$\alpha$ & 66 &  1215.67 & 1215.55 & 1.45\tablenotemark{f} & & & $-$1 \\
\ion{C}{3}*      & 3 & 1175.78 &  1195.01 &  1.06 &   861 &  1.00 & 0,0,0,$-1$ \\
\ion{C}{3}* Broad & 95 & 1175.78 &  1195.01 &  3.13 &  3000 &  1.00 & 0,0,0,$-1$ \\
\ion{Si}{3}     & 4 & 1206.50 &  1226.39 &  1.44 &   861 &  1.00 & 0,0,3,$-1$ \\
Ly$\alpha$ Narrow & 5 & 1215.67 &  1236.55 &  19.3 &   255 &  1.00 & 0,0,0,$-1$ \\
Ly$\alpha$ Intermediate & 6 & 1215.67 &  1236.00 &  67.3 &   861 &  1.00 & 0,0,0,$-1$ \\
Ly$\alpha$ Broad & 7  & 1215.67 &  1235.14 &   176.0 &  3047 &  1.00 & 0,0,0,$-1$ \\
Ly$\alpha$ Medium Broad & 8 & 1215.67 &  1231.59 &   340.0 &  8074 &  1.00 & 0,0,0,$-1$ \\
Ly$\alpha$ Very Broad & 9 & 1215.67 &  1237.64 &   154.0 & 18347 &  1.00 & 0,0,0,$-1$ \\
Ly$\alpha$ Red Bump & 21 & 1215.67 &  1243.04 &  1.73 &   713 &  1.00 & 0,0,0,$-1$ \\
\ion{N}{5} Narrow & 10 & 1238.82 &  1259.75 &  1.22 &   255 &  1.00 & 0,0,5,$-1$ \\
\ion{N}{5} Narrow & 11 & 1242.80 &  1263.80 &  1.22 &   255 &  1.00 & 10,10,10,$-1$ \\
\ion{N}{5} Intermediate & 12 & 1238.82 &  1259.53 &  0.62 &   861 &  1.00 & 6,0,6,$-1$ \\
\ion{N}{5} Intermediate & 13 & 1242.80 &  1263.58 &  0.62 &   861 &  1.00 & 12,12,12,$-1$ \\
\ion{N}{5} Broad & 14   & 1238.82 &  1258.66 &  23.3 &  3047 &  1.00 & 7,0,7,$-1$ \\
\ion{N}{5} Broad & 15   & 1242.80 &  1262.70 &  23.3 &  3047 &  1.00 & 14,14,14,$-1$ \\
\ion{N}{5} Medium Broad & 16 & 1238.82 &  1255.04 &  6.85 &  8074 &  1.00 & 8,0,8,$-1$ \\
\ion{N}{5} Medium Broad & 17 & 1242.80 &  1259.08 &  6.85 &  8074 &  1.00 & 16,16,16,$-1$ \\
\ion{N}{5} Very Broad & 18 & 1240.01 & 1263.24 &  40.8 & 18347 &  1.00 & 9,0,9,$-1$ \\
\ion{S}{2}  & 19   & 1253.81 &  1273.00 &  0.09 &   841 &  1.00 & 0,0,0,$-1$ \\
\ion{Si}{2}   & 20 & 1260.42 &  1281.00 &  5.39 &  3500 &  1.00 & 0,0,0,$-1$ \\
\ion{O}{1} Narrow & 48  & 1302.17 &  1323.52 &  0.10 &   359 &  1.00 & 0,0,0,$-1$ \\
\ion{Si}{2} Narrow & 49 & 1304.37 &  1327.01 &  0.20 &   359 &  1.00 & 0,0,48,$-1$ \\
\ion{Si}{2}+\ion{O}{1} Broad & 50 & 1303.27 & 1329.35 &  13.1 &  4659 &  1.00 & 0,0,0,$-1$ \\
Unknown      & 47 & $\ldots$ &  1317.76 &  1.24 &  1687 &  1.00 & 0,0,0,$-1$ \\
\ion{C}{2} Intermediate & 94 & 1334.53 &  1358.00 &  0.83 &   861 &  1.00 & 0,0,34,$-1$ \\
\ion{C}{2} Broad & 51 & 1334.53 &  1357.36 &  2.28 &  3000 &  1.00 & 0,0,$-1$,$-1$ \\
\ion{Si}{4} Narrow & 96 & 1393.76 &  1416.74 &  0.41 &   303 &  1.00 & 0,0,0,$-1$ \\
\ion{Si}{4} Narrow & 97 & 1402.77 &  1425.90 &  0.20 &   303 &  1.00 & 96,96,96,$-1$ \\
\ion{Si}{4} Intermediate & 23 & 1393.76 &  1415.97 &  1.67 &   842 &  1.00 & 0,0,0,$-1$ \\
\ion{Si}{4} Intermediate & 24 & 1402.77 &  1425.13 &  0.84 &   842 &  1.00 & 23,23,23,$-1$ \\
\ion{Si}{4} Broad & 25    & 1393.76 &  1416.74 &  9.29 &  4889 &  1.00 & 0,0,0,$-1$ \\
\ion{Si}{4} Broad & 26    & 1402.77 &  1425.90 &  9.29 &  4889 &  1.00 & 25,25,25,$-1$ \\
\ion{Si}{4} Medium Broad & 27 & 1393.76 &  1415.28 &  12.5 &  8704 &  1.00 & 0,0,0,$-1$ \\
\ion{Si}{4} Medium Broad & 28 & 1402.77 &  1424.43 &  12.5 &  8704 &  1.00 & 27,27,27,$-1$ \\
\ion{Si}{4} Very Broad & 29 & 1398.26 &  1409.20 &  37.1 & 18148 &  0.92 & 0,0,0,$-1$ \\
\ion{O}{4}] & 52 & 1400.00 &  1425.92 &  6.00 &  1858 &  1.00 & 0,0,0,$-1$ \\
\ion{N}{4}] Intermediate & 39 & 1486.00 &  1510.54 &  3.30 &  1050 &  1.00 & 0,0,0,$-1$ \\
\ion{N}{4}] Broad & 22   & 1486.00 &  1512.13 &  1.83 &  3800 &  1.00 & 0,0,$-1$,$-1$ \\
\ion{C}{4} Narrow & 30  & 1548.20 &  1574.34 &  7.86 &   303 &  1.00 & 0,0,0,$-1$ \\
\ion{C}{4} Narrow & 31  & 1550.77 &  1576.96 &  7.86 &   303 &  1.00 & 32,32,32,$-1$ \\
\ion{C}{4} Intermediate & 32 & 1548.20 &  1574.32 &  21.4 &   861 &  1.00 & 0,0,0,$-1$ \\
\ion{C}{4} Intermediate & 33 & 1550.77 &  1576.93 &  21.4 &   861 &  1.00 & 34,34,34,$-1$ \\
\ion{C}{4} Broad & 34 & 1548.20 &  1574.28 &  62.3 &  3366 &  1.00 & 0,0,0,$-1$ \\
\ion{C}{4} Broad & 35 & 1550.77 &  1576.90 &  62.3 &  3366 &  1.00 & 36,36,36,$-1$ \\
\ion{C}{4} Medium Broad & 36 & 1548.20 &  1569.00 &  98.5 &  8345 &  1.00 & 0,0,0,$-1$ \\
\ion{C}{4} Medium Broad & 37 & 1550.77 &  1571.61 &  98.5 &  8345 &  1.00 & 0,0,0,$-1$ \\
\ion{C}{4} Very Broad & 38 & 1550.48 &  1577.99 &   286.0 & 16367 &  1.00 & 0,0,0,$-1$ \\
\ion{C}{4} Blue Bump & 93 & 1550.48 &  1553.79 &  7.53 &  1900 &  1.00 & 0,0,0,$-1$ \\
\ion{C}{4} Red Bump & 74  & 1550.48 &  1603.34 &  3.95 &  1708 &  1.00 & 0,0,0,$-1$ \\
\ion{He}{2} Narrow & 40 & 1640.45 &  1667.66 &  1.01 &   303 &  1.00 & 0,0,30,$-1$ \\
\ion{He}{2} Intermediate & 41 & 1640.45 &  1667.40 &  5.22 &   861 &  1.00 & 0,0,32,$-1$ \\
\ion{He}{2} Broad & 42  & 1640.45 &  1669.07 &  6.18 &  3366 &  1.00 & 0,0,34,$-1$ \\
\ion{He}{2} Medium Broad & 43 & 1640.45 &  1667.10 &  41.5 &  8345 &  1.00 & 0,0,36,$-1$ \\
\ion{He}{2} Very Broad & 67 & 1640.45 &  1667.14 &  4.93 & 16367 &  1.00 & 0,0,38,$-1$ \\
\ion{O}{3}] Narrow & 92 & 1660.00 & 1686.82 & 0.87 & 771 & 1.00 & 0,47,0,$-1$ \\
\ion{O}{3}] Narrow & 44 & 1666.00 &  1693.74 &  1.64 &   555 &  1.00 & 0,0,0,$-1$ \\
\ion{O}{3}] Broad & 45 & 1663.00 &  1693.23 &  13.5 &  3639 &  1.00 & 0,0,0,$-1$ \\
Unknown & 68 & $\ldots$ &  1709.64 &  2.91 &  3366 &  1.00 & 0,0,0,$-1$ \\
\ion{N}{3}] Broad & 46    & 1750.00 &  1781.33 &  7.56 &  3801 &  1.00 & 0,0,0,$-1$ \\
\ion{Fe}{2} & 53   & $\ldots$ & $\ldots$ & 13.91\tablenotemark{g} & 1861.0 & $\ldots$ & $-1$,0,0,$-1$ \\
\ion{C}{3}* Broad Absorption & 54 & 1175.78 & 1194.00 & $-$0.01 &  3796 &  0.13 & 0,0,55,55 \\
Ly$\alpha$ Broad Absorption & 55 & 1215.67 &  1235.00 & $-$71.62 & 3796 &  0.13 & 0,0,0,0 \\
\ion{N}{5} Broad Absorption & 56 & 1238.82 &  1258.00 & $-$9.19 &  3796 &  0.13 & 0,0,55,55 \\
\ion{N}{5} Broad Absorption & 57 & 1242.80 &  1262.04 & $-$9.19 &  3796 &  0.13 & 56,56,55,55 \\
\ion{S}{2} Broad Absorption & 58 & 1253.81 & 1266.63 & $-$0.94 &  3796 &  0.13 & 0,0,55,55 \\
\ion{Si}{2} Broad Absorption & 59 & 1260.42 & 1284.47 & $-$0.93 &  3796 &  0.13 & 0,0,55,55 \\
\ion{Si}{2} Broad Absorption & 60 & 1304.37 & 1322.46 & $-$0.18 &  3796 &  0.13 & 0,0,55,55 \\
\ion{C}{2} Broad Absorption & 61 & 1334.53 &  1355.00 & $-$0.18 &  1761 &  0.26 & 0,0,0,$-1$ \\
\ion{Si}{4} Broad Absorption & 62 & 1393.76 & 1415.20 & $-$1.91 &  3796 &  0.13 & 0,0,55,55 \\
\ion{Si}{4} Broad Absorption & 63 & 1402.77 & 1424.36 & $-$1.91 &  3796 &  0.13 & 62,62,55,55 \\
\ion{C}{4} Broad Absorption & 64 & 1548.20 &  1572.02 & $-$5.18 &  2963 &  0.13 & 0,0,55,55 \\
\ion{C}{4} Broad Absorption & 65 & 1550.77 &  1574.63 & $-$5.18 &  2963 &  0.13 & 64,64,55,55 \\
\ion{P}{5} Broad Absorption & 69 & 1117.98 &  1136.00 & $-$0.23 &  3796 & 0.65 & 0,0,55,0 \\
\ion{P}{5} Broad Absorption & 70 & 1128.01 &  1146.19 & $-$1.06 &  3796 & 0.65 & 0,69,69,69 \\
\ion{Si}{2} Broad Absorption & 71 & 1190.20 &  1212.26 & $-$3.25 & 3796 & 0.13 & 0,55,55,55 \\
\ion{Si}{3} Broad Absorption & 72 & 1206.50 &  1225.63 & $-$0.94 & 3796 & 0.13 & 0,55,55,55 \\
\ion{Si}{2} Broad Absorption & 73 & 1526.71 & 1557.83 & $-$1.92 & 3796 & 0.13 & 0,55,55,55 \\
\ion{C}{3} Broad Absorption & 75 & 1175.78 & 1182.00 & $-$0.31 & 1000 & 1.00 & 0,$-1$,$-1$,$-1$ \\
Ly$\alpha$ Broad Absorption & 76 & 1215.67 &  1221.87 & $-$1.80 &  1000 &  1.00 & $-1$,0,$-1$,$-1$ \\
\ion{N}{5} Broad Absorption & 77 & 1238.82 &  1245.14 & $-$1.11 &  1000 &  1.00 & $-1$,0,$-1$,$-1$ \\
\ion{N}{5} Broad Absorption & 78 & 1242.80 &  1249.14 & $-$1.11 &  1000 &  1.00 & 77,$-1$,$-1$,$-1$ \\
\ion{S}{2} Broad Absorption & 79 & 1253.81 & 1260.20 & $-$7.13 &  1000 &  1.00 & 76,0,$-1$,$-1$ \\
\ion{Si}{2} Broad Absorption & 80 & 1260.42 & 1266.93 & $-$1.21 &  1000 &  1.00 & 76,0,$-1$,$-1$ \\
\ion{Si}{2} Broad Absorption & 81 & 1304.37 & 1310.65 &  0.00 &  1000 &  1.00 & 76,0,$-1$,$-1$ \\
\ion{C}{2} Broad Absorption & 82 & 1334.53 &  1341.81 & $-$0.02 &  1000 &  1.00 & 76,0,$-1$,$-1$ \\
\ion{Si}{4} Broad Absorption & 83 & 1393.76 & 1405.00 & $-$0.41 &  1000 &  1.00 & 76,0,$-1$,$-1$ \\
\ion{Si}{4} Broad Absorption & 84 & 1402.77 & 1414.09 & $-$0.41 &  1000 &  1.00 & 83,83,$-1$,$-1$ \\
\ion{C}{4} Broad Absorption & 85 & 1548.20 &  1556.09 & $-$0.74 &  1000 &  1.00 & 76,0,$-1$,$-1$ \\
\ion{C}{4} Broad Absorption & 86 & 1550.77 &  1558.68 & $-$0.74 &  1000 &  1.00 & 85,85,$-1$,$-1$ \\
\ion{P}{5} Broad Absorption & 87 & 1117.98 &  1136.00 & $-$0.09 &  1000 &  1.00 & 76,0,$-1$,$-1$ \\
\ion{P}{5} Broad Absorption & 88 & 1128.01 &  1146.19 & $-$0.01 &  1000 &  1.00 & 76,87,$-1$,$-1$ \\
\ion{Si}{2} Broad Absorption & 89 & 1190.42 & 1199.38 &  0.00 &  1000 &  1.00 & 76,0,$-1$,$-1$ \\
\ion{Si}{3} Broad Absorption & 90 & 1206.50 & 1212.61 &  0.00 &  1000 &  1.00 & 76,0,$-1$,$-1$ \\
\ion{Si}{2} Broad Absorption & 91 & 1526.71 & 1541.27 & $-$0.19 &  1000 &  1.00 & 76,0,$-1$,$-1$ \\
\enddata
\tablenotetext{a}{Component number in the {\tt specfit} model. Each component has multiple parameters as given in the table.}
\tablenotetext{b}{Flux is in $\rm 10^{-14}~erg~cm^{-2}~s^{-1}$.}
\tablenotetext{c}{Asymmetry is defined as the ratio of the half width at half maximum of the red side of an asymmetric Gaussian to the half width at half maximum of the blue side.}
\tablenotetext{d}{Key for Freedom Flags: For each component, flags are given for each parameter---$\lambda_{obs}$, Flux, FWHM, and Asymmetry. Values correspond to the following: $-1$ (Fixed), 0 (Free),
$\rm N > 0$ (Linked to corresponding parameter of Component N).}
\tablenotetext{e}{Flux is in $\rm 10^{-14}~erg~cm^{-2}~s^{-1}~\mbox{\AA}^{-1}$.}
\tablenotetext{f}{Column density is in $10^{20}~\rm cm^{-2}$.}
\tablenotetext{g}{The \ion{Fe}{2} model is normalized to $1.0 \times 10^{-16}~\rm erg~cm^{-2}~s^{-1}~\mbox{\AA}^{-1}$ at 1700 \AA.}
\end{deluxetable*}

\clearpage
\newpage
\startlongtable
\begin{deluxetable}{lcccc}
\tablecaption{Wavelength Intervals for Correcting Narrow Absorption Features in the NGC 5548 Spectra\label{tab:correction_intervals}}
\tablehead{
\colhead{Feature} & \colhead{$\lambda_o$} & 
\colhead{$z$} & \colhead{$\lambda_1$} & \colhead{$\lambda_2$}\\
\colhead{ } & \colhead{(\AA)} & \colhead{} & \colhead{(\AA)} & \colhead{(\AA)} }
\startdata
  Blue end &  1132.00 & 0.0      &  1124.00 &  1136.50 \\
 \ion{Fe}{3} 1 &  1122.52 & 0.017175 &  1136.99 &  1137.60 \\
\ion{Fe}{3}* 1 &  1124.87 & 0.017175 &  1139.31 &  1140.01 \\
   \ion{P}{5}r 1 &  1126.72 & 0.017175 &  1140.98 &  1145.65 \\
     \ion{P}{2} &  1152.82 & 0.0      &  1152.28 &  1153.32 \\
\ion{Si}{2} ISM &  1190.42 & 0.0      &  1189.53 &  1192.35 \\
\ion{Si}{2} ISM &  1193.29 & 0.0      &  1192.61 &  1193.84 \\
   Unknown &  1197.00 & 0.0      &  1196.56 &  1197.23 \\
\ion{N}{1} triplet &  1200.00 & 0.0      &  1198.92 &  1201.16 \\
  \ion{S}{3} 1 &  1190.20 & 0.017175 &  1205.50 &  1207.06 \\
\ion{S}{3}, \ion{Si}{1} &  1190.20 & 0.017175 &  1208.55 &  1210.61 \\
Geocoronal Ly$\alpha$ &  1215.67 & 0.0      &  1213.0 & 1219.0 \\
 \ion{Si}{2}* 1 &  1197.39 & 0.017175 &  1212.95 &  1213.49 \\
   Unknown &  1219.00 & 0.0      &  1218.79 &  1220.45 \\
 \ion{Si}{3} 1 &  1206.50 & 0.017175 &  1221.76 &  1223.08 \\
\ion{Si}{3} 3 5 &  1206.50 & 0.017175 &  1224.20 &  1225.84 \\
   Unknown &  1227.00 & 0.0      &  1227.34 &  1227.87 \\
   Unknown &  1228.50 & 0.0      &  1228.46 &  1229.05 \\
Ly$\alpha$ &  1215.67 & 0.017175 &  1230.40 &  1236.85 \\
\ion{N}{5} \ion{Mg}{2}  &  1238.82 & 0.0      &  1237.88 &  1240.50 \\
  \ion{N}{5} ISM &  1242.80 & 0.0      &  1242.23 &  1242.95 \\
 \ion{S}{2} ISM &  1250.58 & 0.0      &  1249.58 &  1250.81 \\
 \ion{S}{2} ISM &  1253.81 & 0.0      &  1253.21 &  1254.10 \\
   \ion{N}{5}b 1 &  1238.82 & 0.017175 &  1254.21 &  1256.37 \\
  \ion{N}{5}b 2 &  1238.82 & 0.017175 &  1256.50 &  1263.42 \\
   \ion{N}{5}r 5 &  1242.80 & 0.017175 &  1262.63 &  1263.56 \\
   \ion{N}{5}r 6 &  1242.80 & 0.017175 &  1263.62 &  1264.34 \\
  \ion{Si}{2} 1 &  1260.42 & 0.017175 &  1276.64 &  1277.50 \\
 \ion{Si}{2}* 1 &  1264.74 & 0.017175 &  1281.03 &  1282.20 \\
  \ion{O}{1} ISM &  1302.10 & 0.0      &  1300.71 &  1302.48 \\
 \ion{Si}{2} ISM &  1304.37 & 0.0      &  1303.13 &  1304.65 \\
 \ion{Si}{3}* 1 &  1296.73 & 0.017175 &  1313.50 &  1314.06 \\
 \ion{Si}{3}* 1 &  1298.96 & 0.017175 &  1315.66 &  1316.56 \\
 \ion{Ni}{2} ISM &  1317.22 & 0.0      &  1316.56 &  1317.30 \\
 \ion{Si}{3}* 1 &  1303.32 & 0.017175 &  1320.15 &  1320.80 \\
  \ion{Si}{2} 1 &  1304.37 & 0.017175 &  1321.10 &  1322.00 \\
 \ion{Si}{2}* 1 &  1309.28 & 0.017175 &  1326.18 &  1327.14 \\
   Unknown &  1326.53 & 0.0      &  1326.31 &  1326.83 \\
  \ion{C}{1} ISM &  1328.82 & 0.0      &  1328.22 &  1329.50 \\
 \ion{C}{2} ISM &  1334.53 & 0.0      &  1333.04 &  1336.18 \\
   \ion{C}{2} 1 &  1334.53 & 0.017175 &  1351.56 &  1352.75 \\
  \ion{C}{2}* 1 &  1335.71 & 0.017175 &  1352.75 &  1353.90 \\
 \ion{P}{3}* 1 &  1344.33 & 0.017175 &  1361.31 &  1362.61 \\
\ion{Ni}{2} ISM &  1370.13 & 0.0      &  1369.57 &  1370.39 \\
\ion{Si}{4}b ISM &  1393.76 & 0.0      &  1392.77 &  1394.17 \\
\ion{Si}{4}r ISM &  1402.77 & 0.0      &  1401.75 &  1403.20 \\
 \ion{Si}{4}b 1 &  1393.76 & 0.017175 &  1411.59 &  1412.80 \\
 \ion{Si}{4}b 3 &  1393.76 & 0.017175 &  1413.94 &  1417.11 \\
 \ion{Si}{4}r 1 &  1402.77 & 0.017175 &  1420.77 &  1422.14 \\
 \ion{Si}{4}r 3 &  1402.77 & 0.017175 &  1423.36 &  1426.35 \\
\ion{Ni}{2} ISM &  1454.84 & 0.0      &  1454.31 &  1455.10 \\
\ion{Si}{2} ISM &  1526.71 & 0.0      &  1525.80 &  1527.10 \\
  \ion{Si}{2} 1 &  1526.71 & 0.017175 &  1546.53 &  1547.21 \\
 \ion{C}{4}b ISM &  1548.19 & 0.0      &  1547.21 &  1548.76 \\
 \ion{C}{4}r ISM &  1550.77 & 0.0      &  1549.71 &  1551.16 \\
 \ion{Si}{2}* 1 &  1533.45 & 0.017175 &  1553.26 &  1554.18 \\
  \ion{C}{4}b 1 &  1548.19 & 0.017175 &  1567.02 &  1576.60 \\
  \ion{C}{4}r 6 &  1550.77 & 0.017175 &  1576.84 &  1577.51 \\
 \ion{Fe}{2} ISM &  1608.45 & 0.0      &  1607.54 &  1612.08 \\
  \ion{C}{1} ISM &  1656.93 & 0.0      &  1656.34 &  1657.13 \\
\ion{Al}{2} ISM &  1670.79 & 0.0      &  1669.77 &  1671.74 \\
  \ion{Al}{2} 1 &  1670.79 & 0.017175 &  1692.42 &  1693.15 \\
\ion{Ni}{2} ISM &  1709.60 & 0.0      &  1708.38 &  1709.62 \\
\ion{Ni}{2} ISM &  1741.55 & 0.0      &  1740.69 &  1741.72 \\
 \ion{Ni}{2} ISM &  1751.91 & 0.0      &  1750.84 &  1752.45 \\
\enddata
\end{deluxetable}

\begin{deluxetable*}{lhchhccccc}
\tablecaption{Modeled Continuum and Deblended Emission-Line Light Curves
for NGC 5548\label{tab:light_curves_int}}
\tablehead{
\colhead{HJD\tablenotemark{a}} & 
\nocolhead{$F_\lambda\left({\rm 1158 \mbox{\AA}}\right)$\tablenotemark{b}}&
\colhead{$F_\lambda\left({\rm 1367 \mbox{\AA}}\right)$\tablenotemark{b}}&
\nocolhead{$F_\lambda\left({\rm 1469 \mbox{\AA}}\right)$\tablenotemark{b}}&
\nocolhead{$F_\lambda\left({\rm 1745 \mbox{\AA}}\right)$\tablenotemark{b}}&
\colhead{$F$(Ly$\alpha$)\tablenotemark{c}}  &
\colhead{$F$(\ion{N}{5})\tablenotemark{c}} & 
\colhead{$F$(\siiv)\tablenotemark{c}} &
\colhead{$F$(\civ)\tablenotemark{c}} &
\colhead{$F$(\ion{He}{2})\tablenotemark{c}}}
\startdata
56690.6120 & 32.17 $\pm$ 0.36 & 29.71 $\pm$ 0.34 & 28.54 $\pm$ 0.42 & 25.27 $\pm$ 0.36 & 66.83 $\pm$ 0.76 & 10.14 $\pm$ 0.12 &  9.03 $\pm$ 0.10 & 63.16 $\pm$ 0.90 &  5.60 $\pm$ 0.08 \\
56691.5416 & 35.52 $\pm$ 0.40 & 32.13 $\pm$ 0.36 & 30.60 $\pm$ 0.44 & 26.50 $\pm$ 0.38 & 65.48 $\pm$ 0.73 &  9.80 $\pm$ 0.11 &  8.66 $\pm$ 0.10 & 62.46 $\pm$ 0.88 &  6.13 $\pm$ 0.09 \\
56692.3940 & 37.19 $\pm$ 0.42 & 34.50 $\pm$ 0.39 & 33.20 $\pm$ 0.48 & 29.53 $\pm$ 0.42 & 69.99 $\pm$ 0.78 &  7.89 $\pm$ 0.09 &  8.16 $\pm$ 0.09 & 61.50 $\pm$ 0.87 &  5.82 $\pm$ 0.08 \\
56693.3237 & 38.82 $\pm$ 0.44 & 34.69 $\pm$ 0.39 & 32.88 $\pm$ 0.48 & 28.10 $\pm$ 0.40 & 65.72 $\pm$ 0.73 &  9.00 $\pm$ 0.11 &  8.37 $\pm$ 0.09 & 62.78 $\pm$ 0.88 &  5.79 $\pm$ 0.08 \\
56695.2701 & 41.14 $\pm$ 0.46 & 36.78 $\pm$ 0.41 & 34.86 $\pm$ 0.50 & 29.80 $\pm$ 0.43 & 66.80 $\pm$ 0.74 & 10.09 $\pm$ 0.11 &  8.92 $\pm$ 0.10 & 62.69 $\pm$ 0.89 &  6.67 $\pm$ 0.10 \\
56696.2459 & 45.85 $\pm$ 0.51 & 40.76 $\pm$ 0.46 & 38.56 $\pm$ 0.55 & 32.77 $\pm$ 0.47 & 66.78 $\pm$ 0.74 &  9.72 $\pm$ 0.11 &  7.73 $\pm$ 0.09 & 61.19 $\pm$ 0.87 &  6.12 $\pm$ 0.09 \\
56697.3080 & 48.54 $\pm$ 0.54 & 42.63 $\pm$ 0.48 & 40.13 $\pm$ 0.57 & 33.66 $\pm$ 0.48 & 69.09 $\pm$ 0.77 &  8.74 $\pm$ 0.11 &  7.56 $\pm$ 0.08 & 63.59 $\pm$ 0.89 &  5.52 $\pm$ 0.08 \\
56698.3041 & 51.87 $\pm$ 0.58 & 44.49 $\pm$ 0.50 & 41.48 $\pm$ 0.59 & 33.93 $\pm$ 0.48 & 68.66 $\pm$ 0.76 &  9.78 $\pm$ 0.11 &  7.69 $\pm$ 0.09 & 64.00 $\pm$ 0.90 &  6.30 $\pm$ 0.09 \\
56699.2338 & 49.23 $\pm$ 0.55 & 43.14 $\pm$ 0.48 & 40.56 $\pm$ 0.58 & 33.95 $\pm$ 0.48 & 70.28 $\pm$ 0.78 &  9.73 $\pm$ 0.11 &  7.99 $\pm$ 0.09 & 64.20 $\pm$ 0.91 &  6.75 $\pm$ 0.10 \\
56700.2299 & 48.09 $\pm$ 0.54 & 42.82 $\pm$ 0.48 & 40.53 $\pm$ 0.58 & 34.50 $\pm$ 0.49 & 71.07 $\pm$ 0.79 &  8.47 $\pm$ 0.10 &  7.61 $\pm$ 0.09 & 64.28 $\pm$ 0.91 &  5.80 $\pm$ 0.08 \\
\enddata
\tablecomments{\label{tab:light_curves}
Modeled light curves are in the observed frame.
Flux uncertainties include both statistical and systematic errors.
Table 3 is published in its entirety in machine-readable format.
A portion is shown here for guidance regarding its form and content.
}
\tablenotetext{a}{Midpoint of the observation $({\rm HJD} - 2400000).$}
\tablenotetext{b}{Units of 10$^{-15}$ erg s$^{-1}$ cm$^{-2}$ \mbox{\AA}$^{-1}$.}
\tablenotetext{c}{Units of 10$^{-13}$ erg s$^{-1}$ cm$^{-2}$. }
\end{deluxetable*}

\clearpage
\newpage
\startlongtable
\begin{deluxetable*}{lccccccc}
\tablecaption{Properties of Narrow Intrinsic Absorption Lines in NGC 5548\label{tab:absorption_intervals}}
\tablehead{\colhead{Feature} & \colhead{$\lambda_o$\tablenotemark{a}} & 
\colhead{$\lambda_1$\tablenotemark{b}} & \colhead{$\lambda_2$\tablenotemark{c}} &
\colhead{EW\tablenotemark{d}} & \colhead{$\rm v$\tablenotemark{e}} & \colhead{$\rm c_f$\tablenotemark{f}} & \colhead{log $\rm N_{ion}$\tablenotemark{g}} \\
\colhead{ } & \colhead{(\AA)} & \colhead{(\AA)} & \colhead{(\AA)} & \colhead{($\rm km~s^{-1}$)} & \colhead{(\AA)} & \colhead{ } & \colhead{($\rm cm^{-2}$)} }
\startdata
\ion{P}{5}b 1 & 1117.98 & 1132.30 & 1133.05 & $-0.167 \pm 0.015$ & $-1207$ & 1.00 & $13.57^{+0.42}_{-0.20}$ \\
\ion{Fe}{3} 1 & 1122.52 & 1136.99 & 1137.60 & $-0.037 \pm 0.004$ & $-1205$ & 1.00 & $13.63^{+0.30}_{-0.27}$ \\
\ion{Fe}{3}* 1 & 1124.87 & 1139.31 & 1140.01 & $-0.031 \pm 0.004$ & $-1199$ & 1.00 & $13.73^{+0.20}_{-0.38}$ \\
\ion{P}{5}r 1\tablenotemark{h} & 1126.72 & 1142.49 & 1143.33 & $-0.200 \pm 0.004$ & $\phantom{0}-828$ & 1.00 & $13.57^{+0.42}_{-0.20}$ \\    
\ion{C}{3}* 1 & 1175.26 & 1190.54 & 1190.98 & $-0.122 \pm 0.002$ & $-1195$ & 0.65 & $13.86^{+0.17}_{-0.24}$ \\
\ion{C}{3}* 1 & 1175.71 & 1190.98 & 1191.59 & $-0.183 \pm 0.002$ & $-1189$ & 0.65 & $14.23^{+0.29}_{-0.28}$ \\
\ion{C}{3}* 1 & 1176.37 & 1191.59 & 1192.35 & $-0.130 \pm 0.002$ & $-1190$ & 0.65 & $14.49^{+0.17}_{-0.24}$ \\
\ion{S}{3} 1\tablenotemark{i} & 1190.20 & 1205.50 & 1207.06 & $-0.571 \pm 0.003$ & $-1093$ & 0.65 & $14.37^{+0.10}_{-0.38}$ \\ 
\ion{Si}{2} 1 & 1193.29 & 1208.55 & 1209.38 & $-0.023 \pm 0.002$ & $-1196$ & 0.34 & $12.99^{+0.25}_{-0.49}$ \\
\ion{S}{3}* 1 & 1194.06 & 1209.53 & 1210.05 & $-0.030 \pm 0.002$ & $-1187$ & 0.34 & $14.37^{+0.32}_{-0.27}$ \\
\ion{Si}{2}* 1 & 1194.50 & 1210.05 & 1210.61 & $-0.034 \pm 0.002$ & $-1194$ & 0.34 & $13.25^{+0.25}_{-0.25}$ \\
\ion{Si}{2}* 1 & 1197.39 & 1212.95 & 1213.49 & $-0.026 \pm 0.003$ & $-1166$ & 0.34 & $13.00^{+0.25}_{-1.0}$ \\
\ion{Si}{3} 1 & 1206.50 & 1221.76 & 1223.08 & $-0.124 \pm 0.003$ & $-1176$ & 0.30 & $13.45^{+0.06}_{-0.62}$ \\
\ion{Si}{3} 3 & 1206.50 & 1224.20 & 1225.07 & $-0.048 \pm 0.002$ & $\phantom{0}-638$ & 0.30 & $12.91^{+0.09}_{-0.61}$ \\
\ion{Si}{3} 5 & 1206.50 & 1225.07 & 1225.84 & $-0.028 \pm 0.002$ & $\phantom{0}-454$ & 0.30 & $12.67^{+0.15}_{-0.61}$ \\
Ly$\alpha$ 1 & 1215.67 & 0.017175 & 1232.98 & $-0.845 \pm 0.002$ & $-1156$ & 0.70 & $> 14.60$ \\
Ly$\alpha$ 2 & 1215.67 & 0.017175 & 1233.54 & $-0.301 \pm 0.001$ & $\phantom{0}-785$ & 0.90 & $> 14.07$ \\
Ly$\alpha$ 3 & 1215.67 & 0.017175 & 1234.21 & $-0.515 \pm 0.001$ & $\phantom{0}-659$ & 0.90 & $> 14.37$ \\
Ly$\alpha$ 4 & 1215.67 & 0.017175 & 1235.19 & $-0.791 \pm 0.001$ & $\phantom{0}-470$ & 0.95 & $> 14.56$ \\
Ly$\alpha$ 5 & 1215.67 & 0.017175 & 1236.10 & $-0.182 \pm 0.002$ & $\phantom{0}-299$ & 0.70 & $> 13.97$ \\
Ly$\alpha$ 6 & 1215.67 & 0.017175 & 1236.73 & $-0.105 \pm 0.002$ & $\phantom{0}\phantom{0}-22$ & 0.70 & $13.00^{+0.3}_{-1.0}$ \\
\ion{N}{5}b 1 & 1238.82 & 1254.21 & 1256.46 & $-0.776 \pm 0.003$ & $-1147$ & 0.80 & $> 14.94$ \\
\ion{N}{5}b 2 & 1238.82 & 1256.46 & 1256.95 & $-0.115 \pm 0.001$ & $\phantom{0}-791$ & 0.70 & $> 14.14$ \\
\ion{N}{5}b 3 & 1238.82 & 1256.95 & 1257.82 & $-0.637 \pm 0.001$ & $\phantom{0}-648$ & 0.90 & $> 14.87$ \\
\ion{N}{5}b 4 & 1238.82 & 1257.82 & 1258.57 & $-0.555 \pm 0.001$ & $\phantom{0}-478$ & 0.95 & $> 14.78$ \\
\ion{N}{5}r 2 & 1242.80 & 1260.70 & 1260.98 & $-0.051 \pm 0.001$ & $\phantom{0}-789$ & 0.70 & $> 13.93$ \\
\ion{N}{5}r 3 & 1242.80 & 1260.98 & 1261.88 & $-0.536 \pm 0.001$ & $\phantom{0}-645$ & 0.90 & $> 14.98$ \\
\ion{N}{5}r 4 & 1242.80 & 1261.88 & 1262.63 & $-0.482 \pm 0.001$ & $\phantom{0}-475$ & 0.95 & $> 14.96$ \\
\ion{N}{5}r 5 & 1242.80 & 1262.63 & 1263.56 & $-0.230 \pm 0.002$ & $\phantom{0}-302$ & 0.70 & $> 14.74$ \\
\ion{N}{5}r 6 & 1242.80 & 1263.67 & 1264.34 & $-0.095 \pm 0.002$ & $\phantom{0}\phantom{0}-35$ & 0.70 & $> 14.25$ \\
\ion{Si}{2} 1 & 1260.42 & 1276.54 & 1277.50 & $-0.046 \pm 0.003$ & $-1193$ & 0.28 & $13.07^{+0.17}_{-0.57}$ \\
\ion{Si}{2}* 1 & 1264.74 & 1280.93 & 1282.20 & $-0.062 \pm 0.003$ & $-1166$ & 0.28 & $13.08^{+0.45}_{-0.58}$ \\
\ion{Si}{3}* 1 & 1296.73 & 1313.40 & 1314.60 & $-0.001 \pm 0.004$ & $-1600$ & 0.90 & $13.00^{+0.3}_{-1.0}$ \\
\ion{Si}{3}* 1 & 1298.96 & 1315.66 & 1316.56 & $-0.026 \pm 0.003$ & $-1185$ & 0.90 & $13.00^{+0.30}_{-1.0}$ \\
\ion{Si}{3}* 1 & 1303.32 & 1320.15 & 1320.80 & $-0.008 \pm 0.003$ & $-1195$ & 0.90 & $12.70^{+0.30}_{-0.7}$ \\
\ion{Si}{2} 1 & 1304.37 & 1321.10 & 1322.00 & $-0.015 \pm 0.003$ & $-1175$ & 0.92 & $13.09^{+0.15}_{-0.59}$ \\
\ion{Si}{2}* 1 & 1309.28 & 1326.18 & 1327.14 & $-0.020 \pm 0.003$ & $-1179$ & 0.92 & $13.09^{+0.15}_{-0.59}$ \\
\ion{C}{2} 1 & 1334.53 & 1351.56 & 1352.75 & $-0.078 \pm 0.003$ & $-1185$ & 1.00 & $13.61^{+0.38}_{-0.11}$ \\ 
\ion{C}{2}* 1 & 1335.71 & 1352.75 & 1353.90 & $-0.100 \pm 0.003$ & $-1194$ & 1.00 & $13.61^{+0.38}_{-0.11}$ \\
\ion{P}{3}* 1 & 1344.33 & 1361.31 & 1362.61 & $-0.009 \pm 0.004$ & $-1270$ & 1.00 & $13.20^{+0.76}_{-0.21}$ \\
\ion{Si}{4}b 1 & 1393.76 & 1411.59 & 1412.93 & $-0.260 \pm 0.003$ & $-1181$ & 0.50 & $14.11^{+0.11}_{-0.06}$ \\
\ion{Si}{4}b 3 & 1393.76 & 1413.78 & 1415.20 & $-0.178 \pm 0.003$ & $\phantom{0}-644$ & 1.00 & $13.32^{+0.05}_{-0.05}$ \\
\ion{Si}{4}b 4 & 1393.76 & 1415.20 & 1416.04 & $-0.071 \pm 0.003$ & $\phantom{0}-455$ & 1.00 & $12.90^{+0.10}_{-0.09}$ \\
\ion{Si}{4}b 5 & 1393.76 & 1416.04 & 1416.47 & $-0.040 \pm 0.002$ & $\phantom{0}-305$ & 1.00 & $12.66^{+0.07}_{-0.07}$ \\
\ion{Si}{4}b 6 & 1393.76 & 1416.47 & 1416.89 & $-0.021 \pm 0.002$ & $\phantom{0}-228$ & 1.00 & $12.37^{+0.11}_{-0.15}$ \\
\ion{Si}{4}r 1 & 1402.77 & 1420.68 & 1422.05 & $-0.210 \pm 0.003$ & $-1179$ & 0.50 & $14.18^{+0.04}_{-0.13}$ \\
\ion{Si}{4}r 3 & 1402.77 & 1423.22 & 1424.39 & $-0.069 \pm 0.003$ & $\phantom{0}-631$ & 1.00 & $13.18^{+0.11}_{-0.12}$ \\   
\ion{Si}{4}r 4 & 1402.77 & 1424.39 & 1425.24 & $-0.021 \pm 0.003$ & $\phantom{0}-451$ & 1.00 & $12.67^{+0.20}_{-0.38}$ \\
\ion{Si}{4}r 5 & 1402.77 & 1425.24 & 1425.67 & $-0.016 \pm 0.002$ & $\phantom{0}-295$ & 1.00 & $12.54^{+0.15}_{-0.21}$ \\
\ion{Si}{4}r 6 & 1402.77 & 1425.67 & 1426.09 & $-0.011 \pm 0.002$ & $\phantom{0}-216$ & 1.00 & $12.38^{+0.20}_{-0.25}$ \\
\ion{Si}{2} 1 & 1526.71 & 1546.53 & 1547.21 & $-0.023 \pm 0.003$ & $-1190$ & 0.70 & $13.08^{+0.16}_{-0.58}$ \\
\ion{Si}{2}* 1 & 1533.45 & 1553.26 & 1554.18 & $-0.039 \pm 0.003$ & $-1193$ & 0.70 & $13.30^{+0.14}_{-0.16}$ \\
\ion{C}{4}b 1 & 1548.19 & 1567.55 & 1570.26 & $-0.724 \pm 0.004$ & $-1162$ & 0.70 & $> 14.62$ \\
\ion{C}{4}b 2 & 1548.19 & 1570.26 & 1570.84 & $-0.169 \pm 0.002$ & $\phantom{0}-793$ & 0.70 & $> 14.05$ \\
\ion{C}{4}b 4 & 1548.19 & 1571.87 & 1572.91 & $-0.704 \pm 0.001$ & $\phantom{0}-478$ & 0.95 & $> 14.54$ \\
\ion{C}{4}r 4 & 1550.77 & 1574.40 & 1575.51 & $-0.551 \pm 0.002$ & $\phantom{0}-478$ & 0.95 & $> 14.62$ \\
\ion{C}{4}r 5 & 1550.77 & 1575.51 & 1576.73 & $-0.131 \pm 0.003$ & $\phantom{0}-322$ & 0.50 & $> 14.32$ \\
\ion{C}{4}r 6 & 1550.77 & 1576.84 & 1577.51 & $-0.005 \pm 0.003$ & $\phantom{0}\phantom{0}\phantom{0}\phantom{-}0$ & 0.50 & $12.50^{+0.5}_{-0.5}$ \\
\enddata
\tablenotetext{a}{Rest wavelength.}
\tablenotetext{b}{Starting wavelength for equivalent width integration.}
\tablenotetext{c}{Ending wavelength for equivalent width integration.}
\tablenotetext{d}{Equivalent width of the absorption feature.}
\tablenotetext{e}{Velocity of the feature relative to the systemic redshift of the host galaxy, $z = 0.017175$.}
\tablenotetext{f}{Covering factor of the absorption feature.}
\tablenotetext{g}{Column density of the absorption feature.}
\tablenotetext{h}{Blended with Galactic \ion{Fe}{2} $\lambda1143$.}
\tablenotetext{i}{Blended with Galactic \ion{Si}{3} $\lambda1206$.}
\end{deluxetable*}

\newpage
\begin{deluxetable*}{lchhchhhchchhhhhhhhhhhhhhhhhhhhhhhhhhhhhhhhhhhhh}
\tablecaption{Light Curves for Absorption Lines in NGC 5548\label{tab:absorp_lc_sample}}
\tablehead{\colhead{THJD\tablenotemark{a}} &
 \colhead{EW(Ly$\alpha$ Broad)} & 
 \nocolhead{EW(\ion{N}{5} Broad)} & 
 \nocolhead{EW(\ion{Si}{4} Broad)} & 
 \colhead{EW(\ion{C}{4} Broad)} & 
 \nocolhead{EW(\ion{Al}{2} $\lambda 1670$,\#1)} &
 \nocolhead{EW(\ion{C}{3}* $\lambda 1175$,\#1)} &
 \nocolhead{EW(\ion{C}{3}* $\lambda 1176$,\#1)} &
 \colhead{EW(\ion{C}{2} $\lambda 1334$,\#1)} &
 \nocolhead{EW(\ion{C}{2}* $\lambda 1335$,\#1)} &
 \colhead{EW(\ion{C}{4} $\lambda 1548$,\#1)} &
 \nocolhead{EW(\ion{C}{4} $\lambda 1548$,\#2)} &
 \nocolhead{EW(\ion{C}{4} $\lambda 1548$,\#4)} &
 \nocolhead{EW(\ion{C}{4} $\lambda 1550$,\#4)} &
 \nocolhead{EW(\ion{C}{4} $\lambda 1550$,\#5)} &
 \nocolhead{EW(\ion{C}{4} $\lambda 1550$,\#6)} &
 \nocolhead{EW(Ly$\alpha$,\#1)} &
 \nocolhead{EW(Ly$\alpha$,\#2)} &
 \nocolhead{EW(Ly$\alpha$,\#3)} &
 \nocolhead{EW(Ly$\alpha$,\#4)} &
 \nocolhead{EW(Ly$\alpha$,\#5)} &
 \nocolhead{EW(Ly$\alpha$,\#6)} &
 \nocolhead{EW(\ion{N}{5} $\lambda 1238$,\#1)} &
 \nocolhead{EW(\ion{N}{5} $\lambda 1238$,\#2)} &
 \nocolhead{EW(\ion{N}{5} $\lambda 1238$,\#3)} &
 \nocolhead{EW(\ion{N}{5} $\lambda 1238$,\#4)} &
 \nocolhead{EW(\ion{N}{5} $\lambda 1242$,\#2)} &
 \nocolhead{EW(\ion{N}{5} $\lambda 1242$,\#3)} &
 \nocolhead{EW(\ion{N}{5} $\lambda 1242$,\#4)} &
 \nocolhead{EW(\ion{N}{5} $\lambda 1242$,\#5)} &
 \nocolhead{EW(\ion{N}{5} $\lambda 1242$,\#6)} &
 \nocolhead{EW(\ion{P}{5} $\lambda 1126$,\#1)} &
 \nocolhead{EW(\ion{S}{3} $\lambda 1190$,\#1)} &
 \nocolhead{EW(\ion{S}{3}* $\lambda 1194$,\#1)} &
 \nocolhead{EW(\ion{Si}{3} $\lambda 1206$,\#1)} &
 \nocolhead{EW(\ion{Si}{3} $\lambda 1206$,\#3)} &
 \nocolhead{EW(\ion{Si}{2} $\lambda 1193$,\#1)} &
 \nocolhead{EW(\ion{Si}{2} $\lambda 1260$,\#1)} &
 \nocolhead{EW(\ion{Si}{2} $\lambda 1304$,\#1)} &
 \nocolhead{EW(\ion{Si}{2} $\lambda 1526$,\#1)} &
 \nocolhead{EW(\ion{Si}{2}* $\lambda 1194$,\#1)} &
 \nocolhead{EW(\ion{Si}{2}* $\lambda 1264$,\#1)} &
 \nocolhead{EW(\ion{Si}{2}* $\lambda 1309$,\#1)} &
 \nocolhead{EW(\ion{Si}{2}* $\lambda 1533$,\#1)} &
 \nocolhead{EW(\ion{Si}{4} $\lambda 1393$,\#1)} &
 \nocolhead{EW(\ion{Si}{4} $\lambda 1393$,\#3)} &
 \nocolhead{EW(\ion{Si}{4} $\lambda 1402$,\#1)} &
 \nocolhead{EW(\ion{Si}{4} $\lambda 1402$,\#3)}
 \\
 \colhead{(d)}  &
 \colhead{(\AA)} &
 \nocolhead{(\AA)} &
 \nocolhead{(\AA)} &
 \colhead{(\AA)} &
 \nocolhead{(\AA)} &
 \nocolhead{(\AA)} &
 \nocolhead{(\AA)} &
 \colhead{(\AA)} &
 \nocolhead{(\AA)} &
 \colhead{(\AA)} &
 \nocolhead{(\AA)} &
 \nocolhead{(\AA)} &
 \nocolhead{(\AA)} &
 \nocolhead{(\AA)} &
 \nocolhead{(\AA)} &
 \nocolhead{(\AA)} &
 \nocolhead{(\AA)} &
 \nocolhead{(\AA)} &
 \nocolhead{(\AA)} &
 \nocolhead{(\AA)} &
 \nocolhead{(\AA)} &
 \nocolhead{(\AA)} &
 \nocolhead{(\AA)} &
 \nocolhead{(\AA)} &
 \nocolhead{(\AA)} &
 \nocolhead{(\AA)} &
 \nocolhead{(\AA)} &
 \nocolhead{(\AA)} &
 \nocolhead{(\AA)} &
 \nocolhead{(\AA)} &
 \nocolhead{(\AA)} &
 \nocolhead{(\AA)} &
 \nocolhead{(\AA)} &
 \nocolhead{(\AA)} &
 \nocolhead{(\AA)} &
 \nocolhead{(\AA)} &
 \nocolhead{(\AA)} &
 \nocolhead{(\AA)} &
 \nocolhead{(\AA)} &
 \nocolhead{(\AA)} &
 \nocolhead{(\AA)} &
 \nocolhead{(\AA)} &
 \nocolhead{(\AA)} &
 \nocolhead{(\AA)} &
 \nocolhead{(\AA)} &
 \nocolhead{(\AA)} &
 \nocolhead{(\AA)}}
\startdata
56690.6120 &  3.978 $\pm$ 0.089 &  3.532 $\pm$ 0.171 &  1.445 $\pm$ 0.209 &  1.620 $\pm$ 0.055 & $-$0.059 $\pm$ 0.027 & $-$0.215 $\pm$ 0.021 & $-$0.180 $\pm$ 0.024 & $-$0.140 $\pm$ 0.035 & $-$0.143 $\pm$ 0.035 & $-$0.794 $\pm$ 0.016 & $-$0.168 $\pm$ 0.007 & $-$0.686 $\pm$ 0.006 \\
56691.5416 &  3.725 $\pm$ 0.087 &  3.194 $\pm$ 0.146 &  1.225 $\pm$ 0.203 &  1.460 $\pm$ 0.055 & $-$0.040 $\pm$ 0.026 & $-$0.200 $\pm$ 0.020 & $-$0.147 $\pm$ 0.023 & $-$0.105 $\pm$ 0.034 & $-$0.128 $\pm$ 0.034 & $-$0.757 $\pm$ 0.016 & $-$0.184 $\pm$ 0.007 & $-$0.708 $\pm$ 0.006 \\
56692.3940 &  3.744 $\pm$ 0.067 &  3.009 $\pm$ 0.153 &  1.694 $\pm$ 0.198 &  1.388 $\pm$ 0.053 & $-$0.051 $\pm$ 0.026 & $-$0.210 $\pm$ 0.018 & $-$0.113 $\pm$ 0.022 & $-$0.065 $\pm$ 0.033 & $-$0.102 $\pm$ 0.035 & $-$0.670 $\pm$ 0.016 & $-$0.160 $\pm$ 0.007 & $-$0.683 $\pm$ 0.006 \\
56693.3237 &  3.545 $\pm$ 0.068 &  3.095 $\pm$ 0.159 &  2.211 $\pm$ 0.202 &  1.397 $\pm$ 0.055 &  0.008 $\pm$ 0.026 & $-$0.216 $\pm$ 0.019 & $-$0.147 $\pm$ 0.022 & $-$0.101 $\pm$ 0.033 & $-$0.104 $\pm$ 0.032 & $-$0.760 $\pm$ 0.016 & $-$0.182 $\pm$ 0.007 & $-$0.685 $\pm$ 0.006 \\
56695.2701 &  3.490 $\pm$ 0.099 &  3.198 $\pm$ 0.155 &  1.579 $\pm$ 0.192 &  1.550 $\pm$ 0.054 & $-$0.018 $\pm$ 0.025 & $-$0.209 $\pm$ 0.018 & $-$0.170 $\pm$ 0.020 & $-$0.082 $\pm$ 0.031 & $-$0.112 $\pm$ 0.031 & $-$0.735 $\pm$ 0.015 & $-$0.155 $\pm$ 0.007 & $-$0.690 $\pm$ 0.006 \\
56696.2459 &  3.186 $\pm$ 0.061 &  2.900 $\pm$ 0.139 &  1.362 $\pm$ 0.181 &  1.153 $\pm$ 0.052 &  0.002 $\pm$ 0.025 & $-$0.181 $\pm$ 0.017 & $-$0.141 $\pm$ 0.020 & $-$0.027 $\pm$ 0.031 & $-$0.111 $\pm$ 0.029 & $-$0.680 $\pm$ 0.016 & $-$0.167 $\pm$ 0.007 & $-$0.704 $\pm$ 0.006 \\
56697.3080 &  3.559 $\pm$ 0.060 &  3.030 $\pm$ 0.149 &  1.293 $\pm$ 0.177 &  1.646 $\pm$ 0.053 &  0.023 $\pm$ 0.024 & $-$0.202 $\pm$ 0.016 & $-$0.132 $\pm$ 0.019 & $-$0.022 $\pm$ 0.030 & $-$0.128 $\pm$ 0.029 & $-$0.604 $\pm$ 0.016 & $-$0.143 $\pm$ 0.007 & $-$0.685 $\pm$ 0.006 \\
56698.3041 &  3.340 $\pm$ 0.058 &  2.960 $\pm$ 0.141 &  1.190 $\pm$ 0.173 &  1.179 $\pm$ 0.050 &  0.024 $\pm$ 0.023 & $-$0.172 $\pm$ 0.016 & $-$0.129 $\pm$ 0.019 & $-$0.014 $\pm$ 0.029 & $-$0.058 $\pm$ 0.031 & $-$0.624 $\pm$ 0.015 & $-$0.164 $\pm$ 0.006 & $-$0.696 $\pm$ 0.006 \\
56699.2338 &  3.369 $\pm$ 0.058 &  2.853 $\pm$ 0.160 &  1.214 $\pm$ 0.173 &  1.291 $\pm$ 0.050 & $-$0.003 $\pm$ 0.024 & $-$0.180 $\pm$ 0.017 & $-$0.089 $\pm$ 0.020 & $-$0.076 $\pm$ 0.029 & $-$0.033 $\pm$ 0.029 & $-$0.619 $\pm$ 0.015 & $-$0.138 $\pm$ 0.007 & $-$0.688 $\pm$ 0.006 \\
56700.2299 &  3.261 $\pm$ 0.058 &  2.742 $\pm$ 0.146 &  0.628 $\pm$ 0.153 &  1.061 $\pm$ 0.049 & $-$0.036 $\pm$ 0.024 & $-$0.178 $\pm$ 0.017 & $-$0.151 $\pm$ 0.019 & $-$0.039 $\pm$ 0.030 & $-$0.010 $\pm$ 0.030 & $-$0.701 $\pm$ 0.015 & $-$0.162 $\pm$ 0.006 & $-$0.706 $\pm$ 0.005 \\
\enddata
\tablenotetext{a}{Midpoint of each observation (Heliocentric Julian Date$-$2400000).}
\tablecomments{
Tabulated equivalent widths are in the observed frame.
Enumerations following the line designations in the column headings refer to the
narrow absorption line components as numbered in
Figure \ref{fig:narrow_abs_norm}.
Table 5 is published in its entirety in machine-readable format.
A portion is shown here for guidance regarding its form and content.
}
\end{deluxetable*}

\end{document}